\documentclass[aps,prl,twocolumn,nofootinbib,superscriptaddress,preprintnumbers,floatfix,10pt]{revtex4-1}
\pdfoutput=1
\usepackage[bookmarks=false]{hyperref}
\usepackage{graphicx}
\usepackage{amsmath}
\usepackage{xspace}
\usepackage[normalem]{ulem}

\newcommand{\refcite}[1]{Ref.~\cite{#1}}
\newcommand{\refscite}[1]{Refs.~\cite{#1}}
\newcommand{\eq}[1]{Eq.~\eqref{eq:#1}}
\newcommand{\eqs}[2]{Eqs.~\eqref{eq:#1} and \eqref{eq:#2}}
\renewcommand{\sec}[1]{Sec.~\ref{sec:#1}}
\newcommand{\secs}[2]{Secs.~\ref{sec:#1} and \ref{sec:#2}}

\newcommand{\fig}[1]{Fig.~\ref{fig:#1}}
\newcommand{\figs}[2]{Figs.~\ref{fig:#1} and \ref{fig:#2}}
\newcommand{\tab}[1]{Table~\ref{tab:#1}}

\newcommand{\abs}[1]{\lvert#1\rvert}
\newcommand{\Abs}[1]{\bigl\lvert#1\bigr\rvert}
\newcommand{\ord}[1]{\mathcal{O}(#1)}

\newcommand{\ORD}[1]{\mathcal{O}\biggl(#1\biggr)}

\newcommand{\df}{\mathrm{d}}
\newcommand{\img}{\mathrm{i}}
\newcommand{\bC}{\overline{C}{}}

\newcommand{\la}{\lambda}

\newcommand{\Tau}{\mathcal{T}}

\newcommand{\MeV}{\mathrm{MeV}}
\newcommand{\GeV}{\mathrm{GeV}}

\newcommand{\nn}{\nonumber}

\newcommand{\lqcd}{\Lambda_\mathrm{QCD}}
\newcommand{\Eg}{\ensuremath{E_\gamma}\xspace}

\newcommand{\cF}{\mathcal{F}}

\newcommand{\cH}{\mathcal{H}}
\newcommand{\hP}{\widehat{P}}

\newcommand{\hm}{\widehat{m}}
\newcommand{\hla}{\widehat{\lambda}}

\newcommand{\Cincl}{\ensuremath{C_7^{\rm incl}}\xspace}
\newcommand{\mbS}{\ensuremath{m_b^{1S}}\xspace}
\newcommand{\mtbar}{\overline{m}_b}
\newcommand{\mbbar}{\overline{m}_b}
\newcommand{\mcbar}{\overline{m}_c}
\newcommand{\cC}{\mathcal{C}}

\newcommand{\MSbar}{$\overline{\text{MS}}$\xspace}

\newcommand{\Li}{\mathrm{Li}}
\newcommand{\incl}{\mathrm{incl}}
\newcommand{\eff}{\mathrm{eff}}
\newcommand{\weak}{\mathrm{weak}}

\newcommand{\sing}{\mathrm{s}}
\newcommand{\nons}{\mathrm{ns}}

\newcommand{\inte}[1]{\int\! \df #1 \,}

\newcommand{\one}{{(1)}}
\newcommand{\two}{{(2)}}
\newcommand{\cG}{\mathcal{G}}

\newcommand{\babar}{\mbox{\ensuremath{{\displaystyle B}\!{\scriptstyle A}{\displaystyle B}\!{\scriptstyle AR}}}\xspace}

\makeatletter
\g@addto@macro\bfseries{\boldmath}
\makeatother

\allowdisplaybreaks[2]

\providecommand{\papersection[1]}{\paragraph{#1}}

\begin{document}


\preprint{\vbox{\hbox{DESY 20-115}\hbox{MIT-CTP 5220}}}

\title{Precision Global Determination of the $B\to X_s\gamma$ Decay Rate}

\author{Florian U.~Bernlochner}
\affiliation{Physikalisches Institut, Rheinische Friedrich-Wilhelms-Universit\"at Bonn, D-53113 Bonn, Germany}

\author{Heiko Lacker}
\affiliation{Humboldt University of Berlin, D-12489 Berlin, Germany}

\author{Zoltan Ligeti}
\affiliation{Lawrence Berkeley National Laboratory, University of California, Berkeley, California 94720, USA}

\author{\\Iain W.~Stewart}
\affiliation{Center for Theoretical Physics, Massachusetts Institute of Technology, Cambridge, Massachusetts 02139, USA}

\author{Frank J.~Tackmann}
\affiliation{Deutsches Elektronen-Synchrotron (DESY), D-22607 Hamburg, Germany}

\author{Kerstin Tackmann}
\affiliation{Deutsches Elektronen-Synchrotron (DESY), D-22607 Hamburg, Germany}

\collaboration{SIMBA Collaboration}
\noaffiliation

\date{July 8, 2020}

\begin{abstract}

We perform the first global fit to inclusive $B\to X_s\gamma$ measurements using
a model-independent treatment of the nonperturbative $b$-quark distribution
function, with next-to-next-to-leading logarithmic resummation and
$\ord{\alpha_s^2}$ fixed-order contributions.
The normalization of the
$B\to X_s\gamma$ decay rate, given by $\abs{\Cincl V_{tb} V_{ts}^*}^2$,
is sensitive to physics beyond the Standard Model (SM). We determine
$\abs{\Cincl V_{tb} V_{ts}^*} = (14.77 \pm 0.51_{\rm fit} \pm 0.59_{\rm theory}
\pm 0.08_{\rm param})\times 10^{-3}$,
in good agreement with the SM prediction, and the $b$-quark mass
$\mbS = (4.750 \pm 0.027_{\rm fit} \pm 0.033_{\rm theory} \pm 0.003_{\rm param})\,\GeV$.
Our results suggest that the uncertainties in the extracted $B\to X_s\gamma$ rate
have been underestimated by up to a factor of two, leaving more room for beyond-SM
contributions.

\end{abstract}

\maketitle

\papersection{Introduction}

The flavor-changing neutral-current $b\to s\gamma$ transition is well known for
its high sensitivity to contributions beyond the Standard Model (SM).
The main goal of our global analysis of the $B\to X_s\gamma$ decay rate
is to obtain a precise constraint on the short-distance physics it probes,
which can then be compared to predictions in the SM~\cite{Bertolini:1986th,
Grinstein:1987vj, Misiak:2006zs, Misiak:2015xwa} or beyond~\cite{Grinstein:1987pu,
Hou:1987kf, Misiak:2017bgg}. In our approach, this amounts to extracting a
precise value of the Wilson coefficient $\abs{C_7^\incl}$ from the measurements.

Since $b\to s\gamma$ is a two-body decay at tree level, the photon energy
spectrum, $\df\Gamma/\df\Eg$, peaks only a few hundred MeV below the kinematic limit
$\Eg \lesssim m_B/2$. In this peak region, the measurements are most precise,
but the theory predictions depend on a nonperturbative function, $\cF(k)$,
often called the shape function, which encodes the distribution of the
residual momentum $k$ of the $b$-quark in a $B$ meson~\cite{Neubert:1993um, Bigi:1993ex}.
A key aspect of our analysis is a model-independent treatment of $\cF(k)$ based
on expanding it in a suitable basis~\cite{Ligeti:2008ac}.
This approach can incorporate any given shape function model,
by using it as the generating function for the basis expansion, and thus goes
beyond existing approaches that use specific models~\cite{Benson:2004sg,
Lange:2005yw, Andersen:2005mj, Gambino:2007rp, Aglietti:2007ik}.

While $\cF(k)$ primarily affects the shape of the decay spectrum, its normalization
is determined by $\abs{C_7^{\rm incl}}^2$, up to small corrections. Thus, with our
treatment of $\cF(k)$, we can perform a global fit to
the measurements of $\df\Gamma/\df\Eg$, including the precisely measured peak
region, to simultaneously determine $\cF(k)$ and a precise value of
$\abs{C_7^\incl}$.
Our global fit is the first to exploit the full available experimental
information on the spectrum~\cite{Aubert:2007my,
Limosani:2009qg, Lees:2012ufa, Lees:2012wg}, together with the most precise
theoretical knowledge of its perturbative contributions.
This provides a more robust approach than the current method of using
theoretical predictions for the $B\to X_s\gamma$ rate with a fixed cut at
$\Eg > 1.6\,\GeV$~\cite{Misiak:2015xwa} and corresponding
extrapolated measurements~\cite{Amhis:2019ckw}.

\papersection{The $B\to X_s\gamma$ Spectrum}

Using SCET~\cite{Bauer:2000ew, Bauer:2000yr, Bauer:2001ct, Bauer:2001yt},
we can write the photon energy spectrum in a factorized form,
\begin{align} \label{eq:dGamma}
\frac{\df\Gamma}{\df E_\gamma} 
&= 2\Gamma_0\, \frac{(2E_\gamma)^3}{\hm_b^3}
\biggl[ \int\! \df k\, \widehat P(k)\, \cF(m_B - 2\Eg - k)
\nn \\ & \quad
+ \frac{1}{\hm_b} \sum_{a} (\widehat P_a\otimes g_a)(m_B-2\Eg)
\biggr]
\,,\end{align}
where
\begin{equation}
\Gamma_0 = \frac{G_F^2\, \hm_b^5}{8\pi^3}\, \frac{\alpha_\mathrm{em}}{4\pi}\, \abs{V_{tb} V_{ts}^*}^2
\,,\end{equation}
and $\widehat m_b$ denotes a short-distance $b$-quark mass, for which we use the
$1S$ scheme~\cite{Hoang:1998ng, Hoang:1998hm, Hoang:1999zc}.

The first term in \eq{dGamma} is the dominant contribution, where $\cF(k)$
contains the leading nonperturbative shape function plus a
combination of subleading shape functions specific for $B\to X_s\gamma$.
The function $\hP(k)$ encodes the perturbatively calculable $b\to s\gamma$
spectrum, with $k \sim m_b - 2\Eg$. It receives contributions from
different operators in the effective electroweak Hamiltonian,
\begin{align} \label{eq:Phat}
\widehat P(k) &= \Abs{C_7^\incl}^2\, \Bigl[W_{77}^\sing(k) + W_{77}^\nons(k) \Bigr]
\\\nn &\quad
+ 2\, \mathrm{Re}\big(C_7^\incl\big) \sum_{i\neq7} \cC_i\, W^\nons_{7i}(k)
+ \sum_{i,j\neq7} \cC_i \cC_j\, W^\nons_{ij}(k)
\,.\end{align}
Here, $W_{77}^\sing(k)$ contains the universal ``singular'' contributions
proportional to $\alpha_s^i \ln^j(k/m_b)/k$ and $\alpha_s^i\,
\delta(k)$, which dominate in the peak region where $k$ is
small~\cite{supplement}.
It is included following \refcite{Ligeti:2008ac} to NNLL$'$ order, which
includes next-to-next-to-leading-logarithmic (NNLL) resummation and all singular
terms at $\ord{\alpha_s^2}$~\cite{Korchemsky:1992xv, Gardi:2005yi,
Bauer:2000yr, Blokland:2005uk, Bauer:2003pi, Becher:2005pd, Becher:2006qw,
Balzereit:1998yf, Neubert:2004dd, Fleming:2007xt}

The coefficient $C_7^\incl$ is dominated by the Wilson coefficient $\bC_7(\mu)$
in the electroweak Hamiltonian,
\begin{equation} \label{eq:C7inclshort}
C_7^{\rm incl} = \bC_7(\mu) + \sum_{i\neq7} \bC_i(\mu) \bigl[ s_i(\mu,\hm_b)
 + r_i(\mu,\hm_b,\hm_c)\bigr]
.\!\end{equation}
The $s_i$ terms are defined to cancel the $\mu$ dependence of $\bC_7(\mu)$ and
to satisfy $s_i(\hm_b,\hm_b)=0$.  The $\bC_i\, r_i$ terms
contain all virtual corrections proportional to $\bC_{i \neq 7}$ that give rise to singular contributions.
In particular, they contain the sizable corrections from virtual $c\bar c$ loops, and the resulting sensitivity to the charm quark mass, $\hm_c$, which are
a dominant theory uncertainty in the decay rate.
Since in our approach these contributions are included in $C_7^{\rm incl}$, they only affect its
SM prediction, but not
its determination from the experimental data.
The results of Refs.~\cite{Misiak:2006zs, Misiak:2015xwa, Misiak:2006ab, Czakon:2015exa}
yield the NNLO SM prediction~\cite{*[{See the supplemental material at the end of the paper}] [{}] supplement},
\begin{equation} \label{eq:C7inclnnlo}
  \Abs{C_7^{\rm incl}}_{\rm SM}
   = 0.3624 \pm  0.0128_{c\bar c}  \pm 0.0080_{\rm scale}
\,.\end{equation}

The remaining $W_{ij}^\nons(k)$ terms in \eq{Phat} are ``nonsingular'' contributions
with $\cC_i = \bC_i(\hm_b)$~\cite{supplement}.
They start at $\ord{\alpha_s}$ and are suppressed by at least $k/m_b$ relative to
$W_{77}^\sing(k)$, and are therefore subleading in the peak region.
They are included to full $\ord{\alpha_s^2}$ for $ij = 77,\, 78$~\cite{Melnikov:2005bx, Ewerth:2008nv, Asatrian:2010rq},
while the remaining ones are known and included to $\ord{\alpha_s^2\beta_0}$~\cite{Ligeti:1999ea, Ferroglia:2010xe, Misiak:2010tk}.
Since $W_{77}^\sing(k)$ dominates in the peak region, the normalization of the spectrum is determined by $\abs{C_7^{\rm incl}}$, enabling its precise extraction.

The second term in \eq{dGamma} is subdominant, and describes so-called resolved
and unresolved contributions, where $\widehat P_a$ are perturbative coefficients
starting at $\ord{\alpha_s}$, and the $g_a$ are additional subleading shape
functions~\cite{Lee:2004ja}. The uncertainties from resolved contributions are
much smaller than suggested by earlier estimates~\cite{Benzke:2010js}, and are
not relevant at the current level of accuracy~\cite{supplement} (see also
\refcite{Gunawardana:2019gep}). The only marginally relevant contribution is
related to the known $\ord{1/\hm_c^2}$ correction to the total
rate~\cite{Voloshin:1996gw, Ligeti:1997tc, Grant:1997ec}, and is included in our
analysis via a subleading $\ord{\lqcd^2}$ shape function $g_{27}(k)$.

The nonperturbative shape function $\cF(k)$ is dominated by the leading-order shape function, so we assume it is positive.
We introduce a dimension-1 parameter $\lambda$, and expand $\cF(k)$ as~\cite{Ligeti:2008ac},
\begin{equation} \label{eq:expdef}
\cF(k)
 = \frac{1}{\lambda}\,\biggl[ \sum_{n = 0}^\infty \tilde c_n\, f_n\Bigl( \frac{k}{\lambda} \Bigr) \biggr]^2
\,,\end{equation}
where $f_n(x)$ are a suitably chosen complete set of orthonormal functions on $[0,\infty)$.
The normalization condition $\int_0^\infty \df k\, \cF(k) = 1$ implies 
\begin{equation} \label{eq:cnconstraint}
\sum_{n=0}^\infty \tilde c_n^2 = 1
\,.\end{equation}

In practice, the expansion for $\cF(k)$ must be truncated at a finite order $N$.
Therefore, the form of $\cF(k)$ used for the fit is given by the following approximation
\begin{equation} \label{eq:truncatedexpansion}
\cF(k) = \sum_{m, n = 0}^N\! c_m\, c_n\, F_{mn}(k)
\,,\end{equation}
where 
\begin{equation} \label{eq:Fmn_def}
F_{mn}(k) = \frac{1}{\lambda}\, f_m\Bigl(\frac{k}{\lambda}\Bigr)\,
  f_n\Bigl(\frac{k}{\lambda}\Bigr)
\,.\end{equation}
The effect of the truncation in \eq{truncatedexpansion} is approximated by the
modified coefficients $c_n$, which differ from the $\tilde c_n$ in \eq{truncatedexpansion}.
In particular, we always keep the normalization of $\cF(k)$ exact by enforcing
\begin{equation} \label{eq:fitnorm}
\sum_{n=0}^N c_n^2 = 1
\,.\end{equation}

Using the expansion for $\cF(k)$ in \eq{truncatedexpansion} we get
\begin{align} \label{eq:expand}
\frac{\df\Gamma}{\df E_\gamma}
&= 16\Gamma_0\frac{E_\gamma^3}{ \hm_b^3}
  \sum_{m, n = 0}^N\!\! c_m c_n\! \int\! \df k\, \widehat P(k)\,
   F_{mn}\bigl(m_B-2E_\gamma - k\bigr)
\nn \\ & \quad
 + 16\Gamma_0\frac{E_\gamma^3}{\hm_b^3}\, \frac{1}{\hm_b^2} \int\! \df k\,
   \widehat P_{27}(k)\, g_{27}(m_B - 2E_\gamma - k)
\nn\\
&\equiv N_s \sum_{m, n = 0}^N c_m\, c_n\, \frac{\df\Gamma_{77,mn}}{\df E_\gamma}
 \quad + \dotsb
\,.\end{align}
Here, $N_s = \abs{C_7^\incl V_{tb} V_{ts}^*}^2 \hm_b^2$, and
\eq{expand} defines $\df\Gamma_{77,mn}/\df E_\gamma$, which we precompute
from \eq{Phat}. The ellipses denote subleading terms not proportional to
$\abs{C_7^\incl}^2$, which are also written in terms of $N_s$ and $c_n$ as
explained in \cite{supplement}.
Then, $N_s$ and the $c_n$ are fitted from the measured spectra,
with the uncertainties and correlations in the measurements captured in the
uncertainties and correlations of the fit parameters. Using the moment
relations for $\cF(k)$~\cite{supplement}, we obtain $C_7^\incl$ and $\hm_b$,
as well as the heavy-quark parameters $\widehat \lambda_1$ and $\widehat \rho_1$
from the fitted $N_s$ and $c_n$.
The other coefficients $\cC_{i\ne 7}$ are fixed to their SM values~\cite{supplement}.
Of these, only $\cC_1$ and $\cC_2$ are numerically relevant, which are known to be
SM dominated, while $\cC_8$, which is sensitive to new physics, gives only
a small contribution.
We use input values for $\widehat\lambda_2$ and $\widehat \rho_2$,
which are obtained from the $B$ and $D$ meson mass splittings~\cite{supplement}.

\papersection{Fit procedure}

We implement a binned $\chi^2$ fit, with 
\begin{equation} \label{eq:chisq}
\chi^2 = \sum_{i,j}\, \bigl(\Gamma^\mathrm{meas}_i - \Gamma_i \bigr)\,
  \bigl(V^{-1}\bigr)_{ij}\,
  \bigl(\Gamma^\mathrm{meas}_j - \Gamma_j \bigr)
\,.\end{equation}
Here $\Gamma^\mathrm{meas}_i$ is the measured $B\to X_s \gamma$ rate in bin $i$,
$\Gamma_i$ is the integral of \eq{expand} over bin $i$, $V$ is the full experimental
covariance matrix, and the sum runs over all bins of all measurements included in the fit.

\begin{figure}[tb]
\includegraphics[width=\columnwidth]{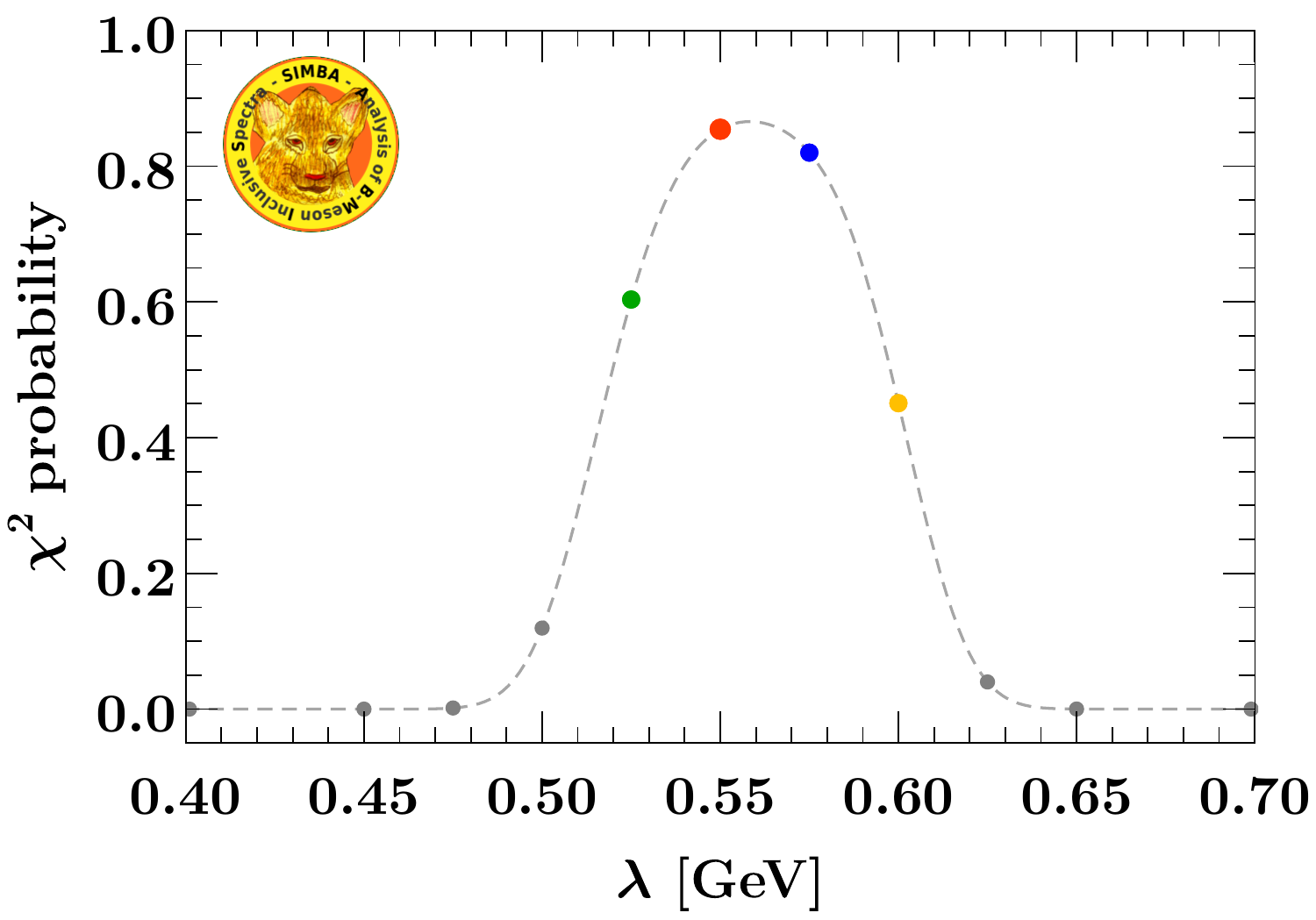}
\caption{%
The pre-fit $\chi^2$ probability for different $\lambda$ corresponding to different
bases. See text for details.}
\label{fig:prefit}
\end{figure}

The orthonormal basis $\{f_n\}$ is constructed~\cite{Ligeti:2008ac} such that
the first $F_{00}(k)$ term in the expansion of $\cF(k)$ can have any
(nonnegative) functional form, while the higher $F_{mn}(k)$ terms provide a
complete expansion generated from it. If $F_{00}(k)$ provides a good
approximation to $\cF(k)$, the expansion converges very quickly due to the
constraint in \eq{cnconstraint}, and consequently a good fit can be obtained
with small $N$, making the best use of the data to constrain $\cF(k)$.
Hence, $F_{00}(k)$ should already provide a reasonable description of the data.
To find such $F_{00}(k)$, we perform a pre-fit to the data using three different
functional forms for $F_{00}(k)$, given in \cite{supplement}, over a wide range
of $\lambda$. We choose the form that provides the best pre-fits. Its $\chi^2$
probability is shown in \fig{prefit} for sufficiently different values of
$\lambda$ such that each can be considered as a different basis. We choose the
best $\lambda = 0.55\,\GeV$ (orange) as our default basis, and use $\lambda =
0.525, 0.575, 0.6\,\GeV$ (green, blue, yellow), which also have good pre-fits,
as alternative bases to test the basis independence.

The truncation in \eq{truncatedexpansion} induces a residual dependence on the functional
form of the basis. To ensure that the corresponding uncertainty is small compared to others,
the truncation order $N$ is chosen based on
the available data, by increasing $N$ until there is no significant improvement in fit quality.
This is done by constructing nested hypothesis tests using the difference in $\chi^2$ between fits of increasing number of coefficients. If the $\chi^2$ improves by more than 1 from the inclusion of an additional coefficient, the higher number of coefficients is retained. To account for the truncation uncertainty, we include one additional coefficient in the fit. It is in this sense that our analysis is model independent within the quoted uncertainties.
The final truncation order is found to be $N = 3$ for each considered basis.
To ensure that the entire fit procedure including the choice of the basis and truncation order is unbiased, it is validated using pseudo-experiments generated around the best fit values, using the full experimental covariance matrices.

\papersection{Results}

We include four differential $B\to X_s\gamma$ measurements~\cite{Aubert:2007my,
Limosani:2009qg, Lees:2012ufa, Lees:2012wg} in the fit. The measurements in
\refcite{Aubert:2007my,
Limosani:2009qg, Lees:2012ufa} include $B\to X_d\gamma$ contributions, which are
subtracted assuming identical shapes for $B\to X_s\gamma$ and $B\to X_d\gamma$ and
that the ratio of branching ratios is $|V_{td}/V_{ts}|^2 = 0.0470$~\cite{pdg201819}.
For \refcite{Lees:2012wg}, we combine the highest six
$E_\gamma$ bins to stay insensitive to possible quark-hadron duality violation
and resonances with masses near $m_{K^*}$.
We use the measurements of \refscite{Limosani:2009qg, Lees:2012ufa} in the
$\Upsilon(4S)$ rest frame and boost the predictions accordingly.
We use the uncorrected measurement from \refcite{Limosani:2009qg} and
apply the experimental resolution matrix~\cite{Tony} to the predictions.

\begin{figure}[tb]
\includegraphics[width=\columnwidth]{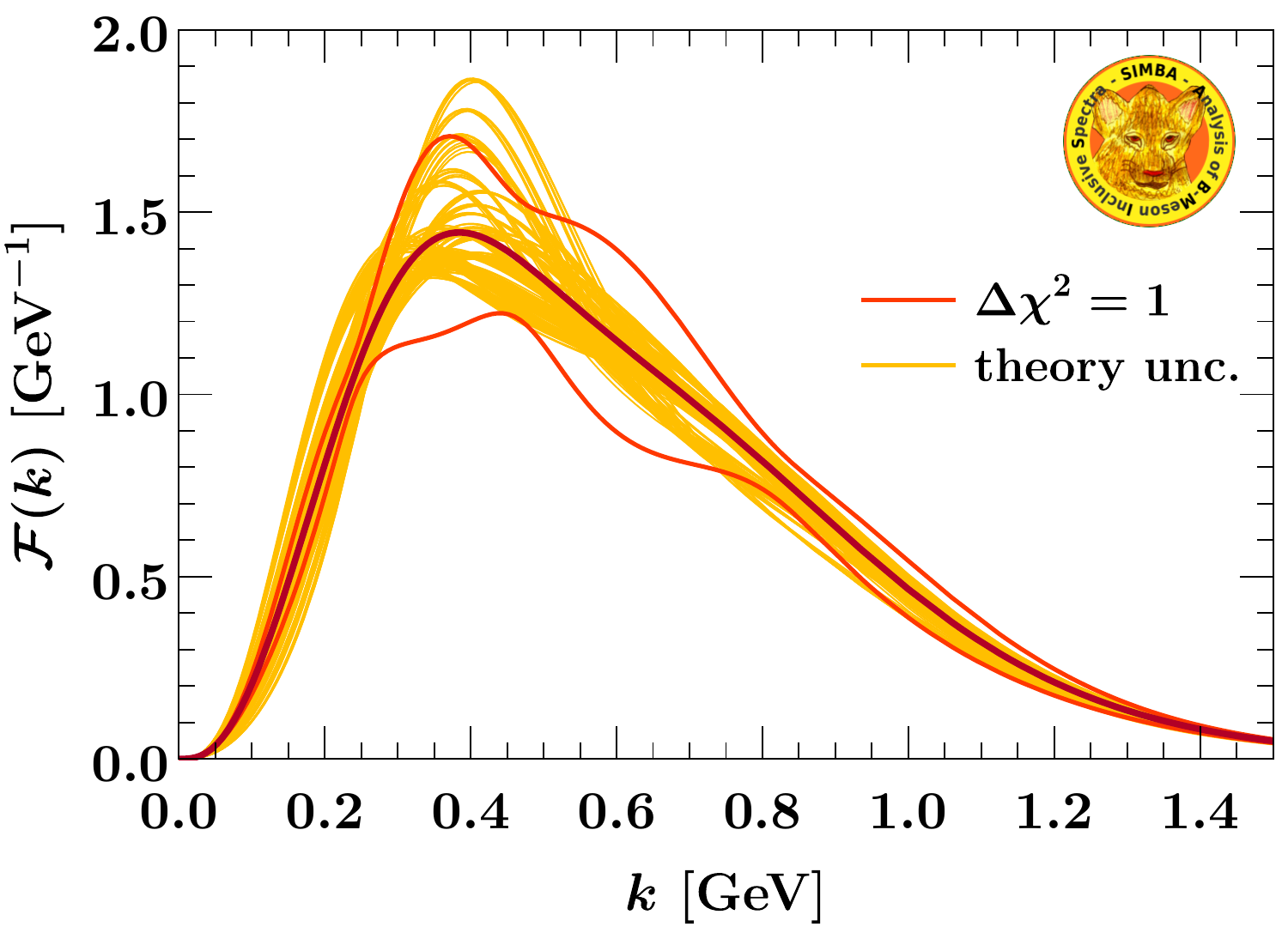}
\caption{The fitted shape function $\cF(k)$ with
central result (dark red) and fit uncertainties (dark orange lines).
The yellow curves show the variation of the fitted shape when
varying the perturbative inputs as discussed in the text.
}
\label{fig:BtoXsgammaSF}
\end{figure}

\begin{figure}[t!]
\includegraphics[width=\columnwidth]{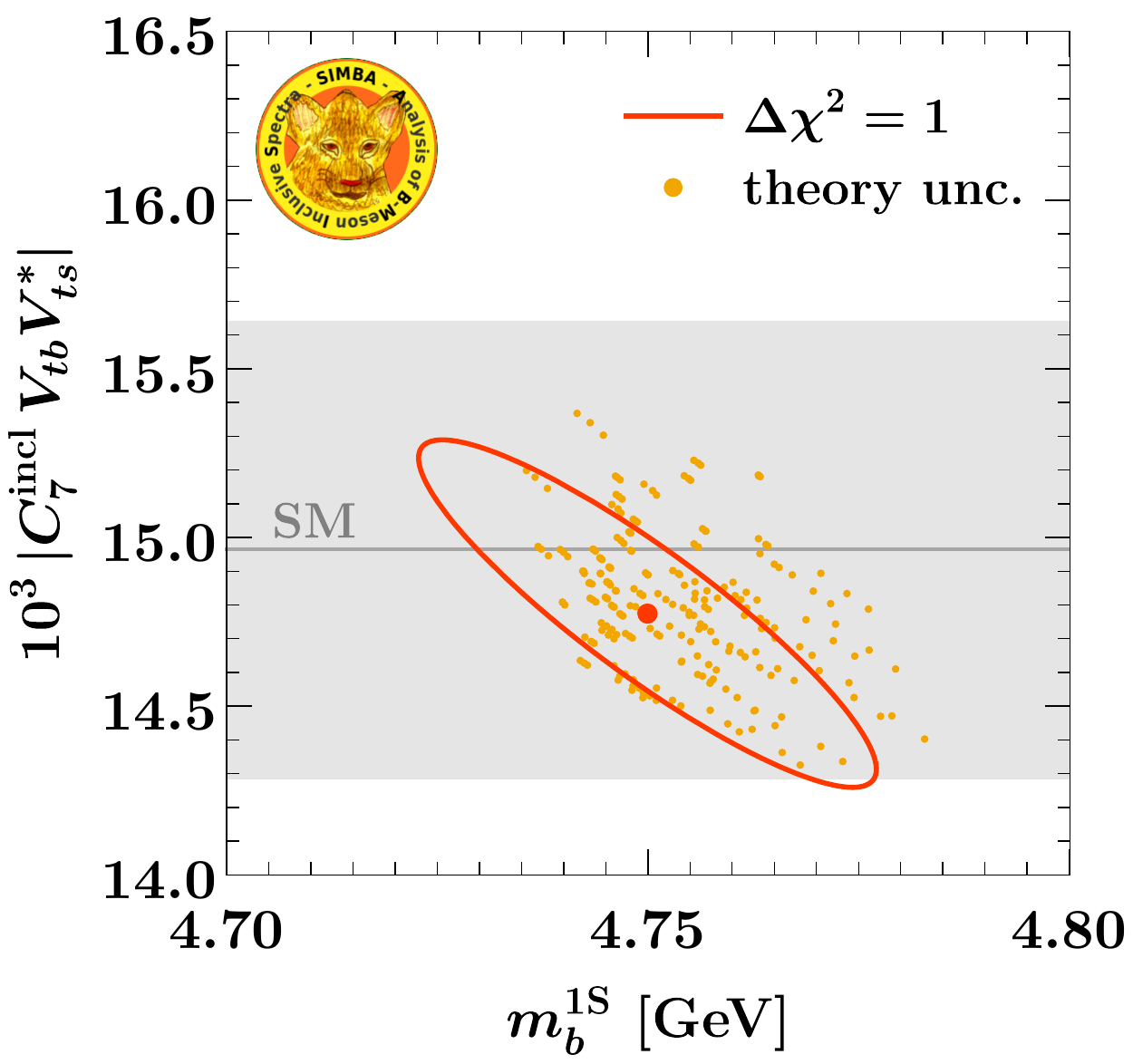}
\caption{Results for $\abs{C_7^\incl V_{tb} V_{ts}^*}$ and \mbS. The central fit
result is shown by the dark orange point and ellipse.
The yellow scattered points show the variation of the fit results when varying
the perturbative inputs as discussed in the text.}
\label{fig:BtoXsgammaresults}
\end{figure}

The fit results for $N_s$ and $c_{0-3}$ including their correlations
are  given in~\cite{supplement}.
The resulting shape function is shown in \fig{BtoXsgammaSF}, and the results
for $\abs{\Cincl}$ and $\hm_b \equiv \mbS$ are shown in \fig{BtoXsgammaresults}.
We also determine the kinetic energy parameter $\hla_1$ in the invisible
scheme~\cite{Ligeti:2008ac}, with plots analogous to \fig{BtoXsgammaresults}
given in \fig{la1} in~\cite{supplement}.
We find the following results:
\begin{alignat}{9}\label{eq:results}
\abs{\Cincl V_{tb} V_{ts}^*}
&= (14.77 \pm 0.51_{\rm fit} \pm 0.59_{\rm theory}
\nn \\ & \qquad\qquad
\pm 0.08_{\rm param}) \times 10^{-3}
\,, \nn \\
\mbS &= (4.750 \pm 0.027_{\rm fit} \pm 0.033_{\rm theory}
\nn \\ & \qquad\qquad
\pm 0.003_{\rm param})\,\GeV
\,, \nn \\
\hla_1
&= (-0.210 \pm 0.046_{\rm fit} \pm 0.040_{\rm theory}
\nn \\ & \qquad\qquad\,\,\,\,
\pm 0.056_{\rm param})\,\GeV^2
\,.\end{alignat}
The first uncertainty with subscript ``fit'' is evaluated from the $\Delta
\chi^2 = 1$ variation around the best fit point. It incorporates the
experimental uncertainties as well as the uncertainty due to the unknown shape
function, which is simultaneously constrained in the fit. The theory
and parametric uncertainties are evaluated by repeating the fit with
different theory inputs~\cite{supplement}.
The theory uncertainties are due to unknown higher-order perturbative
corrections to the shape of the spectrum in the peak region, which are evaluated
by a large set of resummation profile scale variations. The
results for all variations are shown by the yellow lines in \fig{BtoXsgammaSF}
and scatter points in \fig{BtoXsgammaresults}. To be conservative, the theory
uncertainty quoted in \eq{results} is obtained from the largest absolute
deviation for a given quantity (ignoring the apparent asymmetry in the
variations). The parametric uncertainty is only relevant for $\hla_1$, for
which it comes entirely from $\widehat\rho_2$.

Varying the residual $c\bar c$-loop contributions in the theory inputs
for the fit, equivalent to the $c\bar c$ uncertainty in \eq{C7inclnnlo},
changes the extracted $\abs{C_7^\incl}$ by $\pm 0.2\%$ and \mbS by $\pm 1\,\MeV$,
showing that by far the dominant dependence on and uncertainty from these
contributions is factorized into $C_7^\incl$. The uncertainty due to the
numerical value of $\hm_c^2/\hm_b^2$ contributes most of the parametric uncertainty
of $\abs{C_7^\incl}$ in \eq{results}.

From \eq{C7inclnnlo} and $\abs{V_{tb} V_{ts}^*} = (41.29 \pm 0.74)\times 10^{-3}$~\cite{pdg201819},
we find the SM value $\abs{C_7^\incl V_{tb} V_{ts}^*} = (14.96 \pm 0.68)\times 10^{-3}$, with the
uncertainty dominated by $\abs{C_7^\incl}$ in \eq{C7inclnnlo}.
This is shown by the gray band in \fig{BtoXsgammaresults},
and is in excellent agreement with our extracted value.

Converting our result for \mbS to the \MSbar scheme at three loops including
charm-mass effects~\cite{Hoang:2000fm}, we find
\begin{equation}
\mbbar(\mbbar) = (4.224 \pm 0.040 \pm 0.013)\,\GeV
\,,\end{equation}
where the first uncertainty comes from the total uncertainty in \mbS  in
\eq{results}, and the second one is the conversion uncertainty.
This result agrees with the world average of
$\mbbar(\mbbar) = (4.18^{+0.03}_{-0.02})\,\GeV$~\cite{pdg201819}.

\begin{figure}[t!]
\includegraphics[width=\columnwidth]{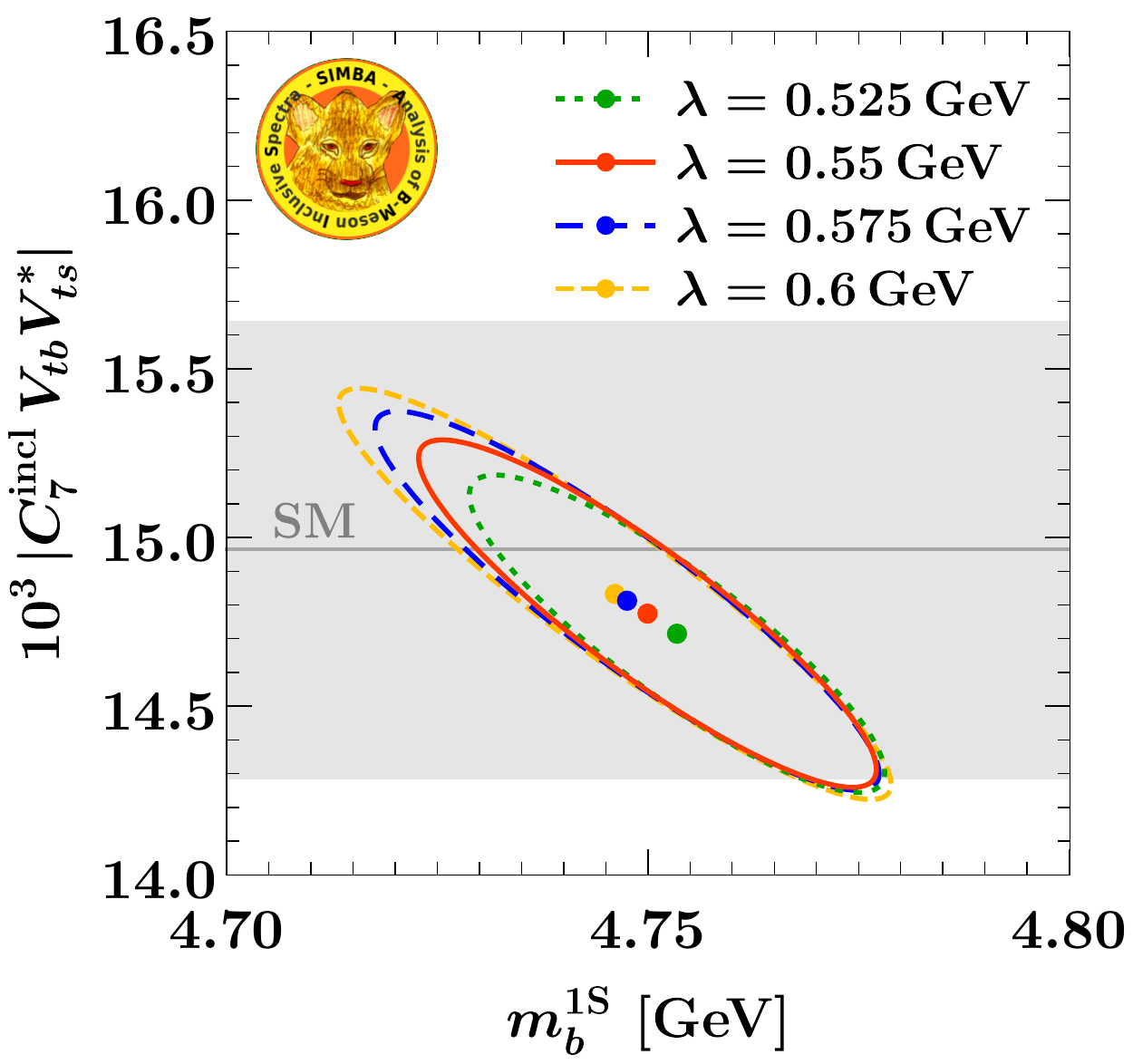}
\caption{Comparison of the fit results for $\abs{C_7^\incl V_{tb} V_{ts}^*}$
and \mbS for four different bases.
The results are consistent within a fraction of the fit uncertainties.
}
\label{fig:diflambda}
\end{figure}

In \fig{diflambda}, we demonstrate the basis independence
by comparing the results for $\abs{\Cincl}$ and \mbS for the four basis
choices in \fig{prefit}. The results using these bases are
consistent within a fraction of the fit uncertainties.
This would not be the case without including an additional coefficient
($c_3$) to account for the truncation uncertainty.

\papersection{Conclusions}

We presented the first global analysis of inclusive $B\to X_s\gamma$
measurements to determine $\abs{\Cincl}$ within a framework that allows a
model-independent and data-driven treatment of the nonperturbative $b$-quark
distribution function $\cF(k)$. The value extracted from \eq{results},
$\abs{\Cincl} = 0.3578 \pm 0.0199$, is consistent with the SM prediction in
\eq{C7inclnnlo}.

In comparison, in the past, the SM prediction for the rate in the $E_\gamma >
1.6\,\GeV$ region, ${\cal B}(B\to X_s\gamma) = (3.36 \pm 0.23)\times
10^{-4}$~\cite{Misiak:2015xwa}, was compared with its measurement, ${\cal
B}(B\to X_s\gamma) = (3.32 \pm 0.15)\times 10^{-4}$~\cite{Amhis:2019ckw}, which
have 6.8\% and 4.5\% uncertainties, respectively. The latter relies on an
extrapolation to the $1.6\,\GeV$ cut and on corresponding uncertainty estimates,
which entail insufficient variations of the nonperturbative shape-function models
and perturbative uncertainties that affect the spectrum. In addition, correlations
in these uncertainties in calculating and measuring the rate for $E_\gamma >
1.6\,\GeV$ cannot be fully assessed. In contrast, in our approach, $\Cincl$ is
reliably calculable in the SM or in models beyond it, and the relevant hadronic
physics and its uncertainties are determined from the data, together with the
extraction of $\abs{\Cincl}$. Hence, our approach is more reliable, as it makes
optimal use of the data, uncertainties from nonperturbative parameters
and perturbative inputs are clearly traceable, and no double counting can occur.

The uncertainty in our extracted $\abs{\Cincl V_{tb} V_{ts}^*}^2$ from
\eq{results} is 10.6\%, about twice the uncertainty in HFLAV's result for the
$E_\gamma > 1.6\,\GeV$ rate. If we neglect the theory uncertainties as well as
the truncation uncertainty (by repeating the fit only including up to $c_2$), we
would obtain a smaller uncertainty of 5.5\%, close to that of HFLAV's result.
This suggests that HFLAV's uncertainty is underestimated by about a factor of
two, which leaves more room for new physics. More importantly, the precision of
testing the SM is currently limited by the extraction of $\abs{C_7^\incl}$ from
data, and can be improved significantly with high-precision Belle~II
measurements.

\begin{acknowledgments}
\paragraph{Acknowledgments}
We thank Antonio Limosani for information about the detector resolution in the Belle measurement, and Francesca di Lodovico for information about the correlations in the \babar\ inclusive measurement.
We thank Anna Sophia Lacker for the artwork for the SIMBA logo.
We thank the DESY and LBL theory groups, KIT, and the Aspen Center for Physics (supported by the NSF Grant PHY-1607611) for hospitality while portions of this work were carried out.
This work was also supported in part by the Offices of
High Energy and Nuclear Physics of the U.S.\ Department of Energy under 
DE-AC02-05CH11231 and DE-SC0011090, the Simons Foundation through grant 327942,
the DFG Emmy-Noether Grants TA 867/1-1 and BE 6075/1-1,
and the Helmholtz Association Grant W2/W3-116.
\end{acknowledgments}

\bibliography{../simba,../simba_exp}

\begin{thebibliography}{94}%
\makeatletter
\providecommand \@ifxundefined [1]{%
 \@ifx{#1\undefined}
}%
\providecommand \@ifnum [1]{%
 \ifnum #1\expandafter \@firstoftwo
 \else \expandafter \@secondoftwo
 \fi
}%
\providecommand \@ifx [1]{%
 \ifx #1\expandafter \@firstoftwo
 \else \expandafter \@secondoftwo
 \fi
}%
\providecommand \natexlab [1]{#1}%
\providecommand \enquote  [1]{``#1''}%
\providecommand \bibnamefont  [1]{#1}%
\providecommand \bibfnamefont [1]{#1}%
\providecommand \citenamefont [1]{#1}%
\providecommand \href@noop [0]{\@secondoftwo}%
\providecommand \href [0]{\begingroup \@sanitize@url \@href}%
\providecommand \@href[1]{\@@startlink{#1}\@@href}%
\providecommand \@@href[1]{\endgroup#1\@@endlink}%
\providecommand \@sanitize@url [0]{\catcode `\\12\catcode `\$12\catcode
  `\&12\catcode `\#12\catcode `\^12\catcode `\_12\catcode `\%12\relax}%
\providecommand \@@startlink[1]{}%
\providecommand \@@endlink[0]{}%
\providecommand \url  [0]{\begingroup\@sanitize@url \@url }%
\providecommand \@url [1]{\endgroup\@href {#1}{\urlprefix }}%
\providecommand \urlprefix  [0]{URL }%
\providecommand \Eprint [0]{\href }%
\providecommand \doibase [0]{http://dx.doi.org/}%
\providecommand \selectlanguage [0]{\@gobble}%
\providecommand \bibinfo  [0]{\@secondoftwo}%
\providecommand \bibfield  [0]{\@secondoftwo}%
\providecommand \translation [1]{[#1]}%
\providecommand \BibitemOpen [0]{}%
\providecommand \bibitemStop [0]{}%
\providecommand \bibitemNoStop [0]{.\EOS\space}%
\providecommand \EOS [0]{\spacefactor3000\relax}%
\providecommand \BibitemShut  [1]{\csname bibitem#1\endcsname}%
\let\auto@bib@innerbib\@empty
\bibitem [{\citenamefont {Bertolini}\ \emph {et~al.}(1987)\citenamefont
  {Bertolini}, \citenamefont {Borzumati},\ and\ \citenamefont
  {Masiero}}]{Bertolini:1986th}%
  \BibitemOpen
  \bibfield  {author} {\bibinfo {author} {\bibfnamefont {S.}~\bibnamefont
  {Bertolini}}, \bibinfo {author} {\bibfnamefont {F.}~\bibnamefont
  {Borzumati}}, \ and\ \bibinfo {author} {\bibfnamefont {A.}~\bibnamefont
  {Masiero}},\ }\href {\doibase 10.1103/PhysRevLett.59.180} {\bibfield
  {journal} {\bibinfo  {journal} {Phys. Rev. Lett.}\ }\textbf {\bibinfo
  {volume} {59}},\ \bibinfo {pages} {180} (\bibinfo {year} {1987})}\BibitemShut
  {NoStop}%
\bibitem [{\citenamefont {Grinstein}\ \emph {et~al.}(1988)\citenamefont
  {Grinstein}, \citenamefont {Springer},\ and\ \citenamefont
  {Wise}}]{Grinstein:1987vj}%
  \BibitemOpen
  \bibfield  {author} {\bibinfo {author} {\bibfnamefont {B.}~\bibnamefont
  {Grinstein}}, \bibinfo {author} {\bibfnamefont {R.~P.}\ \bibnamefont
  {Springer}}, \ and\ \bibinfo {author} {\bibfnamefont {M.~B.}\ \bibnamefont
  {Wise}},\ }\href {\doibase 10.1016/0370-2693(88)90868-4} {\bibfield
  {journal} {\bibinfo  {journal} {Phys.~Lett.~B}\ }\textbf {\bibinfo {volume}
  {202}},\ \bibinfo {pages} {138} (\bibinfo {year} {1988})}\BibitemShut
  {NoStop}%
\bibitem [{\citenamefont {Misiak}\ \emph {et~al.}(2007)\citenamefont {Misiak},
  \citenamefont {Asatrian}, \citenamefont {Bieri}, \citenamefont {Czakon},
  \citenamefont {Czarnecki} \emph {et~al.}}]{Misiak:2006zs}%
  \BibitemOpen
  \bibfield  {author} {\bibinfo {author} {\bibfnamefont {M.}~\bibnamefont
  {Misiak}}, \bibinfo {author} {\bibfnamefont {H.}~\bibnamefont {Asatrian}},
  \bibinfo {author} {\bibfnamefont {K.}~\bibnamefont {Bieri}}, \bibinfo
  {author} {\bibfnamefont {M.}~\bibnamefont {Czakon}}, \bibinfo {author}
  {\bibfnamefont {A.}~\bibnamefont {Czarnecki}},  \emph {et~al.},\ }\href
  {\doibase 10.1103/PhysRevLett.98.022002} {\bibfield  {journal} {\bibinfo
  {journal} {Phys.~Rev.~Lett.}\ }\textbf {\bibinfo {volume} {98}},\ \bibinfo
  {pages} {022002} (\bibinfo {year} {2007})},\ \Eprint
  {http://arxiv.org/abs/hep-ph/0609232} {hep-ph/0609232} \BibitemShut {NoStop}%
\bibitem [{\citenamefont {Misiak}\ \emph {et~al.}(2015)\citenamefont {Misiak}
  \emph {et~al.}}]{Misiak:2015xwa}%
  \BibitemOpen
  \bibfield  {author} {\bibinfo {author} {\bibfnamefont {M.}~\bibnamefont
  {Misiak}} \emph {et~al.},\ }\href {\doibase 10.1103/PhysRevLett.114.221801}
  {\bibfield  {journal} {\bibinfo  {journal} {Phys. Rev. Lett.}\ }\textbf
  {\bibinfo {volume} {114}},\ \bibinfo {pages} {221801} (\bibinfo {year}
  {2015})},\ \Eprint {http://arxiv.org/abs/1503.01789} {arXiv:1503.01789
  [hep-ph]} \BibitemShut {NoStop}%
\bibitem [{\citenamefont {Grinstein}\ and\ \citenamefont
  {Wise}(1988)}]{Grinstein:1987pu}%
  \BibitemOpen
  \bibfield  {author} {\bibinfo {author} {\bibfnamefont {B.}~\bibnamefont
  {Grinstein}}\ and\ \bibinfo {author} {\bibfnamefont {M.~B.}\ \bibnamefont
  {Wise}},\ }\href {\doibase 10.1016/0370-2693(88)90227-4} {\bibfield
  {journal} {\bibinfo  {journal} {Phys.~Lett.~B}\ }\textbf {\bibinfo {volume}
  {201}},\ \bibinfo {pages} {274} (\bibinfo {year} {1988})}\BibitemShut
  {NoStop}%
\bibitem [{\citenamefont {Hou}\ and\ \citenamefont
  {Willey}(1988)}]{Hou:1987kf}%
  \BibitemOpen
  \bibfield  {author} {\bibinfo {author} {\bibfnamefont {W.-S.}\ \bibnamefont
  {Hou}}\ and\ \bibinfo {author} {\bibfnamefont {R.~S.}\ \bibnamefont
  {Willey}},\ }\href {\doibase 10.1016/0370-2693(88)91870-9} {\bibfield
  {journal} {\bibinfo  {journal} {Phys.~Lett.~B}\ }\textbf {\bibinfo {volume}
  {202}},\ \bibinfo {pages} {591} (\bibinfo {year} {1988})}\BibitemShut
  {NoStop}%
\bibitem [{\citenamefont {Misiak}\ and\ \citenamefont
  {Steinhauser}(2017)}]{Misiak:2017bgg}%
  \BibitemOpen
  \bibfield  {author} {\bibinfo {author} {\bibfnamefont {M.}~\bibnamefont
  {Misiak}}\ and\ \bibinfo {author} {\bibfnamefont {M.}~\bibnamefont
  {Steinhauser}},\ }\href {\doibase 10.1140/epjc/s10052-017-4776-y} {\bibfield
  {journal} {\bibinfo  {journal} {Eur. Phys. J.}\ }\textbf {\bibinfo {volume}
  {C77}},\ \bibinfo {pages} {201} (\bibinfo {year} {2017})},\ \Eprint
  {http://arxiv.org/abs/1702.04571} {arXiv:1702.04571 [hep-ph]} \BibitemShut
  {NoStop}%
\bibitem [{\citenamefont {Neubert}(1994)}]{Neubert:1993um}%
  \BibitemOpen
  \bibfield  {author} {\bibinfo {author} {\bibfnamefont {M.}~\bibnamefont
  {Neubert}},\ }\href {\doibase 10.1103/PhysRevD.49.4623} {\bibfield  {journal}
  {\bibinfo  {journal} {Phys.~Rev.~D}\ }\textbf {\bibinfo {volume} {49}},\
  \bibinfo {pages} {4623} (\bibinfo {year} {1994})},\ \Eprint
  {http://arxiv.org/abs/hep-ph/9312311} {hep-ph/9312311} \BibitemShut {NoStop}%
\bibitem [{\citenamefont {Bigi}\ \emph {et~al.}(1994)\citenamefont {Bigi},
  \citenamefont {Shifman}, \citenamefont {Uraltsev},\ and\ \citenamefont
  {Vainshtein}}]{Bigi:1993ex}%
  \BibitemOpen
  \bibfield  {author} {\bibinfo {author} {\bibfnamefont {I.~I.~Y.}\
  \bibnamefont {Bigi}}, \bibinfo {author} {\bibfnamefont {M.~A.}\ \bibnamefont
  {Shifman}}, \bibinfo {author} {\bibfnamefont {N.~G.}\ \bibnamefont
  {Uraltsev}}, \ and\ \bibinfo {author} {\bibfnamefont {A.~I.}\ \bibnamefont
  {Vainshtein}},\ }\href {\doibase 10.1142/S0217751X94000996} {\bibfield
  {journal} {\bibinfo  {journal} {Int. J. Mod. Phys.}\ }\textbf {\bibinfo
  {volume} {A9}},\ \bibinfo {pages} {2467} (\bibinfo {year} {1994})},\ \Eprint
  {http://arxiv.org/abs/hep-ph/9312359} {hep-ph/9312359} \BibitemShut {NoStop}%
\bibitem [{\citenamefont {Ligeti}\ \emph {et~al.}(2008)\citenamefont {Ligeti},
  \citenamefont {Stewart},\ and\ \citenamefont {Tackmann}}]{Ligeti:2008ac}%
  \BibitemOpen
  \bibfield  {author} {\bibinfo {author} {\bibfnamefont {Z.}~\bibnamefont
  {Ligeti}}, \bibinfo {author} {\bibfnamefont {I.~W.}\ \bibnamefont {Stewart}},
  \ and\ \bibinfo {author} {\bibfnamefont {F.~J.}\ \bibnamefont {Tackmann}},\
  }\href {\doibase 10.1103/PhysRevD.78.114014} {\bibfield  {journal} {\bibinfo
  {journal} {Phys.~Rev.~D}\ }\textbf {\bibinfo {volume} {78}},\ \bibinfo
  {pages} {114014} (\bibinfo {year} {2008})},\ \Eprint
  {http://arxiv.org/abs/0807.1926} {arXiv:0807.1926 [hep-ph]} \BibitemShut
  {NoStop}%
\bibitem [{\citenamefont {Benson}\ \emph {et~al.}(2005)\citenamefont {Benson},
  \citenamefont {Bigi},\ and\ \citenamefont {Uraltsev}}]{Benson:2004sg}%
  \BibitemOpen
  \bibfield  {author} {\bibinfo {author} {\bibfnamefont {D.}~\bibnamefont
  {Benson}}, \bibinfo {author} {\bibfnamefont {I.~I.}\ \bibnamefont {Bigi}}, \
  and\ \bibinfo {author} {\bibfnamefont {N.}~\bibnamefont {Uraltsev}},\ }\href
  {\doibase 10.1016/j.nuclphysb.2004.12.035} {\bibfield  {journal} {\bibinfo
  {journal} {Nucl. Phys.}\ }\textbf {\bibinfo {volume} {B710}},\ \bibinfo
  {pages} {371} (\bibinfo {year} {2005})},\ \Eprint
  {http://arxiv.org/abs/hep-ph/0410080} {hep-ph/0410080} \BibitemShut {NoStop}%
\bibitem [{\citenamefont {Lange}\ \emph {et~al.}(2005)\citenamefont {Lange},
  \citenamefont {Neubert},\ and\ \citenamefont {Paz}}]{Lange:2005yw}%
  \BibitemOpen
  \bibfield  {author} {\bibinfo {author} {\bibfnamefont {B.~O.}\ \bibnamefont
  {Lange}}, \bibinfo {author} {\bibfnamefont {M.}~\bibnamefont {Neubert}}, \
  and\ \bibinfo {author} {\bibfnamefont {G.}~\bibnamefont {Paz}},\ }\href
  {\doibase 10.1103/PhysRevD.72.073006} {\bibfield  {journal} {\bibinfo
  {journal} {Phys. Rev.}\ }\textbf {\bibinfo {volume} {D72}},\ \bibinfo {pages}
  {073006} (\bibinfo {year} {2005})},\ \Eprint
  {http://arxiv.org/abs/hep-ph/0504071} {hep-ph/0504071} \BibitemShut {NoStop}%
\bibitem [{\citenamefont {Andersen}\ and\ \citenamefont
  {Gardi}(2006)}]{Andersen:2005mj}%
  \BibitemOpen
  \bibfield  {author} {\bibinfo {author} {\bibfnamefont {J.~R.}\ \bibnamefont
  {Andersen}}\ and\ \bibinfo {author} {\bibfnamefont {E.}~\bibnamefont
  {Gardi}},\ }\href {\doibase 10.1088/1126-6708/2006/01/097} {\bibfield
  {journal} {\bibinfo  {journal} {JHEP}\ }\textbf {\bibinfo {volume} {01}},\
  \bibinfo {pages} {097} (\bibinfo {year} {2006})},\ \Eprint
  {http://arxiv.org/abs/hep-ph/0509360} {hep-ph/0509360} \BibitemShut {NoStop}%
\bibitem [{\citenamefont {Gambino}\ \emph {et~al.}(2007)\citenamefont
  {Gambino}, \citenamefont {Giordano}, \citenamefont {Ossola},\ and\
  \citenamefont {Uraltsev}}]{Gambino:2007rp}%
  \BibitemOpen
  \bibfield  {author} {\bibinfo {author} {\bibfnamefont {P.}~\bibnamefont
  {Gambino}}, \bibinfo {author} {\bibfnamefont {P.}~\bibnamefont {Giordano}},
  \bibinfo {author} {\bibfnamefont {G.}~\bibnamefont {Ossola}}, \ and\ \bibinfo
  {author} {\bibfnamefont {N.}~\bibnamefont {Uraltsev}},\ }\href {\doibase
  10.1088/1126-6708/2007/10/058} {\bibfield  {journal} {\bibinfo  {journal}
  {JHEP}\ }\textbf {\bibinfo {volume} {10}},\ \bibinfo {pages} {058} (\bibinfo
  {year} {2007})},\ \Eprint {http://arxiv.org/abs/0707.2493} {arXiv:0707.2493
  [hep-ph]} \BibitemShut {NoStop}%
\bibitem [{\citenamefont {Aglietti}\ \emph {et~al.}(2009)\citenamefont
  {Aglietti}, \citenamefont {Di~Lodovico}, \citenamefont {Ferrera},\ and\
  \citenamefont {Ricciardi}}]{Aglietti:2007ik}%
  \BibitemOpen
  \bibfield  {author} {\bibinfo {author} {\bibfnamefont {U.}~\bibnamefont
  {Aglietti}}, \bibinfo {author} {\bibfnamefont {F.}~\bibnamefont
  {Di~Lodovico}}, \bibinfo {author} {\bibfnamefont {G.}~\bibnamefont
  {Ferrera}}, \ and\ \bibinfo {author} {\bibfnamefont {G.}~\bibnamefont
  {Ricciardi}},\ }\href {\doibase 10.1140/epjc/s10052-008-0817-x} {\bibfield
  {journal} {\bibinfo  {journal} {Eur. Phys. J.}\ }\textbf {\bibinfo {volume}
  {C59}},\ \bibinfo {pages} {831} (\bibinfo {year} {2009})},\ \Eprint
  {http://arxiv.org/abs/0711.0860} {arXiv:0711.0860 [hep-ph]} \BibitemShut
  {NoStop}%
\bibitem [{\citenamefont {Aubert}\ \emph {et~al.}(2008)\citenamefont {Aubert}
  \emph {et~al.}}]{Aubert:2007my}%
  \BibitemOpen
  \bibfield  {author} {\bibinfo {author} {\bibfnamefont {B.}~\bibnamefont
  {Aubert}} \emph {et~al.} (\bibinfo {collaboration} {BaBar}),\ }\href
  {\doibase 10.1103/PhysRevD.77.051103} {\bibfield  {journal} {\bibinfo
  {journal} {Phys. Rev.}\ }\textbf {\bibinfo {volume} {D77}},\ \bibinfo {pages}
  {051103} (\bibinfo {year} {2008})},\ \Eprint {http://arxiv.org/abs/0711.4889}
  {arXiv:0711.4889 [hep-ex]} \BibitemShut {NoStop}%
\bibitem [{\citenamefont {Limosani}\ \emph {et~al.}(2009)\citenamefont
  {Limosani} \emph {et~al.}}]{Limosani:2009qg}%
  \BibitemOpen
  \bibfield  {author} {\bibinfo {author} {\bibfnamefont {A.}~\bibnamefont
  {Limosani}} \emph {et~al.} (\bibinfo {collaboration} {Belle}),\ }\href
  {\doibase 10.1103/PhysRevLett.103.241801} {\bibfield  {journal} {\bibinfo
  {journal} {Phys. Rev. Lett.}\ }\textbf {\bibinfo {volume} {103}},\ \bibinfo
  {pages} {241801} (\bibinfo {year} {2009})},\ \Eprint
  {http://arxiv.org/abs/0907.1384} {arXiv:0907.1384 [hep-ex]} \BibitemShut
  {NoStop}%
\bibitem [{\citenamefont {Lees}\ \emph
  {et~al.}(2012{\natexlab{a}})\citenamefont {Lees} \emph
  {et~al.}}]{Lees:2012ufa}%
  \BibitemOpen
  \bibfield  {author} {\bibinfo {author} {\bibfnamefont {J.~P.}\ \bibnamefont
  {Lees}} \emph {et~al.} (\bibinfo {collaboration} {BaBar}),\ }\href {\doibase
  10.1103/PhysRevD.86.112008} {\bibfield  {journal} {\bibinfo  {journal} {Phys.
  Rev.}\ }\textbf {\bibinfo {volume} {D86}},\ \bibinfo {pages} {112008}
  (\bibinfo {year} {2012}{\natexlab{a}})},\ \Eprint
  {http://arxiv.org/abs/1207.5772} {arXiv:1207.5772 [hep-ex]} \BibitemShut
  {NoStop}%
\bibitem [{\citenamefont {Lees}\ \emph
  {et~al.}(2012{\natexlab{b}})\citenamefont {Lees} \emph
  {et~al.}}]{Lees:2012wg}%
  \BibitemOpen
  \bibfield  {author} {\bibinfo {author} {\bibfnamefont {J.}~\bibnamefont
  {Lees}} \emph {et~al.} (\bibinfo {collaboration} {BaBar Collaboration}),\
  }\href {\doibase 10.1103/PhysRevD.86.052012} {\bibfield  {journal} {\bibinfo
  {journal} {Phys.~Rev.~D}\ }\textbf {\bibinfo {volume} {86}},\ \bibinfo
  {pages} {052012} (\bibinfo {year} {2012}{\natexlab{b}})},\ \Eprint
  {http://arxiv.org/abs/1207.2520} {arXiv:1207.2520 [hep-ex]} \BibitemShut
  {NoStop}%
\bibitem [{\citenamefont {Amhis}\ \emph {et~al.}(2019)\citenamefont {Amhis}
  \emph {et~al.}}]{Amhis:2019ckw}%
  \BibitemOpen
  \bibfield  {author} {\bibinfo {author} {\bibfnamefont {Y.~S.}\ \bibnamefont
  {Amhis}} \emph {et~al.} (\bibinfo {collaboration} {HFLAV}),\ }\href@noop {}
  {\  (\bibinfo {year} {2019})},\ \Eprint {http://arxiv.org/abs/1909.12524}
  {arXiv:1909.12524 [hep-ex]} \BibitemShut {NoStop}%
\bibitem [{\citenamefont {Bauer}\ \emph {et~al.}(2000)\citenamefont {Bauer},
  \citenamefont {Fleming},\ and\ \citenamefont {Luke}}]{Bauer:2000ew}%
  \BibitemOpen
  \bibfield  {author} {\bibinfo {author} {\bibfnamefont {C.~W.}\ \bibnamefont
  {Bauer}}, \bibinfo {author} {\bibfnamefont {S.}~\bibnamefont {Fleming}}, \
  and\ \bibinfo {author} {\bibfnamefont {M.~E.}\ \bibnamefont {Luke}},\ }\href
  {\doibase 10.1103/PhysRevD.63.014006} {\bibfield  {journal} {\bibinfo
  {journal} {Phys.~Rev.~D}\ }\textbf {\bibinfo {volume} {63}},\ \bibinfo
  {pages} {014006} (\bibinfo {year} {2000})},\ \Eprint
  {http://arxiv.org/abs/hep-ph/0005275} {hep-ph/0005275} \BibitemShut {NoStop}%
\bibitem [{\citenamefont {Bauer}\ \emph {et~al.}(2001)\citenamefont {Bauer},
  \citenamefont {Fleming}, \citenamefont {Pirjol},\ and\ \citenamefont
  {Stewart}}]{Bauer:2000yr}%
  \BibitemOpen
  \bibfield  {author} {\bibinfo {author} {\bibfnamefont {C.~W.}\ \bibnamefont
  {Bauer}}, \bibinfo {author} {\bibfnamefont {S.}~\bibnamefont {Fleming}},
  \bibinfo {author} {\bibfnamefont {D.}~\bibnamefont {Pirjol}}, \ and\ \bibinfo
  {author} {\bibfnamefont {I.~W.}\ \bibnamefont {Stewart}},\ }\href {\doibase
  10.1103/PhysRevD.63.114020} {\bibfield  {journal} {\bibinfo  {journal}
  {Phys.~Rev.~D}\ }\textbf {\bibinfo {volume} {63}},\ \bibinfo {pages} {114020}
  (\bibinfo {year} {2001})},\ \Eprint {http://arxiv.org/abs/hep-ph/0011336}
  {hep-ph/0011336} \BibitemShut {NoStop}%
\bibitem [{\citenamefont {Bauer}\ and\ \citenamefont
  {Stewart}(2001)}]{Bauer:2001ct}%
  \BibitemOpen
  \bibfield  {author} {\bibinfo {author} {\bibfnamefont {C.~W.}\ \bibnamefont
  {Bauer}}\ and\ \bibinfo {author} {\bibfnamefont {I.~W.}\ \bibnamefont
  {Stewart}},\ }\href {\doibase 10.1016/S0370-2693(01)00902-9} {\bibfield
  {journal} {\bibinfo  {journal} {Phys.~Lett.~B}\ }\textbf {\bibinfo {volume}
  {516}},\ \bibinfo {pages} {134} (\bibinfo {year} {2001})},\ \Eprint
  {http://arxiv.org/abs/hep-ph/0107001} {hep-ph/0107001} \BibitemShut {NoStop}%
\bibitem [{\citenamefont {Bauer}\ \emph {et~al.}(2002)\citenamefont {Bauer},
  \citenamefont {Pirjol},\ and\ \citenamefont {Stewart}}]{Bauer:2001yt}%
  \BibitemOpen
  \bibfield  {author} {\bibinfo {author} {\bibfnamefont {C.~W.}\ \bibnamefont
  {Bauer}}, \bibinfo {author} {\bibfnamefont {D.}~\bibnamefont {Pirjol}}, \
  and\ \bibinfo {author} {\bibfnamefont {I.~W.}\ \bibnamefont {Stewart}},\
  }\href {\doibase 10.1103/PhysRevD.65.054022} {\bibfield  {journal} {\bibinfo
  {journal} {Phys.~Rev.~D}\ }\textbf {\bibinfo {volume} {65}},\ \bibinfo
  {pages} {054022} (\bibinfo {year} {2002})},\ \Eprint
  {http://arxiv.org/abs/hep-ph/0109045} {hep-ph/0109045} \BibitemShut {NoStop}%
\bibitem [{\citenamefont {Hoang}\ \emph
  {et~al.}(1999{\natexlab{a}})\citenamefont {Hoang}, \citenamefont {Ligeti},\
  and\ \citenamefont {Manohar}}]{Hoang:1998ng}%
  \BibitemOpen
  \bibfield  {author} {\bibinfo {author} {\bibfnamefont {A.~H.}\ \bibnamefont
  {Hoang}}, \bibinfo {author} {\bibfnamefont {Z.}~\bibnamefont {Ligeti}}, \
  and\ \bibinfo {author} {\bibfnamefont {A.~V.}\ \bibnamefont {Manohar}},\
  }\href {\doibase 10.1103/PhysRevLett.82.277} {\bibfield  {journal} {\bibinfo
  {journal} {Phys.~Rev.~Lett.}\ }\textbf {\bibinfo {volume} {82}},\ \bibinfo
  {pages} {277} (\bibinfo {year} {1999}{\natexlab{a}})},\ \Eprint
  {http://arxiv.org/abs/hep-ph/9809423} {hep-ph/9809423} \BibitemShut {NoStop}%
\bibitem [{\citenamefont {Hoang}\ \emph
  {et~al.}(1999{\natexlab{b}})\citenamefont {Hoang}, \citenamefont {Ligeti},\
  and\ \citenamefont {Manohar}}]{Hoang:1998hm}%
  \BibitemOpen
  \bibfield  {author} {\bibinfo {author} {\bibfnamefont {A.~H.}\ \bibnamefont
  {Hoang}}, \bibinfo {author} {\bibfnamefont {Z.}~\bibnamefont {Ligeti}}, \
  and\ \bibinfo {author} {\bibfnamefont {A.~V.}\ \bibnamefont {Manohar}},\
  }\href {\doibase 10.1103/PhysRevD.59.074017} {\bibfield  {journal} {\bibinfo
  {journal} {Phys.~Rev.~D}\ }\textbf {\bibinfo {volume} {59}},\ \bibinfo
  {pages} {074017} (\bibinfo {year} {1999}{\natexlab{b}})},\ \Eprint
  {http://arxiv.org/abs/hep-ph/9811239} {hep-ph/9811239} \BibitemShut {NoStop}%
\bibitem [{\citenamefont {Hoang}\ and\ \citenamefont
  {Teubner}(1999)}]{Hoang:1999zc}%
  \BibitemOpen
  \bibfield  {author} {\bibinfo {author} {\bibfnamefont {A.}~\bibnamefont
  {Hoang}}\ and\ \bibinfo {author} {\bibfnamefont {T.}~\bibnamefont
  {Teubner}},\ }\href {\doibase 10.1103/PhysRevD.60.114027} {\bibfield
  {journal} {\bibinfo  {journal} {Phys.~Rev.~D}\ }\textbf {\bibinfo {volume}
  {60}},\ \bibinfo {pages} {114027} (\bibinfo {year} {1999})},\ \Eprint
  {http://arxiv.org/abs/hep-ph/9904468} {hep-ph/9904468} \BibitemShut {NoStop}%
\bibitem [{sup()}]{supplement}%
  \BibitemOpen
  \href@noop {} {}\BibitemShut {NoStop}%
\bibitem [{\citenamefont {Korchemsky}\ and\ \citenamefont
  {Marchesini}(1993)}]{Korchemsky:1992xv}%
  \BibitemOpen
  \bibfield  {author} {\bibinfo {author} {\bibfnamefont {G.~P.}\ \bibnamefont
  {Korchemsky}}\ and\ \bibinfo {author} {\bibfnamefont {G.}~\bibnamefont
  {Marchesini}},\ }\href {\doibase 10.1016/0550-3213(93)90167-N} {\bibfield
  {journal} {\bibinfo  {journal} {Nucl. Phys.}\ }\textbf {\bibinfo {volume}
  {B406}},\ \bibinfo {pages} {225} (\bibinfo {year} {1993})},\ \Eprint
  {http://arxiv.org/abs/hep-ph/9210281} {hep-ph/9210281} \BibitemShut {NoStop}%
\bibitem [{\citenamefont {Gardi}(2005)}]{Gardi:2005yi}%
  \BibitemOpen
  \bibfield  {author} {\bibinfo {author} {\bibfnamefont {E.}~\bibnamefont
  {Gardi}},\ }\href {\doibase 10.1088/1126-6708/2005/02/053} {\bibfield
  {journal} {\bibinfo  {journal} {JHEP}\ }\textbf {\bibinfo {volume} {02}},\
  \bibinfo {pages} {053} (\bibinfo {year} {2005})},\ \Eprint
  {http://arxiv.org/abs/hep-ph/0501257} {hep-ph/0501257} \BibitemShut {NoStop}%
\bibitem [{\citenamefont {Blokland}\ \emph {et~al.}(2005)\citenamefont
  {Blokland}, \citenamefont {Czarnecki}, \citenamefont {Misiak}, \citenamefont
  {Slusarczyk},\ and\ \citenamefont {Tkachov}}]{Blokland:2005uk}%
  \BibitemOpen
  \bibfield  {author} {\bibinfo {author} {\bibfnamefont {I.~R.}\ \bibnamefont
  {Blokland}}, \bibinfo {author} {\bibfnamefont {A.}~\bibnamefont {Czarnecki}},
  \bibinfo {author} {\bibfnamefont {M.}~\bibnamefont {Misiak}}, \bibinfo
  {author} {\bibfnamefont {M.}~\bibnamefont {Slusarczyk}}, \ and\ \bibinfo
  {author} {\bibfnamefont {F.}~\bibnamefont {Tkachov}},\ }\href {\doibase
  10.1103/PhysRevD.72.033014} {\bibfield  {journal} {\bibinfo  {journal}
  {Phys.~Rev.~D}\ }\textbf {\bibinfo {volume} {72}},\ \bibinfo {pages} {033014}
  (\bibinfo {year} {2005})},\ \Eprint {http://arxiv.org/abs/hep-ph/0506055}
  {hep-ph/0506055} \BibitemShut {NoStop}%
\bibitem [{\citenamefont {Bauer}\ and\ \citenamefont
  {Manohar}(2004)}]{Bauer:2003pi}%
  \BibitemOpen
  \bibfield  {author} {\bibinfo {author} {\bibfnamefont {C.~W.}\ \bibnamefont
  {Bauer}}\ and\ \bibinfo {author} {\bibfnamefont {A.~V.}\ \bibnamefont
  {Manohar}},\ }\href {\doibase 10.1103/PhysRevD.70.034024} {\bibfield
  {journal} {\bibinfo  {journal} {Phys.~Rev.~D}\ }\textbf {\bibinfo {volume}
  {70}},\ \bibinfo {pages} {034024} (\bibinfo {year} {2004})},\ \Eprint
  {http://arxiv.org/abs/hep-ph/0312109} {hep-ph/0312109} \BibitemShut {NoStop}%
\bibitem [{\citenamefont {Becher}\ and\ \citenamefont
  {Neubert}(2006{\natexlab{a}})}]{Becher:2005pd}%
  \BibitemOpen
  \bibfield  {author} {\bibinfo {author} {\bibfnamefont {T.}~\bibnamefont
  {Becher}}\ and\ \bibinfo {author} {\bibfnamefont {M.}~\bibnamefont
  {Neubert}},\ }\href {\doibase 10.1016/j.physletb.2006.01.006} {\bibfield
  {journal} {\bibinfo  {journal} {Phys. Lett.}\ }\textbf {\bibinfo {volume}
  {B633}},\ \bibinfo {pages} {739} (\bibinfo {year} {2006}{\natexlab{a}})},\
  \Eprint {http://arxiv.org/abs/hep-ph/0512208} {hep-ph/0512208} \BibitemShut
  {NoStop}%
\bibitem [{\citenamefont {Becher}\ and\ \citenamefont
  {Neubert}(2006{\natexlab{b}})}]{Becher:2006qw}%
  \BibitemOpen
  \bibfield  {author} {\bibinfo {author} {\bibfnamefont {T.}~\bibnamefont
  {Becher}}\ and\ \bibinfo {author} {\bibfnamefont {M.}~\bibnamefont
  {Neubert}},\ }\href {\doibase 10.1016/j.physletb.2006.04.046} {\bibfield
  {journal} {\bibinfo  {journal} {Phys. Lett.}\ }\textbf {\bibinfo {volume}
  {B637}},\ \bibinfo {pages} {251} (\bibinfo {year} {2006}{\natexlab{b}})},\
  \Eprint {http://arxiv.org/abs/hep-ph/0603140} {hep-ph/0603140} \BibitemShut
  {NoStop}%
\bibitem [{\citenamefont {Balzereit}\ \emph {et~al.}(1998)\citenamefont
  {Balzereit}, \citenamefont {Mannel},\ and\ \citenamefont
  {Kilian}}]{Balzereit:1998yf}%
  \BibitemOpen
  \bibfield  {author} {\bibinfo {author} {\bibfnamefont {C.}~\bibnamefont
  {Balzereit}}, \bibinfo {author} {\bibfnamefont {T.}~\bibnamefont {Mannel}}, \
  and\ \bibinfo {author} {\bibfnamefont {W.}~\bibnamefont {Kilian}},\ }\href
  {\doibase 10.1103/PhysRevD.58.114029} {\bibfield  {journal} {\bibinfo
  {journal} {Phys. Rev.}\ }\textbf {\bibinfo {volume} {D58}},\ \bibinfo {pages}
  {114029} (\bibinfo {year} {1998})},\ \Eprint
  {http://arxiv.org/abs/hep-ph/9805297} {hep-ph/9805297} \BibitemShut {NoStop}%
\bibitem [{\citenamefont {Neubert}(2005)}]{Neubert:2004dd}%
  \BibitemOpen
  \bibfield  {author} {\bibinfo {author} {\bibfnamefont {M.}~\bibnamefont
  {Neubert}},\ }\href {\doibase 10.1140/epjc/s2005-02141-1} {\bibfield
  {journal} {\bibinfo  {journal} {Eur.~Phys.~J.~C}\ }\textbf {\bibinfo {volume}
  {40}},\ \bibinfo {pages} {165} (\bibinfo {year} {2005})},\ \Eprint
  {http://arxiv.org/abs/hep-ph/0408179} {hep-ph/0408179} \BibitemShut {NoStop}%
\bibitem [{\citenamefont {Fleming}\ \emph {et~al.}(2008)\citenamefont
  {Fleming}, \citenamefont {Hoang}, \citenamefont {Mantry},\ and\ \citenamefont
  {Stewart}}]{Fleming:2007xt}%
  \BibitemOpen
  \bibfield  {author} {\bibinfo {author} {\bibfnamefont {S.}~\bibnamefont
  {Fleming}}, \bibinfo {author} {\bibfnamefont {A.~H.}\ \bibnamefont {Hoang}},
  \bibinfo {author} {\bibfnamefont {S.}~\bibnamefont {Mantry}}, \ and\ \bibinfo
  {author} {\bibfnamefont {I.~W.}\ \bibnamefont {Stewart}},\ }\href {\doibase
  10.1103/PhysRevD.77.114003} {\bibfield  {journal} {\bibinfo  {journal} {Phys.
  Rev.}\ }\textbf {\bibinfo {volume} {D77}},\ \bibinfo {pages} {114003}
  (\bibinfo {year} {2008})},\ \Eprint {http://arxiv.org/abs/0711.2079}
  {arXiv:0711.2079 [hep-ph]} \BibitemShut {NoStop}%
\bibitem [{\citenamefont {Misiak}\ and\ \citenamefont
  {Steinhauser}(2007)}]{Misiak:2006ab}%
  \BibitemOpen
  \bibfield  {author} {\bibinfo {author} {\bibfnamefont {M.}~\bibnamefont
  {Misiak}}\ and\ \bibinfo {author} {\bibfnamefont {M.}~\bibnamefont
  {Steinhauser}},\ }\href {\doibase 10.1016/j.nuclphysb.2006.11.027} {\bibfield
   {journal} {\bibinfo  {journal} {Nucl.~Phys.}\ }\textbf {\bibinfo {volume}
  {B764}},\ \bibinfo {pages} {62} (\bibinfo {year} {2007})},\ \Eprint
  {http://arxiv.org/abs/hep-ph/0609241} {hep-ph/0609241} \BibitemShut {NoStop}%
\bibitem [{\citenamefont {Czakon}\ \emph {et~al.}(2015)\citenamefont {Czakon},
  \citenamefont {Fiedler}, \citenamefont {Huber}, \citenamefont {Misiak},
  \citenamefont {Schutzmeier},\ and\ \citenamefont
  {Steinhauser}}]{Czakon:2015exa}%
  \BibitemOpen
  \bibfield  {author} {\bibinfo {author} {\bibfnamefont {M.}~\bibnamefont
  {Czakon}}, \bibinfo {author} {\bibfnamefont {P.}~\bibnamefont {Fiedler}},
  \bibinfo {author} {\bibfnamefont {T.}~\bibnamefont {Huber}}, \bibinfo
  {author} {\bibfnamefont {M.}~\bibnamefont {Misiak}}, \bibinfo {author}
  {\bibfnamefont {T.}~\bibnamefont {Schutzmeier}}, \ and\ \bibinfo {author}
  {\bibfnamefont {M.}~\bibnamefont {Steinhauser}},\ }\href {\doibase
  10.1007/JHEP04(2015)168} {\bibfield  {journal} {\bibinfo  {journal} {JHEP}\
  }\textbf {\bibinfo {volume} {04}},\ \bibinfo {pages} {168} (\bibinfo {year}
  {2015})},\ \Eprint {http://arxiv.org/abs/1503.01791} {arXiv:1503.01791
  [hep-ph]} \BibitemShut {NoStop}%
\bibitem [{\citenamefont {Melnikov}\ and\ \citenamefont
  {Mitov}(2005)}]{Melnikov:2005bx}%
  \BibitemOpen
  \bibfield  {author} {\bibinfo {author} {\bibfnamefont {K.}~\bibnamefont
  {Melnikov}}\ and\ \bibinfo {author} {\bibfnamefont {A.}~\bibnamefont
  {Mitov}},\ }\href {\doibase 10.1016/j.physletb.2005.06.015} {\bibfield
  {journal} {\bibinfo  {journal} {Phys.~Lett.~B}\ }\textbf {\bibinfo {volume}
  {620}},\ \bibinfo {pages} {69} (\bibinfo {year} {2005})},\ \Eprint
  {http://arxiv.org/abs/hep-ph/0505097} {hep-ph/0505097} \BibitemShut {NoStop}%
\bibitem [{\citenamefont {Ewerth}(2008)}]{Ewerth:2008nv}%
  \BibitemOpen
  \bibfield  {author} {\bibinfo {author} {\bibfnamefont {T.}~\bibnamefont
  {Ewerth}},\ }\href {\doibase 10.1016/j.physletb.2008.09.045} {\bibfield
  {journal} {\bibinfo  {journal} {Phys.~Lett.~B}\ }\textbf {\bibinfo {volume}
  {669}},\ \bibinfo {pages} {167} (\bibinfo {year} {2008})},\ \Eprint
  {http://arxiv.org/abs/0805.3911} {arXiv:0805.3911 [hep-ph]} \BibitemShut
  {NoStop}%
\bibitem [{\citenamefont {Asatrian}\ \emph {et~al.}(2010)\citenamefont
  {Asatrian}, \citenamefont {Ewerth}, \citenamefont {Ferroglia}, \citenamefont
  {Greub},\ and\ \citenamefont {Ossola}}]{Asatrian:2010rq}%
  \BibitemOpen
  \bibfield  {author} {\bibinfo {author} {\bibfnamefont {H.}~\bibnamefont
  {Asatrian}}, \bibinfo {author} {\bibfnamefont {T.}~\bibnamefont {Ewerth}},
  \bibinfo {author} {\bibfnamefont {A.}~\bibnamefont {Ferroglia}}, \bibinfo
  {author} {\bibfnamefont {C.}~\bibnamefont {Greub}}, \ and\ \bibinfo {author}
  {\bibfnamefont {G.}~\bibnamefont {Ossola}},\ }\href {\doibase
  10.1103/PhysRevD.82.074006} {\bibfield  {journal} {\bibinfo  {journal}
  {Phys.~Rev.~D}\ }\textbf {\bibinfo {volume} {82}},\ \bibinfo {pages} {074006}
  (\bibinfo {year} {2010})},\ \Eprint {http://arxiv.org/abs/1005.5587}
  {arXiv:1005.5587 [hep-ph]} \BibitemShut {NoStop}%
\bibitem [{\citenamefont {Ligeti}\ \emph {et~al.}(1999)\citenamefont {Ligeti},
  \citenamefont {Luke}, \citenamefont {Manohar},\ and\ \citenamefont
  {Wise}}]{Ligeti:1999ea}%
  \BibitemOpen
  \bibfield  {author} {\bibinfo {author} {\bibfnamefont {Z.}~\bibnamefont
  {Ligeti}}, \bibinfo {author} {\bibfnamefont {M.~E.}\ \bibnamefont {Luke}},
  \bibinfo {author} {\bibfnamefont {A.~V.}\ \bibnamefont {Manohar}}, \ and\
  \bibinfo {author} {\bibfnamefont {M.~B.}\ \bibnamefont {Wise}},\ }\href
  {\doibase 10.1103/PhysRevD.60.034019} {\bibfield  {journal} {\bibinfo
  {journal} {Phys.~Rev.~D}\ }\textbf {\bibinfo {volume} {60}},\ \bibinfo
  {pages} {034019} (\bibinfo {year} {1999})},\ \Eprint
  {http://arxiv.org/abs/hep-ph/9903305} {hep-ph/9903305} \BibitemShut {NoStop}%
\bibitem [{\citenamefont {Ferroglia}\ and\ \citenamefont
  {Haisch}(2010)}]{Ferroglia:2010xe}%
  \BibitemOpen
  \bibfield  {author} {\bibinfo {author} {\bibfnamefont {A.}~\bibnamefont
  {Ferroglia}}\ and\ \bibinfo {author} {\bibfnamefont {U.}~\bibnamefont
  {Haisch}},\ }\href {\doibase 10.1103/PhysRevD.82.094012} {\bibfield
  {journal} {\bibinfo  {journal} {Phys. Rev.}\ }\textbf {\bibinfo {volume}
  {D82}},\ \bibinfo {pages} {094012} (\bibinfo {year} {2010})},\ \Eprint
  {http://arxiv.org/abs/1009.2144} {arXiv:1009.2144 [hep-ph]} \BibitemShut
  {NoStop}%
\bibitem [{\citenamefont {Misiak}\ and\ \citenamefont
  {Poradzinski}(2011)}]{Misiak:2010tk}%
  \BibitemOpen
  \bibfield  {author} {\bibinfo {author} {\bibfnamefont {M.}~\bibnamefont
  {Misiak}}\ and\ \bibinfo {author} {\bibfnamefont {M.}~\bibnamefont
  {Poradzinski}},\ }\href {\doibase 10.1103/PhysRevD.83.014024} {\bibfield
  {journal} {\bibinfo  {journal} {Phys.~Rev.~D}\ }\textbf {\bibinfo {volume}
  {83}},\ \bibinfo {pages} {014024} (\bibinfo {year} {2011})},\ \Eprint
  {http://arxiv.org/abs/1009.5685} {arXiv:1009.5685 [hep-ph]} \BibitemShut
  {NoStop}%
\bibitem [{\citenamefont {Lee}\ and\ \citenamefont
  {Stewart}(2005)}]{Lee:2004ja}%
  \BibitemOpen
  \bibfield  {author} {\bibinfo {author} {\bibfnamefont {K.~S.~M.}\
  \bibnamefont {Lee}}\ and\ \bibinfo {author} {\bibfnamefont {I.~W.}\
  \bibnamefont {Stewart}},\ }\href {\doibase 10.1016/j.nuclphysb.2005.05.004}
  {\bibfield  {journal} {\bibinfo  {journal} {Nucl. Phys.}\ }\textbf {\bibinfo
  {volume} {B721}},\ \bibinfo {pages} {325} (\bibinfo {year} {2005})},\ \Eprint
  {http://arxiv.org/abs/hep-ph/0409045} {hep-ph/0409045} \BibitemShut {NoStop}%
\bibitem [{\citenamefont {Benzke}\ \emph {et~al.}(2010)\citenamefont {Benzke},
  \citenamefont {Lee}, \citenamefont {Neubert},\ and\ \citenamefont
  {Paz}}]{Benzke:2010js}%
  \BibitemOpen
  \bibfield  {author} {\bibinfo {author} {\bibfnamefont {M.}~\bibnamefont
  {Benzke}}, \bibinfo {author} {\bibfnamefont {S.~J.}\ \bibnamefont {Lee}},
  \bibinfo {author} {\bibfnamefont {M.}~\bibnamefont {Neubert}}, \ and\
  \bibinfo {author} {\bibfnamefont {G.}~\bibnamefont {Paz}},\ }\href {\doibase
  10.1007/JHEP08(2010)099} {\bibfield  {journal} {\bibinfo  {journal} {JHEP}\
  }\textbf {\bibinfo {volume} {08}},\ \bibinfo {pages} {099} (\bibinfo {year}
  {2010})},\ \Eprint {http://arxiv.org/abs/1003.5012} {arXiv:1003.5012
  [hep-ph]} \BibitemShut {NoStop}%
\bibitem [{\citenamefont {Gunawardana}\ and\ \citenamefont
  {Paz}(2019)}]{Gunawardana:2019gep}%
  \BibitemOpen
  \bibfield  {author} {\bibinfo {author} {\bibfnamefont {A.}~\bibnamefont
  {Gunawardana}}\ and\ \bibinfo {author} {\bibfnamefont {G.}~\bibnamefont
  {Paz}},\ }\href {\doibase 10.1007/JHEP11(2019)141} {\bibfield  {journal}
  {\bibinfo  {journal} {JHEP}\ }\textbf {\bibinfo {volume} {11}},\ \bibinfo
  {pages} {141} (\bibinfo {year} {2019})},\ \Eprint
  {http://arxiv.org/abs/1908.02812} {arXiv:1908.02812 [hep-ph]} \BibitemShut
  {NoStop}%
\bibitem [{\citenamefont {Voloshin}(1997)}]{Voloshin:1996gw}%
  \BibitemOpen
  \bibfield  {author} {\bibinfo {author} {\bibfnamefont {M.}~\bibnamefont
  {Voloshin}},\ }\href {\doibase 10.1016/S0370-2693(97)00173-1} {\bibfield
  {journal} {\bibinfo  {journal} {Phys.~Lett.~B}\ }\textbf {\bibinfo {volume}
  {397}},\ \bibinfo {pages} {275} (\bibinfo {year} {1997})},\ \Eprint
  {http://arxiv.org/abs/hep-ph/9612483} {hep-ph/9612483} \BibitemShut {NoStop}%
\bibitem [{\citenamefont {Ligeti}\ \emph {et~al.}(1997)\citenamefont {Ligeti},
  \citenamefont {Randall},\ and\ \citenamefont {Wise}}]{Ligeti:1997tc}%
  \BibitemOpen
  \bibfield  {author} {\bibinfo {author} {\bibfnamefont {Z.}~\bibnamefont
  {Ligeti}}, \bibinfo {author} {\bibfnamefont {L.}~\bibnamefont {Randall}}, \
  and\ \bibinfo {author} {\bibfnamefont {M.~B.}\ \bibnamefont {Wise}},\ }\href
  {\doibase 10.1016/S0370-2693(97)00304-3} {\bibfield  {journal} {\bibinfo
  {journal} {Phys.~Lett.~B}\ }\textbf {\bibinfo {volume} {402}},\ \bibinfo
  {pages} {178} (\bibinfo {year} {1997})},\ \Eprint
  {http://arxiv.org/abs/hep-ph/9702322} {hep-ph/9702322} \BibitemShut {NoStop}%
\bibitem [{\citenamefont {Grant}\ \emph {et~al.}(1997)\citenamefont {Grant},
  \citenamefont {Morgan}, \citenamefont {Nussinov},\ and\ \citenamefont
  {Peccei}}]{Grant:1997ec}%
  \BibitemOpen
  \bibfield  {author} {\bibinfo {author} {\bibfnamefont {A.~K.}\ \bibnamefont
  {Grant}}, \bibinfo {author} {\bibfnamefont {A.}~\bibnamefont {Morgan}},
  \bibinfo {author} {\bibfnamefont {S.}~\bibnamefont {Nussinov}}, \ and\
  \bibinfo {author} {\bibfnamefont {R.}~\bibnamefont {Peccei}},\ }\href
  {\doibase 10.1103/PhysRevD.56.3151} {\bibfield  {journal} {\bibinfo
  {journal} {Phys.~Rev.~D}\ }\textbf {\bibinfo {volume} {56}},\ \bibinfo
  {pages} {3151} (\bibinfo {year} {1997})},\ \Eprint
  {http://arxiv.org/abs/hep-ph/9702380} {hep-ph/9702380} \BibitemShut {NoStop}%
\bibitem [{\citenamefont {Tanabashi}\ \emph {et~al.}(2018)\citenamefont
  {Tanabashi} \emph {et~al.}}]{pdg201819}%
  \BibitemOpen
  \bibfield  {author} {\bibinfo {author} {\bibfnamefont {M.}~\bibnamefont
  {Tanabashi}} \emph {et~al.} (\bibinfo {collaboration} {Particle Data
  Group}),\ }\href {\doibase 10.1103/PhysRevD.98.030001} {\bibfield  {journal}
  {\bibinfo  {journal} {Phys.~Rev.~D}\ }\textbf {\bibinfo {volume} {98}},\
  \bibinfo {pages} {030001} (\bibinfo {year} {2018})},\ \bibinfo {note} {{and
  2019 update}}\BibitemShut {NoStop}%
\bibitem [{\citenamefont {Limosani}()}]{Tony}%
  \BibitemOpen
  \bibfield  {author} {\bibinfo {author} {\bibfnamefont {A.}~\bibnamefont
  {Limosani}} (\bibinfo {collaboration} {Belle Collaboration}),\ }\href@noop {}
  {}\bibinfo {note} {{private communication}}\BibitemShut {NoStop}%
\bibitem [{\citenamefont {Hoang}(2000)}]{Hoang:2000fm}%
  \BibitemOpen
  \bibfield  {author} {\bibinfo {author} {\bibfnamefont {A.}~\bibnamefont
  {Hoang}},\ }\href@noop {} {\  (\bibinfo {year} {2000})},\ \Eprint
  {http://arxiv.org/abs/hep-ph/0008102} {hep-ph/0008102} \BibitemShut {NoStop}%
\bibitem [{\citenamefont {Lee}\ and\ \citenamefont
  {Stewart}(2006)}]{Lee:2005pk}%
  \BibitemOpen
  \bibfield  {author} {\bibinfo {author} {\bibfnamefont {K.~S.}\ \bibnamefont
  {Lee}}\ and\ \bibinfo {author} {\bibfnamefont {I.~W.}\ \bibnamefont
  {Stewart}},\ }\href {\doibase 10.1103/PhysRevD.74.014005} {\bibfield
  {journal} {\bibinfo  {journal} {Phys.~Rev.~D}\ }\textbf {\bibinfo {volume}
  {74}},\ \bibinfo {pages} {014005} (\bibinfo {year} {2006})},\ \Eprint
  {http://arxiv.org/abs/hep-ph/0511334} {hep-ph/0511334} \BibitemShut {NoStop}%
\bibitem [{\citenamefont {Lee}\ \emph {et~al.}(2007{\natexlab{a}})\citenamefont
  {Lee}, \citenamefont {Ligeti}, \citenamefont {Stewart},\ and\ \citenamefont
  {Tackmann}}]{Lee:2006gs}%
  \BibitemOpen
  \bibfield  {author} {\bibinfo {author} {\bibfnamefont {K.~S.}\ \bibnamefont
  {Lee}}, \bibinfo {author} {\bibfnamefont {Z.}~\bibnamefont {Ligeti}},
  \bibinfo {author} {\bibfnamefont {I.~W.}\ \bibnamefont {Stewart}}, \ and\
  \bibinfo {author} {\bibfnamefont {F.~J.}\ \bibnamefont {Tackmann}},\ }\href
  {\doibase 10.1103/PhysRevD.75.034016} {\bibfield  {journal} {\bibinfo
  {journal} {Phys.~Rev.~D}\ }\textbf {\bibinfo {volume} {75}},\ \bibinfo
  {pages} {034016} (\bibinfo {year} {2007}{\natexlab{a}})},\ \Eprint
  {http://arxiv.org/abs/hep-ph/0612156} {hep-ph/0612156} \BibitemShut {NoStop}%
\bibitem [{\citenamefont {Kapustin}\ and\ \citenamefont
  {Ligeti}(1995)}]{Kapustin:1995nr}%
  \BibitemOpen
  \bibfield  {author} {\bibinfo {author} {\bibfnamefont {A.}~\bibnamefont
  {Kapustin}}\ and\ \bibinfo {author} {\bibfnamefont {Z.}~\bibnamefont
  {Ligeti}},\ }\href {\doibase 10.1016/0370-2693(95)00762-A} {\bibfield
  {journal} {\bibinfo  {journal} {Phys.~Lett.~B}\ }\textbf {\bibinfo {volume}
  {355}},\ \bibinfo {pages} {318} (\bibinfo {year} {1995})},\ \Eprint
  {http://arxiv.org/abs/hep-ph/9506201} {hep-ph/9506201} \BibitemShut {NoStop}%
\bibitem [{\citenamefont {Buchalla}\ \emph {et~al.}(1996)\citenamefont
  {Buchalla}, \citenamefont {Buras},\ and\ \citenamefont
  {Lautenbacher}}]{Buchalla:1995vs}%
  \BibitemOpen
  \bibfield  {author} {\bibinfo {author} {\bibfnamefont {G.}~\bibnamefont
  {Buchalla}}, \bibinfo {author} {\bibfnamefont {A.~J.}\ \bibnamefont {Buras}},
  \ and\ \bibinfo {author} {\bibfnamefont {M.~E.}\ \bibnamefont
  {Lautenbacher}},\ }\href {\doibase 10.1103/RevModPhys.68.1125} {\bibfield
  {journal} {\bibinfo  {journal} {Rev. Mod. Phys.}\ }\textbf {\bibinfo {volume}
  {68}},\ \bibinfo {pages} {1125} (\bibinfo {year} {1996})},\ \Eprint
  {http://arxiv.org/abs/hep-ph/9512380} {hep-ph/9512380} \BibitemShut {NoStop}%
\bibitem [{\citenamefont {Czakon}\ \emph {et~al.}(2007)\citenamefont {Czakon},
  \citenamefont {Haisch},\ and\ \citenamefont {Misiak}}]{Czakon:2006ss}%
  \BibitemOpen
  \bibfield  {author} {\bibinfo {author} {\bibfnamefont {M.}~\bibnamefont
  {Czakon}}, \bibinfo {author} {\bibfnamefont {U.}~\bibnamefont {Haisch}}, \
  and\ \bibinfo {author} {\bibfnamefont {M.}~\bibnamefont {Misiak}},\ }\href
  {\doibase 10.1088/1126-6708/2007/03/008} {\bibfield  {journal} {\bibinfo
  {journal} {JHEP}\ }\textbf {\bibinfo {volume} {03}},\ \bibinfo {pages} {008}
  (\bibinfo {year} {2007})},\ \Eprint {http://arxiv.org/abs/hep-ph/0612329}
  {hep-ph/0612329} \BibitemShut {NoStop}%
\bibitem [{\citenamefont {Greub}\ \emph {et~al.}(1996)\citenamefont {Greub},
  \citenamefont {Hurth},\ and\ \citenamefont {Wyler}}]{Greub:1996tg}%
  \BibitemOpen
  \bibfield  {author} {\bibinfo {author} {\bibfnamefont {C.}~\bibnamefont
  {Greub}}, \bibinfo {author} {\bibfnamefont {T.}~\bibnamefont {Hurth}}, \ and\
  \bibinfo {author} {\bibfnamefont {D.}~\bibnamefont {Wyler}},\ }\href
  {\doibase 10.1103/PhysRevD.54.3350} {\bibfield  {journal} {\bibinfo
  {journal} {Phys.~Rev.~D}\ }\textbf {\bibinfo {volume} {54}},\ \bibinfo
  {pages} {3350} (\bibinfo {year} {1996})},\ \Eprint
  {http://arxiv.org/abs/hep-ph/9603404} {hep-ph/9603404} \BibitemShut {NoStop}%
\bibitem [{\citenamefont {Buras}\ \emph {et~al.}(2001)\citenamefont {Buras},
  \citenamefont {Czarnecki}, \citenamefont {Misiak},\ and\ \citenamefont
  {Urban}}]{Buras:2001mq}%
  \BibitemOpen
  \bibfield  {author} {\bibinfo {author} {\bibfnamefont {A.~J.}\ \bibnamefont
  {Buras}}, \bibinfo {author} {\bibfnamefont {A.}~\bibnamefont {Czarnecki}},
  \bibinfo {author} {\bibfnamefont {M.}~\bibnamefont {Misiak}}, \ and\ \bibinfo
  {author} {\bibfnamefont {J.}~\bibnamefont {Urban}},\ }\href {\doibase
  10.1016/S0550-3213(01)00336-4} {\bibfield  {journal} {\bibinfo  {journal}
  {Nucl. Phys.}\ }\textbf {\bibinfo {volume} {B611}},\ \bibinfo {pages} {488}
  (\bibinfo {year} {2001})},\ \Eprint {http://arxiv.org/abs/hep-ph/0105160}
  {hep-ph/0105160} \BibitemShut {NoStop}%
\bibitem [{\citenamefont {Buras}\ \emph {et~al.}(2002)\citenamefont {Buras},
  \citenamefont {Czarnecki}, \citenamefont {Misiak},\ and\ \citenamefont
  {Urban}}]{Buras:2002tp}%
  \BibitemOpen
  \bibfield  {author} {\bibinfo {author} {\bibfnamefont {A.~J.}\ \bibnamefont
  {Buras}}, \bibinfo {author} {\bibfnamefont {A.}~\bibnamefont {Czarnecki}},
  \bibinfo {author} {\bibfnamefont {M.}~\bibnamefont {Misiak}}, \ and\ \bibinfo
  {author} {\bibfnamefont {J.}~\bibnamefont {Urban}},\ }\href {\doibase
  10.1016/S0550-3213(02)00261-4} {\bibfield  {journal} {\bibinfo  {journal}
  {Nucl.~Phys.}\ }\textbf {\bibinfo {volume} {B631}},\ \bibinfo {pages} {219}
  (\bibinfo {year} {2002})},\ \Eprint {http://arxiv.org/abs/hep-ph/0203135}
  {hep-ph/0203135} \BibitemShut {NoStop}%
\bibitem [{\citenamefont {Bieri}\ \emph {et~al.}(2003)\citenamefont {Bieri},
  \citenamefont {Greub},\ and\ \citenamefont {Steinhauser}}]{Bieri:2003ue}%
  \BibitemOpen
  \bibfield  {author} {\bibinfo {author} {\bibfnamefont {K.}~\bibnamefont
  {Bieri}}, \bibinfo {author} {\bibfnamefont {C.}~\bibnamefont {Greub}}, \ and\
  \bibinfo {author} {\bibfnamefont {M.}~\bibnamefont {Steinhauser}},\ }\href
  {\doibase 10.1103/PhysRevD.67.114019} {\bibfield  {journal} {\bibinfo
  {journal} {Phys.~Rev.~D}\ }\textbf {\bibinfo {volume} {67}},\ \bibinfo
  {pages} {114019} (\bibinfo {year} {2003})},\ \Eprint
  {http://arxiv.org/abs/hep-ph/0302051} {hep-ph/0302051} \BibitemShut {NoStop}%
\bibitem [{\citenamefont {Misiak}\ and\ \citenamefont
  {Steinhauser}(2010)}]{Misiak:2010sk}%
  \BibitemOpen
  \bibfield  {author} {\bibinfo {author} {\bibfnamefont {M.}~\bibnamefont
  {Misiak}}\ and\ \bibinfo {author} {\bibfnamefont {M.}~\bibnamefont
  {Steinhauser}},\ }\href {\doibase 10.1016/j.nuclphysb.2010.07.009} {\bibfield
   {journal} {\bibinfo  {journal} {Nucl. Phys.}\ }\textbf {\bibinfo {volume}
  {B840}},\ \bibinfo {pages} {271} (\bibinfo {year} {2010})},\ \Eprint
  {http://arxiv.org/abs/1005.1173} {arXiv:1005.1173 [hep-ph]} \BibitemShut
  {NoStop}%
\bibitem [{\citenamefont {Korchemsky}\ and\ \citenamefont
  {Sterman}(1994)}]{Korchemsky:1994jb}%
  \BibitemOpen
  \bibfield  {author} {\bibinfo {author} {\bibfnamefont {G.~P.}\ \bibnamefont
  {Korchemsky}}\ and\ \bibinfo {author} {\bibfnamefont {G.~F.}\ \bibnamefont
  {Sterman}},\ }\href {\doibase 10.1016/0370-2693(94)91304-8} {\bibfield
  {journal} {\bibinfo  {journal} {Phys.~Lett.~B}\ }\textbf {\bibinfo {volume}
  {340}},\ \bibinfo {pages} {96} (\bibinfo {year} {1994})},\ \Eprint
  {http://arxiv.org/abs/hep-ph/9407344} {hep-ph/9407344} \BibitemShut {NoStop}%
\bibitem [{\citenamefont {Asatrian}\ \emph
  {et~al.}(2007{\natexlab{a}})\citenamefont {Asatrian}, \citenamefont {Ewerth},
  \citenamefont {Gabrielyan},\ and\ \citenamefont {Greub}}]{Asatrian:2006rq}%
  \BibitemOpen
  \bibfield  {author} {\bibinfo {author} {\bibfnamefont {H.}~\bibnamefont
  {Asatrian}}, \bibinfo {author} {\bibfnamefont {T.}~\bibnamefont {Ewerth}},
  \bibinfo {author} {\bibfnamefont {H.}~\bibnamefont {Gabrielyan}}, \ and\
  \bibinfo {author} {\bibfnamefont {C.}~\bibnamefont {Greub}},\ }\href
  {\doibase 10.1016/j.physletb.2007.02.027} {\bibfield  {journal} {\bibinfo
  {journal} {Phys. Lett. B}\ }\textbf {\bibinfo {volume} {647}},\ \bibinfo
  {pages} {173} (\bibinfo {year} {2007}{\natexlab{a}})},\ \Eprint
  {http://arxiv.org/abs/hep-ph/0611123} {hep-ph/0611123} \BibitemShut {NoStop}%
\bibitem [{\citenamefont {Asatrian}\ \emph
  {et~al.}(2007{\natexlab{b}})\citenamefont {Asatrian}, \citenamefont {Ewerth},
  \citenamefont {Ferroglia}, \citenamefont {Gambino},\ and\ \citenamefont
  {Greub}}]{Asatrian:2006sm}%
  \BibitemOpen
  \bibfield  {author} {\bibinfo {author} {\bibfnamefont {H.~M.}\ \bibnamefont
  {Asatrian}}, \bibinfo {author} {\bibfnamefont {T.}~\bibnamefont {Ewerth}},
  \bibinfo {author} {\bibfnamefont {A.}~\bibnamefont {Ferroglia}}, \bibinfo
  {author} {\bibfnamefont {P.}~\bibnamefont {Gambino}}, \ and\ \bibinfo
  {author} {\bibfnamefont {C.}~\bibnamefont {Greub}},\ }\href {\doibase
  10.1016/j.nuclphysb.2006.11.002} {\bibfield  {journal} {\bibinfo  {journal}
  {Nucl. Phys.}\ }\textbf {\bibinfo {volume} {B762}},\ \bibinfo {pages} {212}
  (\bibinfo {year} {2007}{\natexlab{b}})},\ \Eprint
  {http://arxiv.org/abs/hep-ph/0607316} {hep-ph/0607316} \BibitemShut {NoStop}%
\bibitem [{\citenamefont {Misiak}(2008)}]{Misiak:2008ss}%
  \BibitemOpen
  \bibfield  {author} {\bibinfo {author} {\bibfnamefont {M.}~\bibnamefont
  {Misiak}},\ }in\ \href@noop {} {\emph {\bibinfo {booktitle} {{Heavy Quarks
  and Leptons 2008 (HQ\&L08)}}}}\ (\bibinfo {year} {2008})\ \Eprint
  {http://arxiv.org/abs/0808.3134} {arXiv:0808.3134 [hep-ph]} \BibitemShut
  {NoStop}%
\bibitem [{\citenamefont {Ali}\ and\ \citenamefont
  {Greub}(1991{\natexlab{a}})}]{Ali:1990tj}%
  \BibitemOpen
  \bibfield  {author} {\bibinfo {author} {\bibfnamefont {A.}~\bibnamefont
  {Ali}}\ and\ \bibinfo {author} {\bibfnamefont {C.}~\bibnamefont {Greub}},\
  }\href {\doibase 10.1007/BF01549696} {\bibfield  {journal} {\bibinfo
  {journal} {Z.~Phys.~C}\ }\textbf {\bibinfo {volume} {49}},\ \bibinfo {pages}
  {431} (\bibinfo {year} {1991}{\natexlab{a}})}\BibitemShut {NoStop}%
\bibitem [{\citenamefont {Ali}\ and\ \citenamefont
  {Greub}(1991{\natexlab{b}})}]{Ali:1990vp}%
  \BibitemOpen
  \bibfield  {author} {\bibinfo {author} {\bibfnamefont {A.}~\bibnamefont
  {Ali}}\ and\ \bibinfo {author} {\bibfnamefont {C.}~\bibnamefont {Greub}},\
  }\href {\doibase 10.1016/0370-2693(91)90156-K} {\bibfield  {journal}
  {\bibinfo  {journal} {Phys.~Lett.~B}\ }\textbf {\bibinfo {volume} {259}},\
  \bibinfo {pages} {182} (\bibinfo {year} {1991}{\natexlab{b}})}\BibitemShut
  {NoStop}%
\bibitem [{\citenamefont {Ali}\ and\ \citenamefont {Greub}(1995)}]{Ali:1995bi}%
  \BibitemOpen
  \bibfield  {author} {\bibinfo {author} {\bibfnamefont {A.}~\bibnamefont
  {Ali}}\ and\ \bibinfo {author} {\bibfnamefont {C.}~\bibnamefont {Greub}},\
  }\href {\doibase 10.1016/0370-2693(95)01118-A} {\bibfield  {journal}
  {\bibinfo  {journal} {Phys. Lett.}\ }\textbf {\bibinfo {volume} {B361}},\
  \bibinfo {pages} {146} (\bibinfo {year} {1995})},\ \Eprint
  {http://arxiv.org/abs/hep-ph/9506374} {hep-ph/9506374} \BibitemShut {NoStop}%
\bibitem [{\citenamefont {Pott}(1996)}]{Pott:1995if}%
  \BibitemOpen
  \bibfield  {author} {\bibinfo {author} {\bibfnamefont {N.}~\bibnamefont
  {Pott}},\ }\href {\doibase 10.1103/PhysRevD.54.938} {\bibfield  {journal}
  {\bibinfo  {journal} {Phys. Rev.}\ }\textbf {\bibinfo {volume} {D54}},\
  \bibinfo {pages} {938} (\bibinfo {year} {1996})},\ \Eprint
  {http://arxiv.org/abs/hep-ph/9512252} {hep-ph/9512252} \BibitemShut {NoStop}%
\bibitem [{\citenamefont {Abbate}\ \emph {et~al.}(2011)\citenamefont {Abbate},
  \citenamefont {Fickinger}, \citenamefont {Hoang}, \citenamefont {Mateu},\
  and\ \citenamefont {Stewart}}]{Abbate:2010xh}%
  \BibitemOpen
  \bibfield  {author} {\bibinfo {author} {\bibfnamefont {R.}~\bibnamefont
  {Abbate}}, \bibinfo {author} {\bibfnamefont {M.}~\bibnamefont {Fickinger}},
  \bibinfo {author} {\bibfnamefont {A.~H.}\ \bibnamefont {Hoang}}, \bibinfo
  {author} {\bibfnamefont {V.}~\bibnamefont {Mateu}}, \ and\ \bibinfo {author}
  {\bibfnamefont {I.~W.}\ \bibnamefont {Stewart}},\ }\href {\doibase
  10.1103/PhysRevD.83.074021} {\bibfield  {journal} {\bibinfo  {journal} {Phys.
  Rev.}\ }\textbf {\bibinfo {volume} {D83}},\ \bibinfo {pages} {074021}
  (\bibinfo {year} {2011})},\ \Eprint {http://arxiv.org/abs/1006.3080}
  {arXiv:1006.3080 [hep-ph]} \BibitemShut {NoStop}%
\bibitem [{\citenamefont {Tackmann}(2005)}]{Tackmann:2005ub}%
  \BibitemOpen
  \bibfield  {author} {\bibinfo {author} {\bibfnamefont {F.~J.}\ \bibnamefont
  {Tackmann}},\ }\href {\doibase 10.1103/PhysRevD.72.034036} {\bibfield
  {journal} {\bibinfo  {journal} {Phys.~Rev.~D}\ }\textbf {\bibinfo {volume}
  {72}},\ \bibinfo {pages} {034036} (\bibinfo {year} {2005})},\ \Eprint
  {http://arxiv.org/abs/hep-ph/0503095} {hep-ph/0503095} \BibitemShut {NoStop}%
\bibitem [{\citenamefont {Gremm}\ and\ \citenamefont
  {Kapustin}(1997)}]{Gremm:1996df}%
  \BibitemOpen
  \bibfield  {author} {\bibinfo {author} {\bibfnamefont {M.}~\bibnamefont
  {Gremm}}\ and\ \bibinfo {author} {\bibfnamefont {A.}~\bibnamefont
  {Kapustin}},\ }\href {\doibase 10.1103/PhysRevD.55.6924} {\bibfield
  {journal} {\bibinfo  {journal} {Phys.~Rev.~D}\ }\textbf {\bibinfo {volume}
  {55}},\ \bibinfo {pages} {6924} (\bibinfo {year} {1997})},\ \Eprint
  {http://arxiv.org/abs/hep-ph/9603448} {hep-ph/9603448} \BibitemShut {NoStop}%
\bibitem [{\citenamefont {Bauer}(1998)}]{Bauer:1997fe}%
  \BibitemOpen
  \bibfield  {author} {\bibinfo {author} {\bibfnamefont {C.~W.}\ \bibnamefont
  {Bauer}},\ }\href {\doibase 10.1103/PhysRevD.60.099907,
  10.1103/PhysRevD.57.5611} {\bibfield  {journal} {\bibinfo  {journal} {Phys.
  Rev.}\ }\textbf {\bibinfo {volume} {D57}},\ \bibinfo {pages} {5611} (\bibinfo
  {year} {1998})},\ \bibinfo {note} {[Erratum: Phys. Rev.D60,099907(1999)]},\
  \Eprint {http://arxiv.org/abs/hep-ph/9710513} {hep-ph/9710513} \BibitemShut
  {NoStop}%
\bibitem [{\citenamefont {Kapustin}\ \emph {et~al.}(1995)\citenamefont
  {Kapustin}, \citenamefont {Ligeti},\ and\ \citenamefont
  {Politzer}}]{Kapustin:1995fk}%
  \BibitemOpen
  \bibfield  {author} {\bibinfo {author} {\bibfnamefont {A.}~\bibnamefont
  {Kapustin}}, \bibinfo {author} {\bibfnamefont {Z.}~\bibnamefont {Ligeti}}, \
  and\ \bibinfo {author} {\bibfnamefont {H.~D.}\ \bibnamefont {Politzer}},\
  }\href {\doibase 10.1016/0370-2693(95)00962-K} {\bibfield  {journal}
  {\bibinfo  {journal} {Phys.~Lett.~B}\ }\textbf {\bibinfo {volume} {357}},\
  \bibinfo {pages} {653} (\bibinfo {year} {1995})},\ \Eprint
  {http://arxiv.org/abs/hep-ph/9507248} {hep-ph/9507248} \BibitemShut {NoStop}%
\bibitem [{\citenamefont {Lee}\ \emph {et~al.}(2007{\natexlab{b}})\citenamefont
  {Lee}, \citenamefont {Neubert},\ and\ \citenamefont {Paz}}]{Lee:2006wn}%
  \BibitemOpen
  \bibfield  {author} {\bibinfo {author} {\bibfnamefont {S.~J.}\ \bibnamefont
  {Lee}}, \bibinfo {author} {\bibfnamefont {M.}~\bibnamefont {Neubert}}, \ and\
  \bibinfo {author} {\bibfnamefont {G.}~\bibnamefont {Paz}},\ }\href {\doibase
  10.1103/PhysRevD.75.114005} {\bibfield  {journal} {\bibinfo  {journal}
  {Phys.~Rev.~D}\ }\textbf {\bibinfo {volume} {75}},\ \bibinfo {pages} {114005}
  (\bibinfo {year} {2007}{\natexlab{b}})},\ \Eprint
  {http://arxiv.org/abs/hep-ph/0609224} {hep-ph/0609224} \BibitemShut {NoStop}%
\bibitem [{\citenamefont {Misiak}(2009)}]{Misiak:2009nr}%
  \BibitemOpen
  \bibfield  {author} {\bibinfo {author} {\bibfnamefont {M.}~\bibnamefont
  {Misiak}},\ }\href@noop {} {\bibfield  {journal} {\bibinfo  {journal} {Acta
  Phys. Polon.}\ }\textbf {\bibinfo {volume} {B40}},\ \bibinfo {pages} {2987}
  (\bibinfo {year} {2009})},\ \Eprint {http://arxiv.org/abs/0911.1651}
  {arXiv:0911.1651 [hep-ph]} \BibitemShut {NoStop}%
\bibitem [{\citenamefont {Watanuki}\ \emph {et~al.}(2019)\citenamefont
  {Watanuki} \emph {et~al.}}]{Watanuki:2018xxg}%
  \BibitemOpen
  \bibfield  {author} {\bibinfo {author} {\bibfnamefont {S.}~\bibnamefont
  {Watanuki}} \emph {et~al.} (\bibinfo {collaboration} {Belle}),\ }\href
  {\doibase 10.1103/PhysRevD.99.032012} {\bibfield  {journal} {\bibinfo
  {journal} {Phys. Rev.}\ }\textbf {\bibinfo {volume} {D99}},\ \bibinfo {pages}
  {032012} (\bibinfo {year} {2019})},\ \Eprint
  {http://arxiv.org/abs/1807.04236} {arXiv:1807.04236 [hep-ex]} \BibitemShut
  {NoStop}%
\bibitem [{\citenamefont {Bobeth}\ \emph {et~al.}(2000)\citenamefont {Bobeth},
  \citenamefont {Misiak},\ and\ \citenamefont {Urban}}]{Bobeth:1999mk}%
  \BibitemOpen
  \bibfield  {author} {\bibinfo {author} {\bibfnamefont {C.}~\bibnamefont
  {Bobeth}}, \bibinfo {author} {\bibfnamefont {M.}~\bibnamefont {Misiak}}, \
  and\ \bibinfo {author} {\bibfnamefont {J.}~\bibnamefont {Urban}},\ }\href
  {\doibase 10.1016/S0550-3213(00)00007-9} {\bibfield  {journal} {\bibinfo
  {journal} {Nucl. Phys.}\ }\textbf {\bibinfo {volume} {B574}},\ \bibinfo
  {pages} {291} (\bibinfo {year} {2000})},\ \Eprint
  {http://arxiv.org/abs/hep-ph/9910220} {hep-ph/9910220} \BibitemShut {NoStop}%
\bibitem [{\citenamefont {Misiak}\ and\ \citenamefont
  {Steinhauser}(2004)}]{Misiak:2004ew}%
  \BibitemOpen
  \bibfield  {author} {\bibinfo {author} {\bibfnamefont {M.}~\bibnamefont
  {Misiak}}\ and\ \bibinfo {author} {\bibfnamefont {M.}~\bibnamefont
  {Steinhauser}},\ }\href {\doibase 10.1016/j.nuclphysb.2004.02.006} {\bibfield
   {journal} {\bibinfo  {journal} {Nucl. Phys.}\ }\textbf {\bibinfo {volume}
  {B683}},\ \bibinfo {pages} {277} (\bibinfo {year} {2004})},\ \Eprint
  {http://arxiv.org/abs/hep-ph/0401041} {hep-ph/0401041} \BibitemShut {NoStop}%
\bibitem [{\citenamefont {Buras}\ \emph {et~al.}(1993)\citenamefont {Buras},
  \citenamefont {Jamin}, \citenamefont {Lautenbacher},\ and\ \citenamefont
  {Weisz}}]{Buras:1992tc}%
  \BibitemOpen
  \bibfield  {author} {\bibinfo {author} {\bibfnamefont {A.~J.}\ \bibnamefont
  {Buras}}, \bibinfo {author} {\bibfnamefont {M.}~\bibnamefont {Jamin}},
  \bibinfo {author} {\bibfnamefont {M.~E.}\ \bibnamefont {Lautenbacher}}, \
  and\ \bibinfo {author} {\bibfnamefont {P.~H.}\ \bibnamefont {Weisz}},\ }\href
  {\doibase 10.1016/0550-3213(93)90397-8} {\bibfield  {journal} {\bibinfo
  {journal} {Nucl. Phys.}\ }\textbf {\bibinfo {volume} {B400}},\ \bibinfo
  {pages} {37} (\bibinfo {year} {1993})},\ \Eprint
  {http://arxiv.org/abs/hep-ph/9211304} {hep-ph/9211304} \BibitemShut {NoStop}%
\bibitem [{\citenamefont {Ciuchini}\ \emph {et~al.}(1994)\citenamefont
  {Ciuchini}, \citenamefont {Franco}, \citenamefont {Martinelli},\ and\
  \citenamefont {Reina}}]{Ciuchini:1993vr}%
  \BibitemOpen
  \bibfield  {author} {\bibinfo {author} {\bibfnamefont {M.}~\bibnamefont
  {Ciuchini}}, \bibinfo {author} {\bibfnamefont {E.}~\bibnamefont {Franco}},
  \bibinfo {author} {\bibfnamefont {G.}~\bibnamefont {Martinelli}}, \ and\
  \bibinfo {author} {\bibfnamefont {L.}~\bibnamefont {Reina}},\ }\href
  {\doibase 10.1016/0550-3213(94)90118-X} {\bibfield  {journal} {\bibinfo
  {journal} {Nucl. Phys.}\ }\textbf {\bibinfo {volume} {B415}},\ \bibinfo
  {pages} {403} (\bibinfo {year} {1994})},\ \Eprint
  {http://arxiv.org/abs/hep-ph/9304257} {hep-ph/9304257} \BibitemShut {NoStop}%
\bibitem [{\citenamefont {Chetyrkin}\ \emph {et~al.}(1997)\citenamefont
  {Chetyrkin}, \citenamefont {Misiak},\ and\ \citenamefont
  {M{\"u}nz}}]{Chetyrkin:1996vx}%
  \BibitemOpen
  \bibfield  {author} {\bibinfo {author} {\bibfnamefont {K.~G.}\ \bibnamefont
  {Chetyrkin}}, \bibinfo {author} {\bibfnamefont {M.}~\bibnamefont {Misiak}}, \
  and\ \bibinfo {author} {\bibfnamefont {M.}~\bibnamefont {M{\"u}nz}},\ }\href
  {\doibase 10.1016/S0370-2693(97)00324-9} {\bibfield  {journal} {\bibinfo
  {journal} {Phys.~Lett.~B}\ }\textbf {\bibinfo {volume} {400}},\ \bibinfo
  {pages} {206} (\bibinfo {year} {1997})},\ \bibinfo {note} {[Erratum: Phys.
  Lett.B425,414(1998)]},\ \Eprint {http://arxiv.org/abs/hep-ph/9612313}
  {hep-ph/9612313} \BibitemShut {NoStop}%
\bibitem [{\citenamefont {Gambino}\ \emph {et~al.}(2003)\citenamefont
  {Gambino}, \citenamefont {Gorbahn},\ and\ \citenamefont
  {Haisch}}]{Gambino:2003zm}%
  \BibitemOpen
  \bibfield  {author} {\bibinfo {author} {\bibfnamefont {P.}~\bibnamefont
  {Gambino}}, \bibinfo {author} {\bibfnamefont {M.}~\bibnamefont {Gorbahn}}, \
  and\ \bibinfo {author} {\bibfnamefont {U.}~\bibnamefont {Haisch}},\ }\href
  {\doibase 10.1016/j.nuclphysb.2003.09.024} {\bibfield  {journal} {\bibinfo
  {journal} {Nucl. Phys.}\ }\textbf {\bibinfo {volume} {B673}},\ \bibinfo
  {pages} {238} (\bibinfo {year} {2003})},\ \Eprint
  {http://arxiv.org/abs/hep-ph/0306079} {hep-ph/0306079} \BibitemShut {NoStop}%
\bibitem [{\citenamefont {Gorbahn}\ and\ \citenamefont
  {Haisch}(2005)}]{Gorbahn:2004my}%
  \BibitemOpen
  \bibfield  {author} {\bibinfo {author} {\bibfnamefont {M.}~\bibnamefont
  {Gorbahn}}\ and\ \bibinfo {author} {\bibfnamefont {U.}~\bibnamefont
  {Haisch}},\ }\href {\doibase 10.1016/j.nuclphysb.2005.01.047} {\bibfield
  {journal} {\bibinfo  {journal} {Nucl. Phys.}\ }\textbf {\bibinfo {volume}
  {B713}},\ \bibinfo {pages} {291} (\bibinfo {year} {2005})},\ \Eprint
  {http://arxiv.org/abs/hep-ph/0411071} {hep-ph/0411071} \BibitemShut {NoStop}%
\bibitem [{\citenamefont {Gorbahn}\ \emph {et~al.}(2005)\citenamefont
  {Gorbahn}, \citenamefont {Haisch},\ and\ \citenamefont
  {Misiak}}]{Gorbahn:2005sa}%
  \BibitemOpen
  \bibfield  {author} {\bibinfo {author} {\bibfnamefont {M.}~\bibnamefont
  {Gorbahn}}, \bibinfo {author} {\bibfnamefont {U.}~\bibnamefont {Haisch}}, \
  and\ \bibinfo {author} {\bibfnamefont {M.}~\bibnamefont {Misiak}},\ }\href
  {\doibase 10.1103/PhysRevLett.95.102004} {\bibfield  {journal} {\bibinfo
  {journal} {Phys.~Rev.~Lett.}\ }\textbf {\bibinfo {volume} {95}},\ \bibinfo
  {pages} {102004} (\bibinfo {year} {2005})},\ \Eprint
  {http://arxiv.org/abs/hep-ph/0504194} {hep-ph/0504194} \BibitemShut {NoStop}%
\bibitem [{\citenamefont {Chetyrkin}\ \emph {et~al.}(2000)\citenamefont
  {Chetyrkin}, \citenamefont {Kuhn},\ and\ \citenamefont
  {Steinhauser}}]{Chetyrkin:2000yt}%
  \BibitemOpen
  \bibfield  {author} {\bibinfo {author} {\bibfnamefont {K.~G.}\ \bibnamefont
  {Chetyrkin}}, \bibinfo {author} {\bibfnamefont {J.~H.}\ \bibnamefont {Kuhn}},
  \ and\ \bibinfo {author} {\bibfnamefont {M.}~\bibnamefont {Steinhauser}},\
  }\href {\doibase 10.1016/S0010-4655(00)00155-7} {\bibfield  {journal}
  {\bibinfo  {journal} {Comput. Phys. Commun.}\ }\textbf {\bibinfo {volume}
  {133}},\ \bibinfo {pages} {43} (\bibinfo {year} {2000})},\ \Eprint
  {http://arxiv.org/abs/hep-ph/0004189} {hep-ph/0004189} \BibitemShut {NoStop}%
\bibitem [{\citenamefont {Bauer}\ \emph {et~al.}(2003)\citenamefont {Bauer},
  \citenamefont {Ligeti}, \citenamefont {Luke},\ and\ \citenamefont
  {Manohar}}]{Bauer:2002sh}%
  \BibitemOpen
  \bibfield  {author} {\bibinfo {author} {\bibfnamefont {C.~W.}\ \bibnamefont
  {Bauer}}, \bibinfo {author} {\bibfnamefont {Z.}~\bibnamefont {Ligeti}},
  \bibinfo {author} {\bibfnamefont {M.}~\bibnamefont {Luke}}, \ and\ \bibinfo
  {author} {\bibfnamefont {A.~V.}\ \bibnamefont {Manohar}},\ }\href {\doibase
  10.1103/PhysRevD.67.054012} {\bibfield  {journal} {\bibinfo  {journal}
  {Phys.~Rev.~D}\ }\textbf {\bibinfo {volume} {67}},\ \bibinfo {pages} {054012}
  (\bibinfo {year} {2003})},\ \Eprint {http://arxiv.org/abs/hep-ph/0210027}
  {hep-ph/0210027} \BibitemShut {NoStop}%
\bibitem [{\citenamefont {Urquijo}()}]{Phill}%
  \BibitemOpen
  \bibfield  {author} {\bibinfo {author} {\bibfnamefont {P.}~\bibnamefont
  {Urquijo}},\ }\href@noop {} {}\bibinfo {note} {{private
  communication}}\BibitemShut {NoStop}%
\bibitem [{\citenamefont {Hoang}\ \emph {et~al.}(2010)\citenamefont {Hoang},
  \citenamefont {Jain}, \citenamefont {Scimemi},\ and\ \citenamefont
  {Stewart}}]{Hoang:2009yr}%
  \BibitemOpen
  \bibfield  {author} {\bibinfo {author} {\bibfnamefont {A.~H.}\ \bibnamefont
  {Hoang}}, \bibinfo {author} {\bibfnamefont {A.}~\bibnamefont {Jain}},
  \bibinfo {author} {\bibfnamefont {I.}~\bibnamefont {Scimemi}}, \ and\
  \bibinfo {author} {\bibfnamefont {I.~W.}\ \bibnamefont {Stewart}},\ }\href
  {\doibase 10.1103/PhysRevD.82.011501} {\bibfield  {journal} {\bibinfo
  {journal} {Phys. Rev.}\ }\textbf {\bibinfo {volume} {D82}},\ \bibinfo {pages}
  {011501} (\bibinfo {year} {2010})},\ \Eprint {http://arxiv.org/abs/0908.3189}
  {arXiv:0908.3189 [hep-ph]} \BibitemShut {NoStop}%
\bibitem [{\citenamefont {Grozin}\ \emph {et~al.}(2008)\citenamefont {Grozin},
  \citenamefont {Marquard}, \citenamefont {Piclum},\ and\ \citenamefont
  {Steinhauser}}]{Grozin:2007fh}%
  \BibitemOpen
  \bibfield  {author} {\bibinfo {author} {\bibfnamefont {A.~G.}\ \bibnamefont
  {Grozin}}, \bibinfo {author} {\bibfnamefont {P.}~\bibnamefont {Marquard}},
  \bibinfo {author} {\bibfnamefont {J.~H.}\ \bibnamefont {Piclum}}, \ and\
  \bibinfo {author} {\bibfnamefont {M.}~\bibnamefont {Steinhauser}},\ }\href
  {\doibase 10.1016/j.nuclphysb.2007.08.012} {\bibfield  {journal} {\bibinfo
  {journal} {Nucl. Phys.}\ }\textbf {\bibinfo {volume} {B789}},\ \bibinfo
  {pages} {277} (\bibinfo {year} {2008})},\ \Eprint
  {http://arxiv.org/abs/0707.1388} {arXiv:0707.1388 [hep-ph]} \BibitemShut
  {NoStop}%
\bibitem [{\citenamefont {Hoang}\ \emph {et~al.}(2008)\citenamefont {Hoang},
  \citenamefont {Jain}, \citenamefont {Scimemi},\ and\ \citenamefont
  {Stewart}}]{Hoang:2008yj}%
  \BibitemOpen
  \bibfield  {author} {\bibinfo {author} {\bibfnamefont {A.~H.}\ \bibnamefont
  {Hoang}}, \bibinfo {author} {\bibfnamefont {A.}~\bibnamefont {Jain}},
  \bibinfo {author} {\bibfnamefont {I.}~\bibnamefont {Scimemi}}, \ and\
  \bibinfo {author} {\bibfnamefont {I.~W.}\ \bibnamefont {Stewart}},\ }\href
  {\doibase 10.1103/PhysRevLett.101.151602} {\bibfield  {journal} {\bibinfo
  {journal} {Phys. Rev. Lett.}\ }\textbf {\bibinfo {volume} {101}},\ \bibinfo
  {pages} {151602} (\bibinfo {year} {2008})},\ \Eprint
  {http://arxiv.org/abs/0803.4214} {arXiv:0803.4214 [hep-ph]} \BibitemShut
  {NoStop}%
\end{thebibliography}%


\onecolumngrid
\clearpage

\setcounter{equation}{1000}
\setcounter{figure}{1000}
\setcounter{table}{1000}

\renewcommand{\theequation}{S\the\numexpr\value{equation}-1000\relax}
\renewcommand{\thefigure}{S\the\numexpr\value{figure}-1000\relax}
\renewcommand{\thetable}{S\the\numexpr\value{table}-1000\relax}

\section*{Supplemental material}

\subsection{Additional fit results}

\begin{figure}[t]
\includegraphics[width=0.5\textwidth]{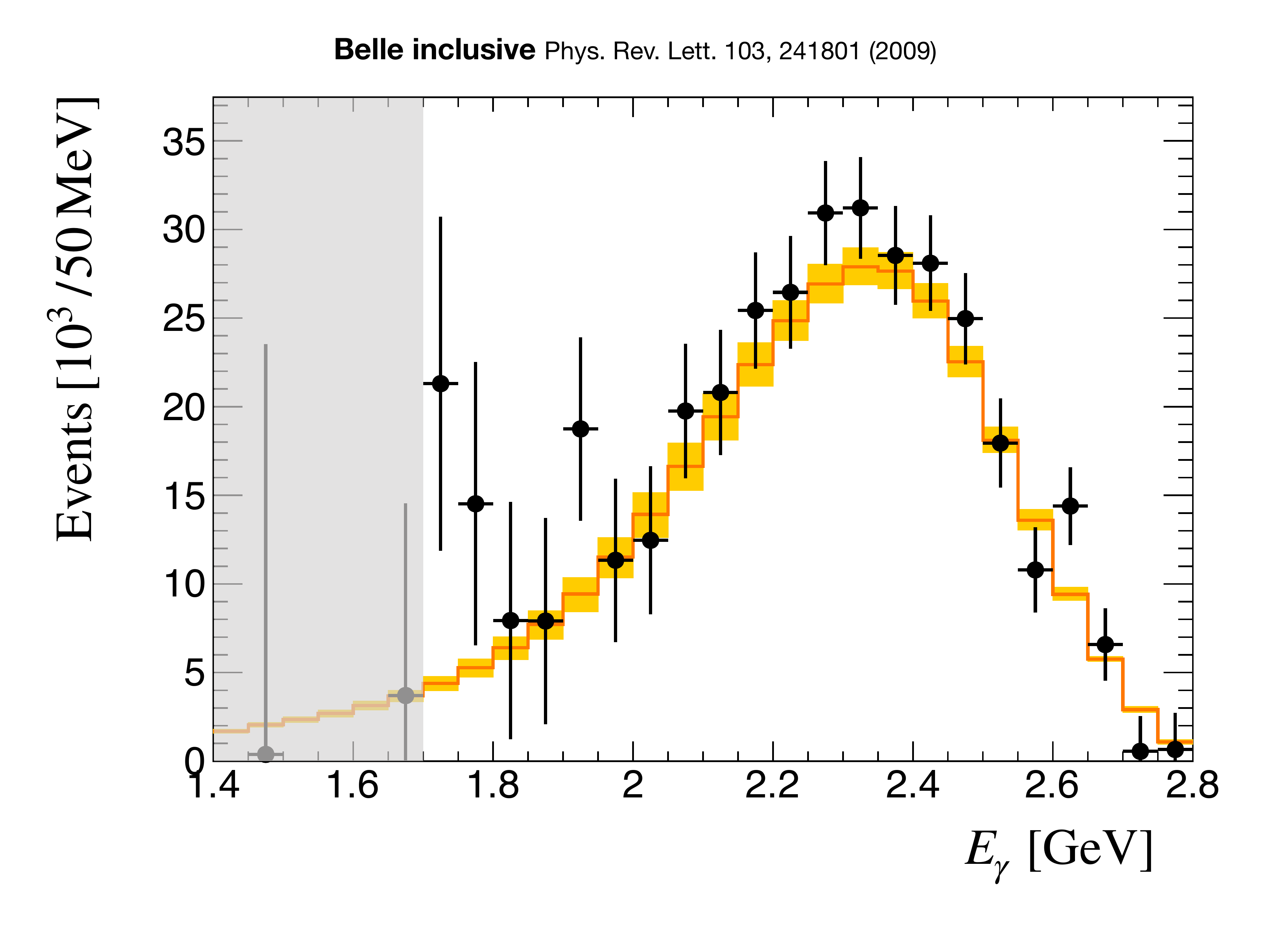}%
\hfill%
\includegraphics[width=0.5\textwidth]{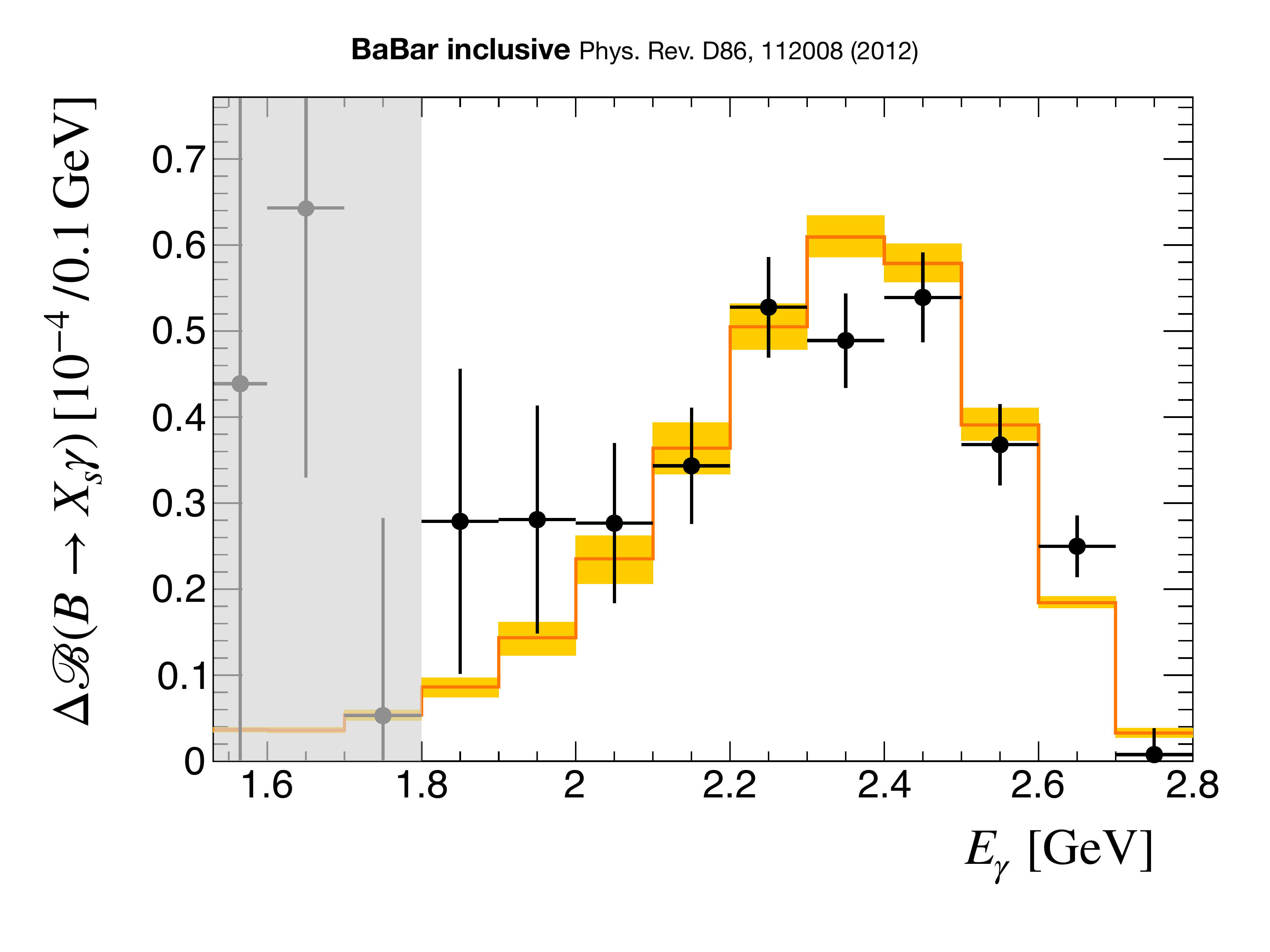}%
\\[-2ex]
\includegraphics[width=0.5\textwidth]{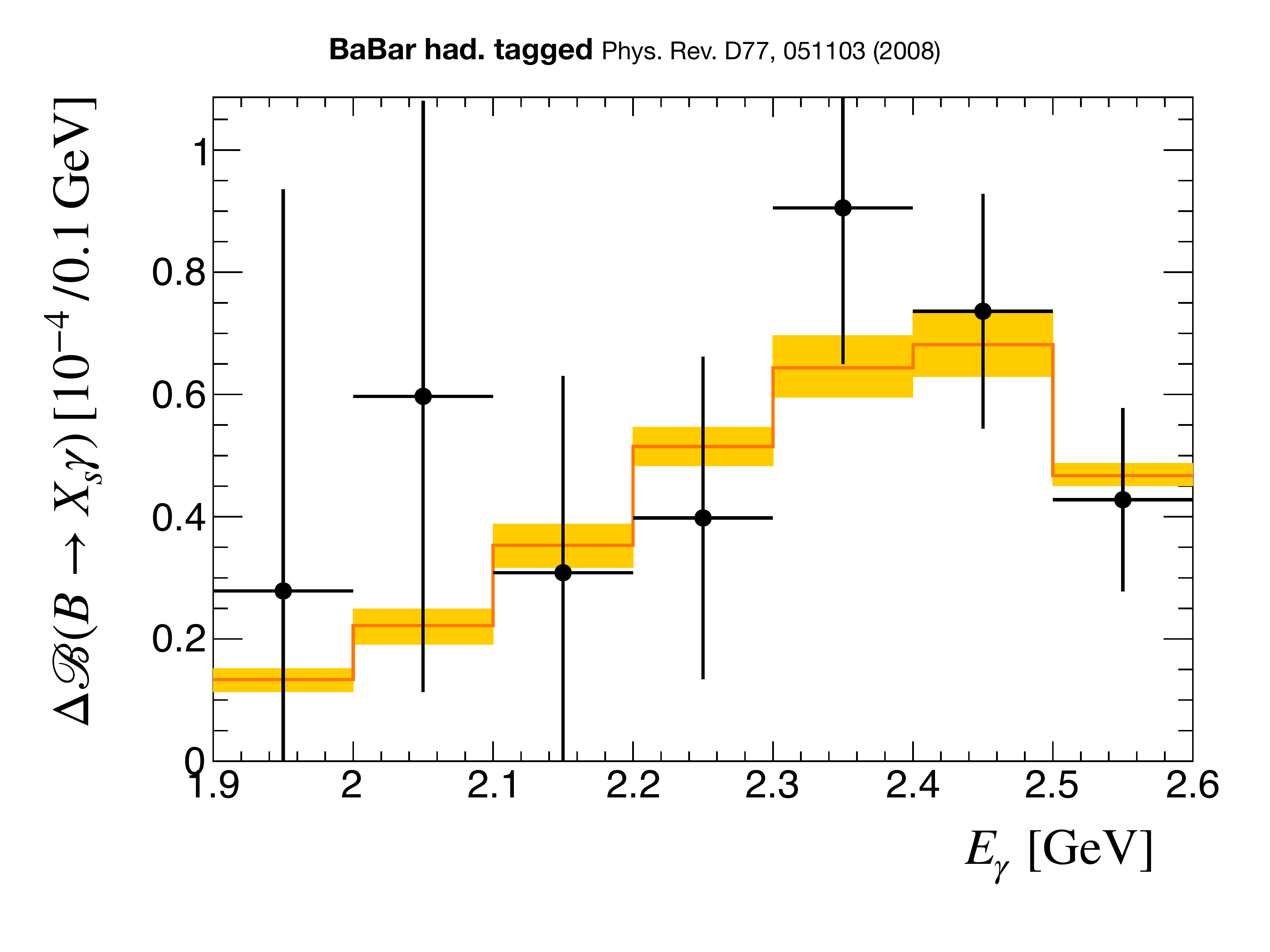}%
\hfill%
\includegraphics[width=0.5\textwidth]{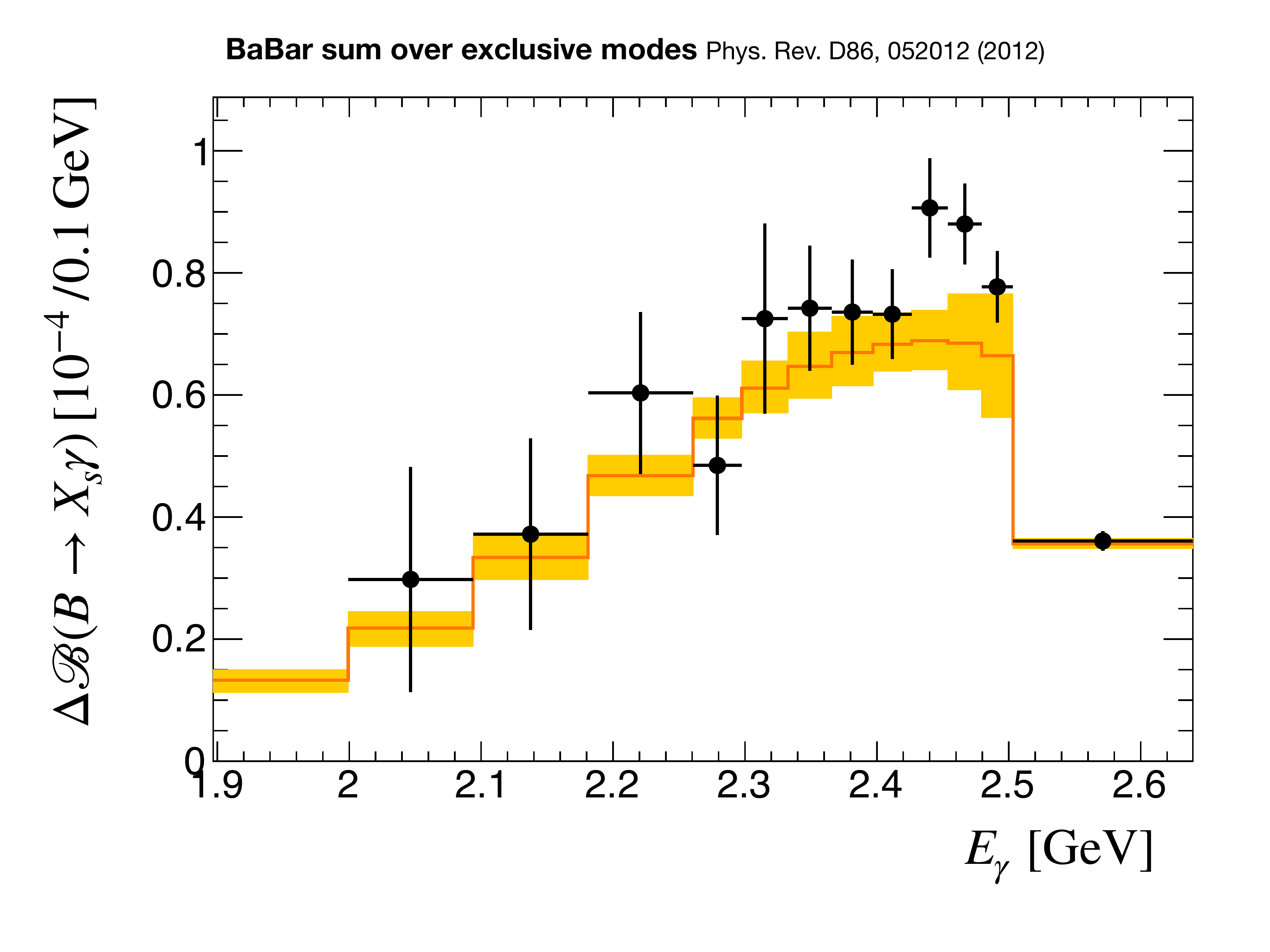}%
\vspace{-3ex}
\caption{Fit results to the measured photon energy spectra~\cite{Aubert:2007my,
Limosani:2009qg, Lees:2012ufa, Lees:2012wg}.
The orange lines are the fitted central values, and the yellow bands correspond
to the $\Delta \chi^2 = 1$ variation. We omit the first 6 bins of the Belle inclusive
spectrum and the first 3 bins of the \babar\ inclusive spectrum, as these
have very large uncertainties and provide no additional information.}
\label{fig:spectra}
\end{figure}

The full expression of \eq{expand} used in the fit including the non-$77$
terms is given by
\begin{align} \label{eq:expand_detail}
\frac{\df\Gamma}{\df E_\gamma}
&= N_s \sum_{m, n = 0}^N c_m\, c_n\, \frac{\df\Gamma_{77,mn}}{\df E_\gamma}
   +  \sqrt{N_s} \sum_{ij = 27,78} N_{ij} \sum_{m, n = 0}^N c_m\, c_n\, \frac{\df\Gamma_{ij,mn}}{\df E_\gamma}
   + \sum_{ij= 22,28,88} N_{ij} \sum_{m, n = 0}^N c_m\, c_n\, \frac{\df\Gamma_{ij,mn}}{\df E_\gamma}
\nn \\ & \quad
   + \sqrt{N_s}\,N_{27}\, \frac{\hla_2}{\hm_b^2} \sum_{n=0}^2 d_n \,  \frac{\df\Gamma_{g27,n}}{\df E_\gamma}
\,, \nn \\
N_{27} &= -2\Bigl(\cC_2 - \frac{\cC_1}{6}\Bigr)\, \abs{V_{tb} V_{ts}^*}\, \hm_b
\,, \qquad
N_{78} = -2\cC_8\, \abs{V_{tb} V_{ts}^*}\, \hm_b
\,, \nn \\
N_{22} &= \Bigl(\cC_2 - \frac{\cC_1}{6}\Bigr)^2 \abs{V_{tb} V_{ts}^*}^2\, \hm_b^2
\,, \qquad
N_{28} = 2 \Bigl(\cC_2 - \frac{\cC_1}{6}\Bigr)\cC_8\, \abs{V_{tb} V_{ts}^*}^2\, \hm_b^2
\,, \qquad
N_{88} = \cC_8^2\, \abs{V_{tb} V_{ts}^*}^2\, \hm_b^2
\,,\end{align}
where the normalization $N_s$ is defined by
\begin{equation} \label{eq:Ns_def}
N_s = \abs{ C_7^\incl V_{tb} V_{ts}^*}^2\, \hm_b^2
\,.\end{equation}
The $\df\Gamma_{ij,mn}$ and $\df\Gamma_{g27,n}$ are precomputed from the basis
expansion of the shape function. The $c_n$ and the normalization $N_s$ are
determined from the fit. For the normalization prefactors of the remaining
non-$77$ nonsingular terms in \eq{expand_detail} we use the SM input values collected in
\sec{inputs}. The overall minus sign in $N_{27}$ and $N_{78}$ arises from
assuming the SM negative sign for $\mathrm{Re}(C_7^\incl) = - \abs{C_7^\incl}$,
and assuming the SM imaginary part of $C_7^\incl$, which is negligible.
The value for $\hm_b$ in the prefactors is obtained during the fit from the
$c_n$ as discussed in \sec{Fmoments}. The coefficients $d_n$ parametrize the
$g_{27}$ subleading shape function that cannot be absorbed into the leading
shape function, cf. \sec{resolved}.

\begin{figure}[t!]
\includegraphics[scale=0.65]{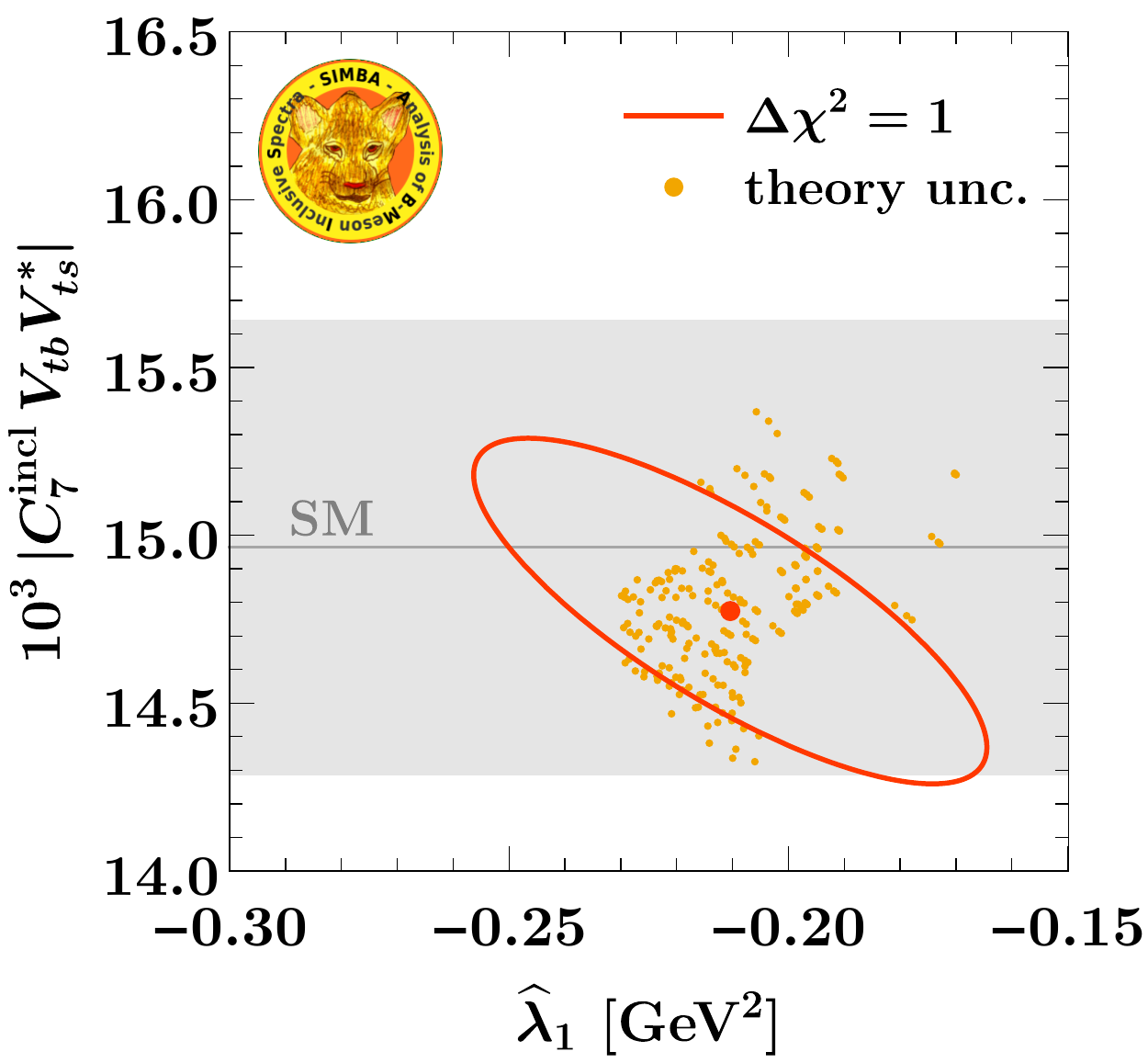}%
\hfill%
\includegraphics[scale=0.65]{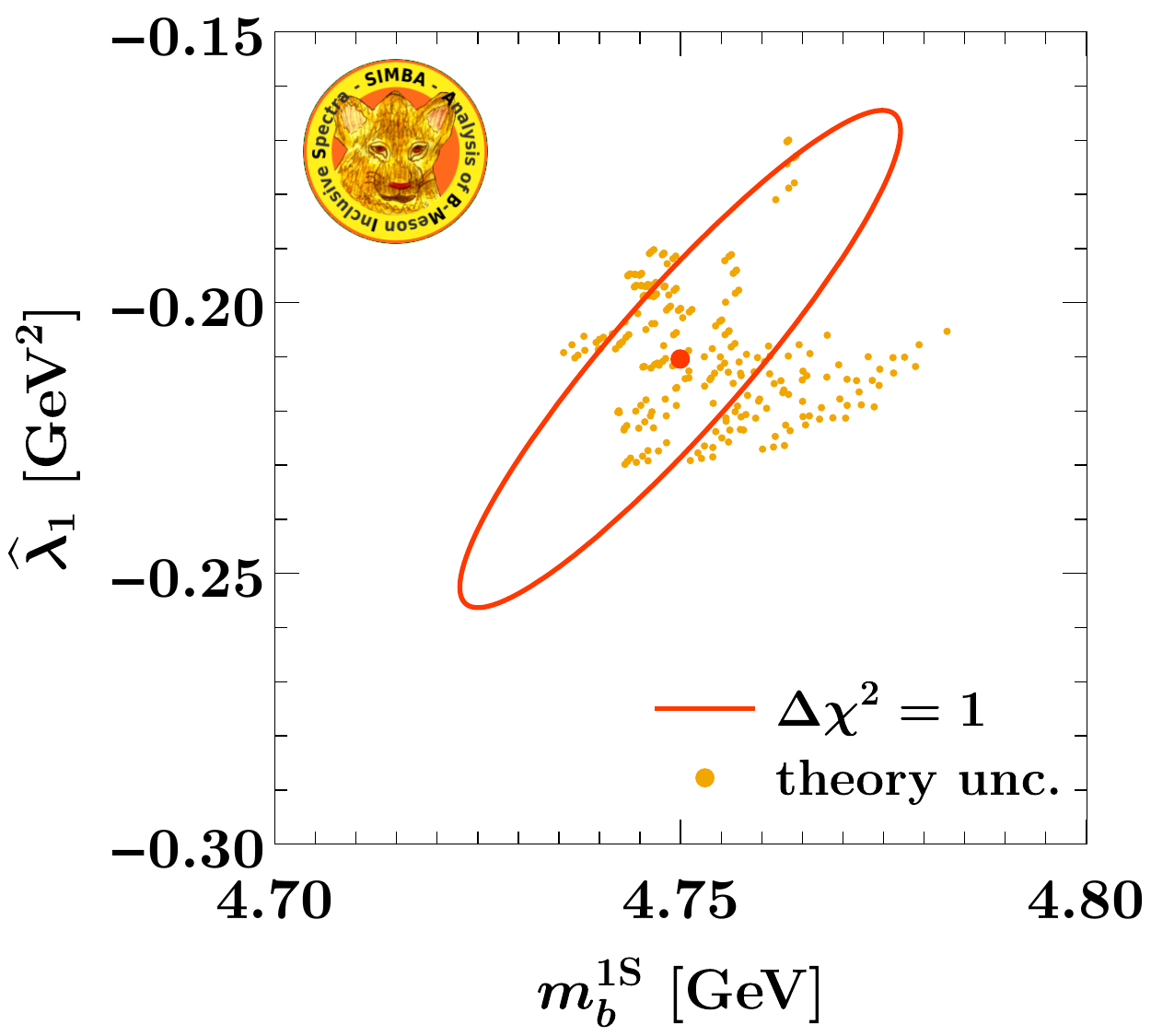}%
\caption{Fit results in the projection of $\abs{C_7^\incl V_{tb}V_{ts}^*}$ vs.\
$\hla_1$ (left) and $\hla_1$ vs.\ $\mbS$ (right), analogous to \fig{BtoXsgammaresults} in the
main text. The orange ellipse shows the $\Delta \chi^2 = 1$ contour.
The yellow points show fit results from varying the perturbative inputs.}
\label{fig:la1}
\end{figure}

The fitted experimental spectra with the fit results overlayed are shown in
\fig{spectra}.  The central value is shown by the orange line, and the yellow
band corresponds to the $\Delta \chi^2 = 1$ uncertainties of the fit.  The fit
results for $N_s$ and $c_n$ and their correlation matrix are given in \tab{cn}.
The final results for $\abs{C_7^\incl}$, $\hm_b \equiv \mbS$, $\hla_1$, and
$\widehat\rho_1$ together with their correlation matrix are given in
\tab{finalcorrelations}. They are obtained from the fitted $N_s$ and $c_n$ by
using \eq{Ns_def} and the moment relations discussed in \sec{Fmoments}. In
\fig{la1} these results and the corresponding theory uncertainties are shown as
well, analogous to \fig{BtoXsgammaresults} in the main text.

%
%
%
\begin{table}[t]
\hfill
\begin{tabular}{cc}
\hline\hline
Parameter & Fit result\\
\hline
$10^3\, N_s$  & $4.925 \pm 0.294$   \\
$c_0$         & $0.9956 \pm 0.0063$ \\
$c_1$         & $0.0641 \pm 0.0361$ \\
$c_2$         & $0.0624 \pm 0.0458$ \\
$c_3$         & $0.0267 \pm 0.0727$ \\
\hline\hline
\end{tabular}
\hfill
\begin{tabular}{l|ccccc}
\hline\hline
& $N_s$ & $c_0$ & $c_1$ & $c_2$ & $c_3$ \\
\hline
$N_s$ & $ 1       $ & $-0.804332$ & $+0.809278$ & $+0.703457$ & $+0.579938$ \\
$c_0$ & $-0.804332$ & $ 1       $ & $-0.860744$ & $-0.980474$ & $-0.738548$ \\
$c_1$ & $+0.809278$ & $-0.860744$ & $ 1       $ & $+0.844262$ & $+0.325844$ \\
$c_2$ & $+0.703457$ & $-0.980474$ & $+0.844262$	& $ 1       $ & $+0.666741$ \\
$c_3$ & $+0.579938$ & $-0.738548$ & $+0.325844$ & $+0.666741$ & $ 1       $ \\
\hline\hline
\end{tabular}
\hspace*{\fill}
\caption{Fit results (left) and correlations (right) for the normalization parameter $N_s$
and the fitted shape-function coefficients $c_{0,1,2,3}$ for the default fit with
$\lambda = 0.55\,\GeV$ and $N = 3$.}
\label{tab:cn}
\end{table}

%
%
%
%
\begin{table}[tb]
\hfill
\begin{tabular}{lc}
\hline\hline
Parameter & Fit result\\
\hline
$10^3\, \abs{C_7^\incl V_{tb}V_{ts}^*}$ & $\,\,       14.77 \pm 0.51$  \\
$\mbS / \GeV$                         & $\,\,\,\,\, 4.750 \pm 0.027$ \\
$\hla_1 / \GeV^2$                      & $          -0.210 \pm 0.046$ \\
$\widehat\rho_1 / \GeV^3$              & $\,\,\,\,\, 0.134 \pm 0.036$ \\
\hline\hline
\end{tabular}
\hfill
\begin{tabular}{l|cccc}
\hline\hline
& $\abs{C_7^\incl V_{tb}V_{ts}^*}$ & $\mbS$ & $\hla_1$ & $\widehat\rho_1$ \\
\hline
$\abs{C_7^\incl V_{tb}V_{ts}^*}$ & $ 1       $ & $-0.895754$ & $-0.788116$ & $+0.685843$ \\
$\mbS$                          & $-0.895754$ & $ 1       $ & $+0.917563$ & $-0.770155$ \\
$\hla_1$                         & $-0.788116$ & $+0.917563$ & $ 1       $ & $-0.953347$ \\
$\widehat\rho_1$                 & $+0.685843$ & $-0.770155$ & $-0.953347$ & $ 1       $ \\
\hline\hline
\end{tabular}
\hspace*{\fill}
\caption{Results (left) and correlations (right) for $\abs{C_7^\incl V_{tb} V_{ts}^*}$,
$\mbS$, $\hla_1$, $\widehat\rho_1$ obtained from the default fit results in \tab{cn}
for $\lambda = 0.55\,\GeV$ and $N = 3$ by inverting the moment relations for $\cF(k)$.
Only the fit uncertainties are included.}
\label{tab:finalcorrelations}
\end{table}

\begin{figure}[t!]
\includegraphics[scale=0.65]{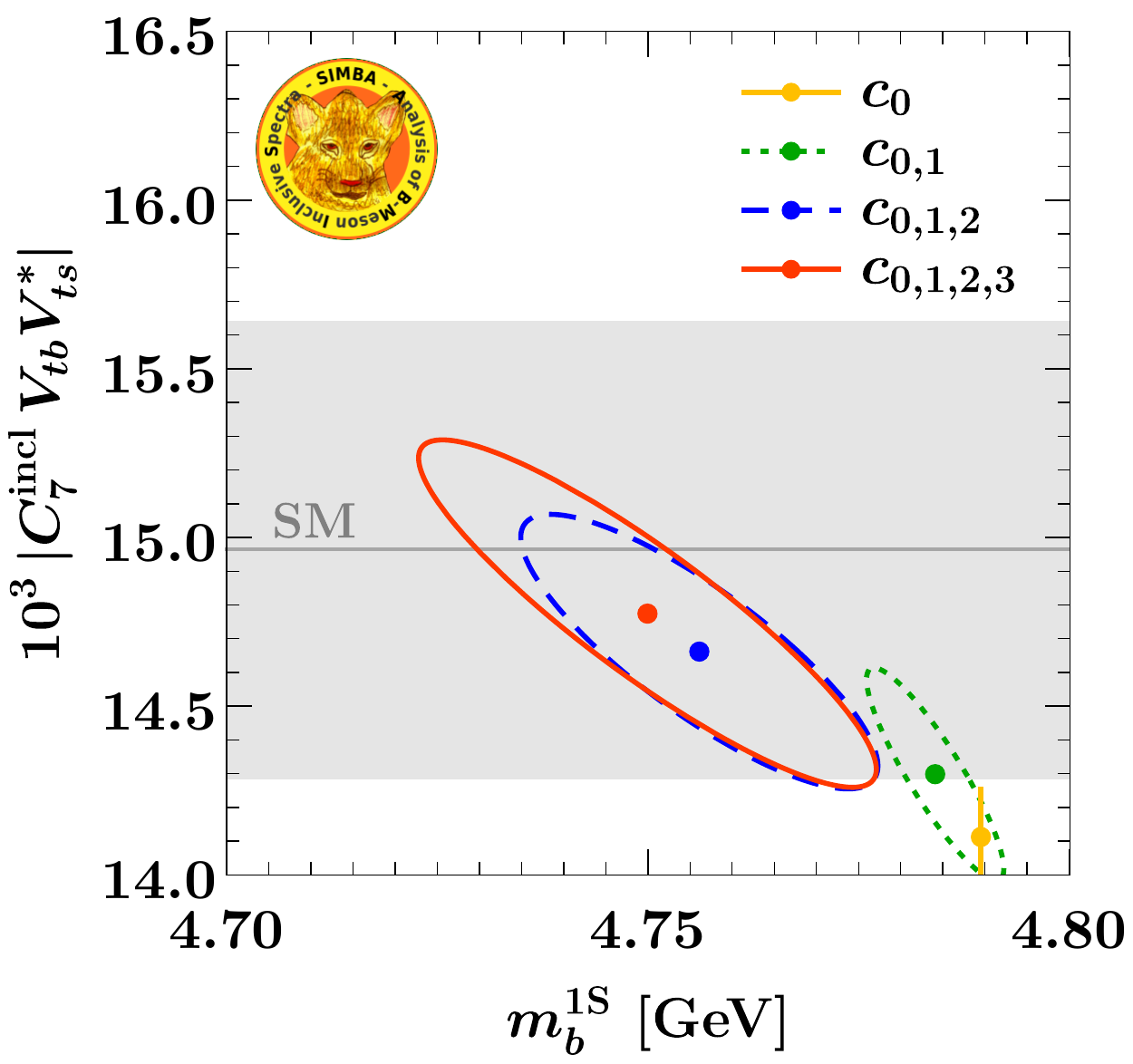}%
\hfill%
\includegraphics[scale=0.65]{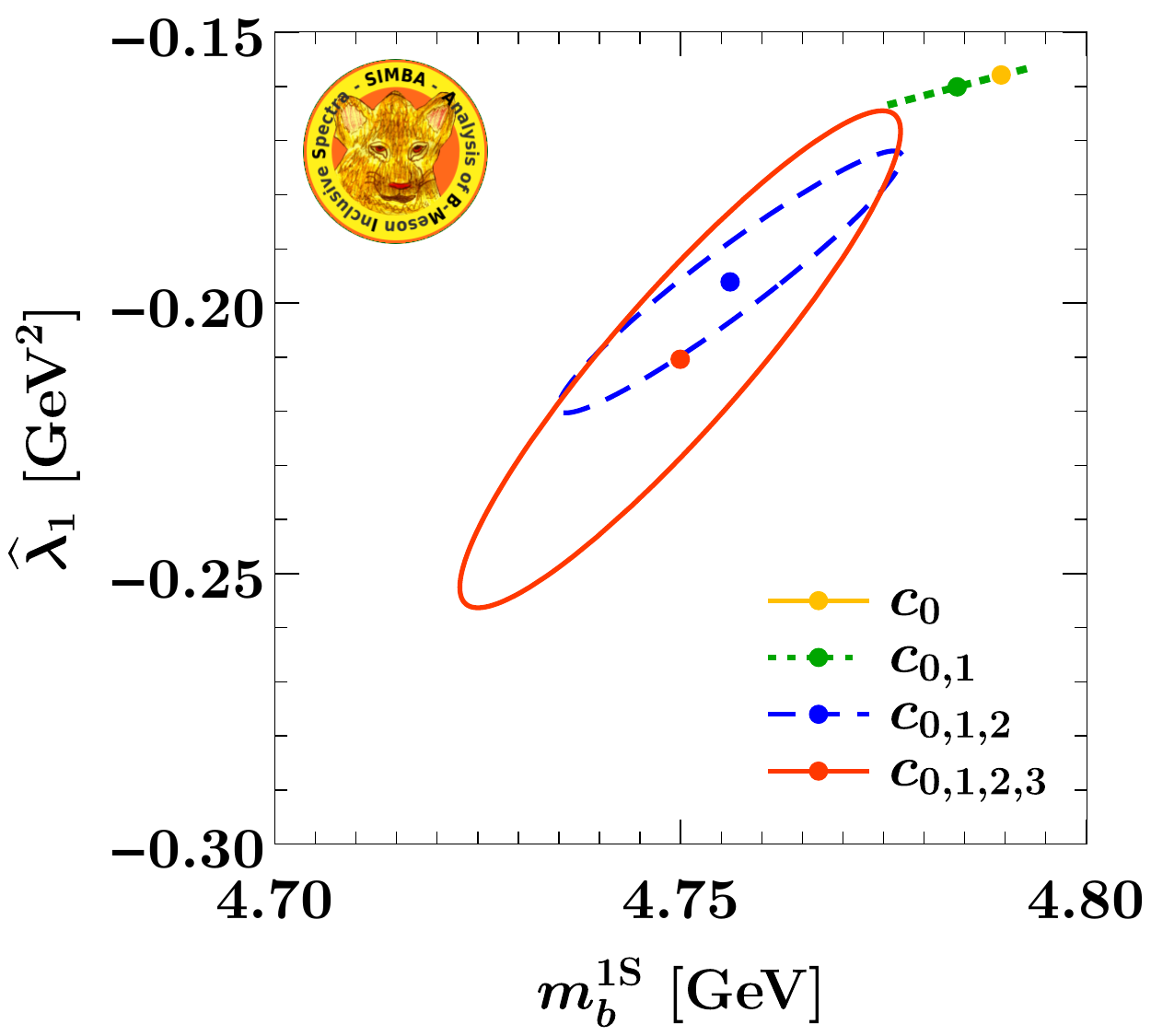}%
\caption{Fit results as a function of the number of fitted basis coefficients in
the projection of $\abs{C_7^\incl V_{tb}V_{ts}^*}$ vs.\ $\mbS$ (left) and
$\hla_1$ vs.\ $\mbS$ (right). For more details see text.} \label{fig:conv}
\end{figure}

In \fig{conv} the convergence of the fit results for our default basis with $\la
= 0.55\,\GeV$ for an increasing number of basis coefficients is shown. As
discussed in the main text, the truncation order is determined using a nested
hypothesis test to determine the appropriate number of coefficients given the
available data sets. For the nominal fit we use 4 coefficients ($c_{0,1,2,3}$).
Note that all fits with fewer coefficients also have acceptable $\chi^2$, so the
fit quality alone is not a sufficient criterion for choosing the number of coefficients.
On the other hand, the results with only $c_0$ and $c_{0,1}$, which effectively
correspond to using a fixed model for the shape function, clearly show a model
bias and underestimated uncertainties. The central values change only moderately
by the inclusion of the fourth coefficient $c_3$. The resulting increase in the
fit uncertainties illustrates the effect of accounting for the truncation
uncertainty by including this additional coefficient. Including $c_3$ is
essential for the results with different basis choices to be consistent as in
\fig{diflambda}. Without including $c_3$, the results still show a clear bias
between different bases.

\subsection{Wilson coefficients and \Cincl}
\label{sec:C7incl}

\subsubsection{Split matching}

The effective Hamiltonian for $B\to X_s \gamma$ is
\begin{equation}\label{eq:Heff}
\cH_\eff = -\frac{4G_F}{\sqrt2}\, V_{tb} V^*_{ts}\, \sum_{i=1}^8 C_i\, O_i
\,.\end{equation}
The dominant contributions are from
\begin{alignat}{9} \label{eq:O1278}
O_1 &= (\bar s\, \gamma_\mu T^a P_L\, c)(\bar c\, \gamma^\mu T^a P_L\, b)
\,, \qquad &
O_2 &= (\bar s\, \gamma_\mu P_L\, c)(\bar c\, \gamma^\mu P_L\, b)
\,,\nn\\
O_7 &= \frac{e}{16\pi^2}\,\mbbar
   \, \bar s\, \sigma_{\mu\nu}F^{\mu\nu} P_R\, b
\,, \qquad &
O_8 &= \frac{g}{16\pi^2}\, \mbbar
  \, \bar s\,\sigma_{\mu\nu} G^{\mu\nu} P_R\, b
\,,\end{alignat}
where $P_{R,L} = (1\pm \gamma_5)/2$, and we neglected the mass of the strange
quark (giving $m_s^2/m_b^2$ suppressed corrections). The $O_{3-6}$ in \eq{Heff}
are four-quark operators generated at one-loop level in the SM. The renormalized
Wilson coefficients $C_i(\mu)$ and operators $O_i(\mu)$ are defined in the
\MSbar scheme, and $\mbbar(\mu)$ is the \MSbar $b$-quark mass.

The Wilson coefficient $C_7^\incl$ arises when we carry out a split matching procedure to separate the scale dependence above and below the scale $\mu_0 \sim m_b$~\cite{Lee:2005pk,Lee:2006gs}. Above $\mu_0$,
we have the matching onto $\cH_\eff$ at the weak scale $\mu_\weak\sim m_W$ and its renormalization group evolution down to $\mu_0\sim m_b$. At $\mu_0$, we have virtual matrix element
corrections from all operators $O_{i\neq 7}$ that are proportional to the tree-level matrix element of the chromomagnetic operator $O_7$.
Together, these effects can be combined to define the effective Wilson coefficient
\begin{equation}\label{eq:C7incl}
C_7^\incl 
= \bC_7(\mu_0) + \sum_i \bC_i(\mu_0) \bigl[ s_i(\mu_0,\hm_b) + r_i(\mu_0,\hm_b,\hm_c) \bigr]
= \cC_7 + \sum_{i\neq7} \bC_i(\mu_0)\,  r_i(\mu_0,\hm_b,\hm_c)
\,,\end{equation}
which is the main short-distance perturbative coefficient that is constrained by the $B\to X_s\gamma$ measurements. Its value is sensitive to beyond Standard-Model physics, while the shape
of the photon spectrum is not~\cite{Kapustin:1995nr}.
The two terms in the last equality in \eq{C7incl} are separately $\mu_0$ independent order by order in $\alpha_s$. The barred coefficients $\bC_i(\mu)$ are defined as
\begin{equation}\label{eq:Cibar}
\bC_i(\mu_0) =
\begin{cases}
  C_i(\mu_0)\,, & \quad i = 1,\ldots,6 \,,\\[2pt] \displaystyle
  C_i^\eff(\mu_0)\, \frac{\mbbar(\mu_0)}{\hm_b}\,, & \quad i = 7,\, 8\,,
\end{cases}
\end{equation}
where $C_{7,8}^\eff(\mu_0)$ are the standard scheme-independent effective Wilson
coefficients~\cite{Buchalla:1995vs}.
The additional factors $\mbbar(\mu_0)/\hm_b$ are included in $\bC_{7,8}$ to convert
to a short-distance $b$-quark mass scheme, $\hm_b$, which improves the convergence
of perturbation theory.

The $r_i(\mu_0,\hm_b,\hm_c)$ in \eq{C7incl} encode the finite virtual
corrections from operators other than $O_7$ that give rise to singular
contributions to the photon energy spectrum, and are responsible for the
difference between $\cC_7$ and $C_7^\incl$. Hence, the split matching procedure
essentially amounts to matching $\cH_\eff$ at $\mu_0$ onto a single $O_7$
chromomagnetic operator, which is subsequently matched onto its corresponding
operator in SCET. In doing so, we treat the charm quark as a heavy quark and
integrate out both charm and bottom quarks at the scale $\mu_0$. As a result,
(most of) the sizable contributions from $c\bar c$-loops proportional to $\bC_{1,2} \bC_7$ and
$\bC_{1,2}^{\,2}$ are contained within $\abs{\Cincl}^2$, including their full $\hm_c$
dependence, namely in the terms $r_{1,2}(\mu_0, \hm_b, \hm_c)$. As already
mentioned in the main body, this organization of the perturbative contributions
has the advantage that the associated theory uncertainty due to the $\hm_c$
dependence only enters in the SM prediction for \Cincl, but does not limit the
accuracy with which \Cincl can be extracted from the experimental data.
This treatment is furthermore motivated by the fact that in the experimental
measurements of $B\to X_s\gamma$, charmed final states are not included in the
signal and are treated as background.

\subsubsection{Perturbative results}

The coefficient $\cC_7$ in \eq{C7incl} is defined to be $\mu_0$ independent and to satisfy $\cC_7=\bC_7(\hm_b)$. Thus it is equal to $\bC_7(\mu_0)$ plus the additional terms from the renormalization group that cancel the $\mu_0$ dependence of $\bC_7(\mu_0)$ and vanish when $\mu_0 = \hm_b$.
Explicitly, up to $\ord{\alpha_s^2}$ with $\hm_b$ in the $1S$ mass scheme, we have
\begin{align} \label{eq:cC7nnlo}
\cC_7
& = \bC_7(\mu_0) + \sum_i \bC_i(\mu_0) s_i(\mu_0,\hm_b) 
 \nn\\
& = \bC_7(\mu_0) + \frac{\alpha_s(\mu_0)}{4\pi}\,
  \ln\frac{\hm_b}{\mu_0} \biggl[ \gamma_m^{(0)}\, \bC_7(\mu_0)
  + \sum_{i} \gamma_{i7}^{(0)}\, \bC_i(\mu_0) \biggr]
\nn\\* & \quad
+ \frac{\alpha_s^2(\mu_0)}{(4\pi)^2}\, \ln\frac{\hm_b}{\mu_0} \biggl\{
   \gamma_m^{(1)}\, \bC_7(\mu_0)
  + \sum_{i} \gamma_{i7}^{(1)}\, \bC_i(\mu_0)
+ \frac{1}{2} \ln\frac{\hm_b}{\mu_0} \biggl[
  \bigl( \gamma_m^{(0)} + \gamma_{77}^{(0)} \bigr)^2\, \bC_7(\mu_0)
  + \bigl( \gamma_m^{(0)} + \gamma_{77}^{(0)} \bigr)
  \sum_{j\neq7} \gamma_{j7}^{(0)}\, \bC_j(\mu_0)
\nn\\* &\qquad
  + \gamma_{87}^{(0)} \gamma_m^{(0)}\, \bC_8(\mu_0)
  + \gamma_{87}^{(0)} \sum_{j\neq7} \gamma_{j8}^{(0)}\, \bC_j(\mu_0)
+ \sum_{j\neq7,8} \sum_k \gamma_{j7}^{(0)}\,
  \gamma_{kj}^{(0)}\, \bC_k(\mu_0) \biggr]
\nn\\ & \qquad
- \sum_{j\neq7,8}
  4C_F\, \gamma_{j7}^{(0)}\, \bC_j(\mu_0) \biggl[
  1 - \frac{C_F\,\pi\alpha_s(\mu_0)}{8}
  - \frac{3}{4} \ln\frac{\hm_b}{\mu_0} \biggr]
- \beta_0 \ln\frac{\hm_b}{\mu_0}
  \biggl[ \gamma_m^{(0)}\, \bC_7(\mu_0)
  + \sum_{i} \gamma_{i7}^{(0)}\, \bC_i(\mu_0) \biggr]
 \biggr\}
\,,\end{align}
where $\beta_0 = (11 C_A - 4 T_F n_f)/3$ and $n_f=5$ is the number of active flavors above the scale $m_b$,
and $C_A=3$, $C_F=4/3$, $T_F=1/2$.
The $\gamma_{ij}^{(k)}$ and $\gamma_m^{(k)}$ are anomalous dimension coefficients defined via
\begin{alignat}{9}
\mu\, \frac{\df}{\df\mu}\, C_j(\mu) &= \sum_i C_i(\mu)\, \gamma_{ij}(\mu)
\,,  \qquad &
\gamma_{ij}(\mu)
&= \frac{\alpha_s(\mu)}{4\pi}\, \gamma_{ij}^{(0)}
 + \frac{\alpha_s^2(\mu)}{(4\pi)^2}\, \gamma_{ij}^{(1)}
 + \ord{\alpha_s^3}
\,, \nn \\
\mu\, \frac{\df}{\df\mu}\, \mbbar(\mu) &= \mbbar(\mu)\, \gamma_m(\mu)
\,, \qquad &
\gamma_m(\mu)
&= \frac{\alpha_s(\mu)}{4\pi}\, \gamma_m^{(0)}
 + \frac{\alpha_s^2(\mu)}{(4\pi)^2}\, \gamma_m^{(1)}
 + \ord{\alpha_s^3}
\,.\end{alignat}
For example, $\gamma_m^{(0)}=-8$ and $\gamma_{77}^{(0)}=32/3$.
The full set of required anomalous dimension coefficients can be found in \refcite{Czakon:2006ss}.

To fully implement the split matching procedure it is convenient to also define
scale-independent coefficients $\cC_{i\neq 7}$, which appear in the nonsingular terms in \eq{Phat}.
Analogous to $\cC_7$ above, they are defined to be $\mu_0$ independent and to satisfy
$\cC_{i} = \bC_i(\mu_0=\hm_b)$. To one-loop order they are given by
\begin{equation} \label{eq:cCi}
\cC_8
 = \bC_8(\mu_0) + \frac{\alpha_s(\mu_0)}{4\pi}\,
  \ln\frac{\hm_b}{\mu_0} \biggl[ \gamma_m^{(0)}\, \bC_8(\mu_0)
  + \sum_{i} \gamma_{i8}^{(0)}\, \bC_i(\mu_0) \biggr]
\,,\qquad
\cC_{j\ne 7,8}
 = \bC_j(\mu_0) + \frac{\alpha_s(\mu_0)}{4\pi}\,
  \ln\frac{\hm_b}{\mu_0} \sum_{i} \gamma_{ij}^{(0)}\, \bC_i(\mu_0)
\,.\end{equation}

The $\mu_0$ dependence of $r_i(\mu_0,\hm_b,\hm_c)$ in \eq{C7incl} is defined
such that it cancels that of the coefficients $\bC_i(\mu_0)$ in \eq{C7incl}, while for
$\mu_0=\widehat m_b$, the $r_i(\hm_b,\hm_b,\hm_c)$ agree with their usual definitions
in the literature. Denoting their $\alpha_s$ expansions at $\mu_0 = \hm_b$ as
\begin{equation}
r_i(\hm_b,\hm_b,\hm_c) = \frac{\alpha_s(\hm_b)}{4\pi}\, r_i^{(1)}
  + \frac{\alpha_s^2(\hm_b)}{(4\pi)^2}\, r_i^{(2)} + \ord{\alpha_s^3}
\,,\end{equation}
we have at NNLO
\begin{align} \label{eq:r8soln}
r_8(\mu_0,\hm_b,\hm_c)
&= \frac{\alpha_s(\mu_0)}{4\pi}\, r_8^{(1)}
  + \frac{\alpha_s^2(\mu_0)}{(4\pi)^2} \biggl[ r_8^{(2)} +
  \ln\frac{\hm_b}{\mu_0}\,  r_8^{(1)} \bigl( \gamma_m^{(0)} + \gamma_{88}^{(0)} - 2\beta_0 \bigr)
\biggr]
 + \ord{\alpha_s^3}
\,,\end{align}
while for $k=1,\ldots,6$,
\begin{align} \label{eq:rksoln}
r_k(\mu_0,\hm_b,\hm_c) &= \frac{\alpha_s(\mu_0)}{4\pi}\, r_k^{(1)}
  + \frac{\alpha_s^2(\mu_0)}{(4\pi)^2}
\biggl[ r_k^{(2)} + \ln\frac{\hm_b}{\mu_0}
  \biggl( \sum_{i\neq7} \gamma_{ki}^{(0)}\, r_i^{(1)}
   - 2\beta_0\, r_k^{(1)}\biggr) \biggr]
 + \ord{\alpha_s^3}
\,.\end{align}
The results of \refscite{Ewerth:2008nv, Asatrian:2010rq} give
\begin{align} \label{eq:r8result}
r_8^{(1)}
&= \frac{C_F}{3} \biggl(11 - \frac{2\pi^2}{3} + 2\img\pi \biggr)
\,,\nn\\[2pt]
\mathrm{Re}\, r_8^{(2)}
&= C_F\, \biggl\{
\biggl(\frac{c^a_{78,0}}{2} + \frac{55}{3} + \frac{34\pi^2}{9} - \frac{8\pi^4}{27} \biggr)\, C_F
+ \biggl(\frac{c^{na}_{78,0}}{2} - \frac{3454}{81} + \frac{176 \pi^2}{81} + \frac{88 \zeta_3}{9} \biggr)\, C_A
\nn\\ &\quad
+ \biggl(\frac{314}{27} - \frac{16 \pi^2}{27} - \frac{8 \zeta_3}{3} \biggr) \beta_0(n_l)
+ \biggl[\frac{976}{81} -\frac{4 \pi}{\sqrt{3}} - \frac{244 \pi^2}{81}
  + 16 \sqrt{3}\, \mathrm{Cl}_2 \Bigl(\frac{\pi}{3}\Bigr)
- \frac{32\zeta_3}{27} \biggr] T_F\, n_h
\biggr\}
\,.\end{align}
A value for $\mathrm{Im}\,r_8^{(2)}$ is not yet known, but it only
contributes to the spectrum at $\ord{\alpha_s^3}$. In \eq{r8result},
$n_h=1$ is the number of flavors with mass $\hm_b$, and $n_l = 4$ is the number of massless flavors, since we neglected for simplicity the $\hm_c$ dependence in $r_8^{(2)}$.
The full $\hm_c$ dependence of $\mathrm{Re}\, r_8^{(2)}$, arising from $c\bar c$ loops inserted into
gluon propagators, is known~\cite{Ewerth:2008nv, Asatrian:2010rq}, 
but the massless approximation is sufficiently accurate for our purposes.

For $r_{1-6}(\mu_0)$ we have the NLO coefficients~\cite{Greub:1996tg,Buras:2001mq,Buras:2002tp}
\begin{align}
r_1^{(1)}
&= - \frac{1}{6} r_2^{(1)}
\,,
&r_2^{(1)}
&= -\frac{1666}{243} + 2a(\rho) + 2b(\rho) - \frac{80}{81}\,\img \pi
\,,\\*
r_3^{(1)}
&= \frac{2392}{243} + \frac{8\pi}{3\sqrt{3}} - a(1) + 2b(1) + \frac{32}{9} X_b + \frac{56}{81}\,\img\pi
\,,
&r_4^{(1)}
&= \frac{145}{243} - \frac{1}{6} r_3^{(1)} + 2b(\rho) + 2b(1) - \frac{40}{81}\,\img \pi
\,,\nn\\*
r_5^{(1)}
&= \frac{6136}{81} - \frac{32\pi}{\sqrt{3}} + 16 r_3^{(1)} - \frac{128}{3} X_b
\,,
& r_6^{(1)}
&= - \frac{310}{27} + 6r_2^{(1)} - \frac{4}{3} r_3^{(1)} + 4r_4^{(1)} + \frac{1}{3} r_5^{(1)} - \frac{104}{27}\,\img\pi
\,.\nn
\end{align}
Here, $\rho = \hm_c^2/\hm_b^2$ and $a(\rho)$, $b(\rho)$, and $X_b$ are given in
\refcite{Buras:2002tp}. Since the Wilson coefficients $C_{3-6}$ are small, the
$r_{3-6}$ terms only have very small impacts, and their NNLO contributions
$r_{3-6}^{(2)}$ can be safely neglected.

The NNLO contributions $r_{1,2}^{(2)}(\mu_0)$ are only fully known in the large
$\beta_0$ approximation, where they are obtained as an expansion in
$m_c/m_b$~\cite{Bieri:2003ue}. They are given by
\begin{align} \label{eq:r22largeb0}
r_1^{(2)} =- r_2^{(2)}/6
\,, \qquad
r_2^{(2)}
& = -\frac{3}{2} \beta_0 \biggl[
   \Bigl({\rm Re}\, r_2^{(2)}\Bigr)_\text{\refcite{Bieri:2003ue}}
   + \img \Bigl({\rm Im}\: r_2^{(2)}\Bigr)_\text{\refcite{Bieri:2003ue}}
\biggr]
+ \dotsb
\,,\end{align}
where the terms in the square brackets are given in Eqs.~(26) and (27) of
\refcite{Bieri:2003ue}, and the ellipses denote other independent color
structures. The full NNLO contributions to $r_{1,2}^{(2)}$ are required to
cancel the $\hm_c$-scheme dependence and have been computed in the limits
$m_c\gg m_b/2$ and $m_c=0$~\cite{Misiak:2006ab, Misiak:2010sk, Czakon:2015exa}.

The SM prediction for \Cincl in \eq{C7inclnnlo} is obtained using the above
results together with the input parameters and numerical values for the Wilson
coefficients given below in \sec{inputs}. Although the two terms in
\eq{C7incl} are formally $\mu_0$ independent, there is still residual $\mu_0$
dependence from the truncation of perturbation theory. We vary $\mu_0$ between
$\hm_b/2$ and $2\hm_b$ to obtain an estimate of the associated perturbative uncertainty,
quoted in \eq{C7inclnnlo} with the subscript ``scale''.
In addition, to estimate the uncertainty from missing
$\ord{\alpha_s^2}$ charm-loop contributions we use the $\alpha_s^2\beta_0$ result
in \eq{r22largeb0} with a  multiplicative prefactor of $1.0\pm 0.5$, yielding
the uncertainty quoted in \eq{C7inclnnlo} with a subscript ``$c\bar c$''.
The parametric uncertainties from input parameters, including the numerical value
of $\hm_c$ itself, are much smaller than these two
sources of uncertainties and can be safely neglected.

\subsection{Perturbative ingredients for the photon energy spectrum}
\label{sec:pertspectrum}

The perturbative components of the photon energy spectrum are described by \eq{Phat},
which we repeat here for convenience
\begin{align} \label{eq:Phat2}
\widehat P(k)
&= \Abs{C_7^\incl}^2\, \Bigl[W_{77}^\sing(k) + W_{77}^\nons(k) \Bigr]
 + 2\, \mathrm{Re}\big(C_7^\incl\big) \sum_{i\neq7} \cC_i\, W^\nons_{7i}(k)
 + \sum_{i,j\neq7} \cC_i \cC_j\, W^\nons_{ij}(k)
\,.\end{align}
Definitions for the Wilson coefficients $C_7^\incl$ and $\cC_i$ are given
above in \sec{C7incl}, $W^\sing_{77}(k)$ contains the dominant singular
contributions and the $W_{ij}^{\nons}(k)$ are the various nonsingular terms. In
our formula for $\df\Gamma/\df E_\gamma$ in \eq{dGamma} we have kept an overall
$E_\gamma^3$ kinematic prefactor. Here one power of $E_\gamma$ arises from the
photon phase-space integration, and for the dominant $77$-like contributions two
more factors of $E_\gamma$ arise from the derivative that acts on the photon
field in each $F^{\mu\nu}$. Since these factors are universal we do not expand
them about the singular limit. This improves the behavior of the decomposition
into singular and nonsingular terms in the tail region where these components
become comparable.

\subsubsection{Singular contributions}

The all-order factorization theorem for the singular contributions $W^\sing_{77}(k)$
is well known~\cite{Korchemsky:1994jb, Bauer:2001yt}.
For our treatment we follow \refcite{Ligeti:2008ac} and
make use of the SCET-based factorization theorem, expressing the perturbative
ingredients in a short-distance scheme. (In the notation
of~\refcite{Ligeti:2008ac}, $\mu_i=\mu_J$ and $\mu_\Lambda = \mu_S$.)
\begin{align} \label{eq:Wsing77_resummed}
W^\sing_{77}(k)
&= h_s(\hm_b, \hm_c, \mu_b)\, U_H(\widehat m_b,\mu_b,\mu_J)
  \int\!\! \df\omega\,\df\omega' \hm_b\, J[\hm_b(k-\omega),\mu_J] \, U_S(\omega-\omega',\mu_J,\mu_S)\, \widehat C_0(\omega',\mu_S)
  \,.
\end{align}
Here $h_s$, $J$, and $\widehat C_0$ are the fixed-order hard, jet, and soft
functions, which we include up to NNLO. The evolution kernels
$U_H$ and $U_S$ sum large logarithms of $k/\hm_b \sim 1 - 2E_\gamma/\hm_b$ to all
orders in perturbation theory, and are included at NNLL order.
The perturbative expressions for $h_s$, $J$, $\widehat C_0$ as well as $U_H$
and $U_S$ together with the required anomalous
dimensions can be found in \refcite{Ligeti:2008ac},
where they were obtained using results from \refscite{Korchemsky:1992xv,
Gardi:2005yi, Bauer:2000yr, Blokland:2005uk, Bauer:2003pi, Becher:2005pd,
Becher:2006qw, Balzereit:1998yf, Neubert:2004dd, Fleming:2007xt}.

In the appropriate region the
logarithmic summation is achieved by choosing $\mu_b\sim \hm_b$, $\mu_S\sim
\hm_b-2E_\gamma \gtrsim \Lambda_{\rm QCD}$, $\mu_J^2\sim \mu_b\mu_S$.  The
dependence of $W_{77}^\sing(k)$ on $\mu_b$, $\mu_J$, and $\mu_S$ cancels between
the fixed-order functions and evolution kernels order by order in resummed
perturbation theory, and will be used to estimate higher-order perturbative
uncertainties. The precise procedure we use to estimate these uncertainty
and to transition into and out of this resummation region is described in more
detail in \sec{pertunc} below.

The hard function $h_s$ arises from matching the QCD chromomagnetic operator $O_7$
onto a corresponding SCET operator at the scale $\mu_b$, which is the second step
in the split matching procedure described in the previous section. To be consistent
with the first step of the split matching, also here we integrate out bottom and
charm quarks at the hard matching scale $\mu_b$. As a result, the hard function
includes all effects of virtual massive charm loops inserted into gluon propagators, which starting at two loops gives rise to its $\hm_c$ dependence given by
\begin{align}
h_s(\hm_b, \hm_c, \mu_b)
&= h_s(\hm_b) + \frac{\alpha_s^2(\mu_b)}{(4\pi)^2}\, C_F\, T_F\, f_{2s}\Bigl(\frac{\hm_c^2}{\hm_b^2}\Bigr)
\nn \\
f_{2s}(\rho)
&= \frac{4}{9} \ln^3\rho + \frac{50}{9} \ln^2\rho
   + \Bigl[-\frac{124}{9} \ln(1-\rho) + \frac{8 \pi^2}{9} + \frac{794}{27}\Bigr] \ln\rho
   - \frac{16}{3} \Li_3(\rho)
   + \Bigl(\frac{8 \ln\rho}{3} - \frac{124}{9}\Bigr) \Li_2(\rho)
\nn \\ & \quad
   + \frac{124 \pi^2}{27} + \frac{5578}{81}
   + \rho \Bigl(\frac{32}{9}\ln\rho + \frac{172}{9}\Bigr)
   + 12 \rho^2 \Bigl[
      -\Li_2(\rho ) + \frac{1}{2}\ln^2\rho -\log (1-\rho ) \log (\rho )+\frac{\pi ^2}{3}
   \Bigr]
\nn \\ & \quad
   - \frac{2}{9} \sqrt{\rho} (35 \rho +81) \Bigl[
      -4\Li_2(\sqrt{\rho}) + \Li_2(\rho) + 2\mathrm{arctanh}(\sqrt{\rho}) \ln\rho + \pi^2
   \Bigr]
\,,\end{align}
where $h_s(\hm_b)$ is the two-loop result for massless quarks given in
\refcite{Ligeti:2008ac} and $f_{2s}(\rho)$ is extracted from the results of
\refcite{Asatrian:2006rq}.
At the same time, all SCET ingredients are defined for $n_f = 3$ massless flavors.

\subsubsection{Nonsingular contributions}

The remaining nonsingular terms $W_{ij}^\nons(k)$ in \eq{Phat2} are included
using fixed-order perturbation theory. These terms are power-suppressed by
$k/\hm_b$ in the $B\to X_s\gamma$ peak region, but loose this suppression in the
tail of the spectrum where $k\sim \hm_b$. By using $C_7^\incl$ and $\cC_i$ in
\eq{Phat2}, the $W_{ij}^\nons$ are also $\mu$ independent order by order in
$\alpha_s$. We use the notation $\mu=\mu_{ns}$ for the residual scale dependence
in all nonsingular terms, and will vary this scale as part of our perturbative
uncertainty estimate. Up to $\ord{\alpha_s^2}$ we write
\begin{equation} \label{eq:W77nons}
\hm_b W_{ij}^\nons(\hm_b x)
= \frac{1}{(1-x)^3} \biggl\{\frac{\alpha_s(\mu_{ns})}{\pi}C_F\, w_{ij}^{\nons\one}(x)
+ \frac{\alpha_s^2(\mu_{ns})}{\pi^2} C_F \Bigl[
w_{ij}^{\nons\two}(x) + \frac12 \beta_0 \ln\frac{\mu_{ns}}{m_b}\, w_{ij}^{\nons\one}(x) + \Delta w_{ij}^{\nons}(x) \Bigr] \biggr\}
\,.\end{equation}
The NLO and NNLO coefficient functions, $w_{ij}^{\nons\one}(x)$ and
$w_{ij}^{\nons\two}(x)$, are determined by taking the full fixed-order results
for $\df\Gamma/\df E_\gamma$ calculated in the literature, reorganizing the
Wilson coefficients as in \eq{Phat2}, and then using \eq{dGamma} with $\cF(m_B -
2E_\gamma - k) = \delta(\hm_b - 2E_\gamma - k)$ and subtracting the fixed-order
singular terms predicted  by $W_{77}^\sing(k)$ at each order. When this
construction is carried out with the results consistently expressed in a
short-distance mass scheme, there is an additional $\ord{\alpha_s^2}$ correction
induced, which is denoted as $\Delta w_{ij}^{\nons}(x)$ in \eq{W77nons}.

Note that the extraction of the nonsingular corrections is somewhat nontrivial.
For example, if we take the full theory result in a short-distance mass scheme
and extract the coefficient of the terms proportional to ${\rm Re}[\bC_7\bC_8^*]$
(setting $\mu = \hm_b$), we find
\begin{align} \label{eq:78nons_example}
\frac{\hm_b}{2\Gamma_0} \frac{\df\Gamma}{\df E_\gamma}
  \Bigg|_{\overline C_7\overline C_8}
&= \frac{\alpha_s(\hm_b)}{4\pi} \Bigl[\mathrm{Re}(r_8^\one)\, \delta(x) + 4C_F\, w_{78}^{\nons\one}(x) \Bigr]
\\ & \quad
+ \frac{\alpha_s^2(\hm_b)}{(4\pi)^2}
\Bigl\{\mathrm{Re}(r_8^\two)\, \delta(x) + \mathrm{Re}(r_8^\one)\, 4 C_F
\bigl[ w_{77}^{\sing\one}(x) +w_{77}^{\nons\one}(x) \bigr]
 + 16 C_F \bigl[ w_{78}^{\nons\two}(x) + \Delta w_{78}^\nons(x) \bigr] \Bigr\}
 \nn\\
& \quad + \ord{\alpha_s^3}
\,.\nn
\end{align}
Here the $\delta(x)$ and $w_{77}^{\sing\one}(x)$ terms are both reproduced by
the singular $E_\gamma^3 \abs{C_7^\incl}^2 W_{77}^\sing$ term. Furthermore, the
$\mathrm{Re}(r_8^\one) w_{77}^{\nons\one}(x)$ term is reproduced by
$E_\gamma^3 \abs{C_7^\incl}^2 W_{77}^\nons$.  Only the remaining terms
contribute to $ W_{78}^\nons$, as indicated by their superscripts.

By far the numerically dominant nonsingular corrections come from $W_{77}^\nons$. Using as input the results
from \refscite{Melnikov:2005bx, Blokland:2005uk, Asatrian:2006sm}, we find that the one-loop and two-loop nonsingular coefficient functions are
\begin{align}
w_{77}^{\nons\one}(x)
&= - \frac{8 - 7 x + 2 x^2}{2} \ln x - \frac34 (1 - x) (5 - 3 x)
\,,\nn\\*
w_{77}^{\nons\two}(x)
&=
\frac12 \Bigl(C_F - \frac12 C_A \Bigr) \biggl\{
     (2 + x - x^2 - x^3) \Bigl[L_1(x) + \frac1x \Bigl(\ln x - \frac23\Bigr) L_3(x) \Bigr]
\nn\\* & \qquad
   + (2 - x + x^2) \bigl[2\Li_3(x) - \Li_2(x) \ln x \bigr]
   - x^2 \bigl[\Li_3(x^2) - \Li_2(x^2)\ln x \bigr]
   + (-15 + 11 x - 3 x^2) \frac{\zeta_3}{2} \biggr\}
\nn\\ & \quad
+ C_F \biggl\{
     (2 + 10 x - x^2) \frac{1}{4x} \Bigl[\Li_3(1-x) + \Li_3(x) - \Li_2(x) \ln x - \frac{1}{2} \ln(1-x)\ln^2 x - \zeta_3 \Bigr]
\nn\\* & \qquad
   + \frac{8 - 32 x + 8 x^2 + 47 x^3 - 46 x^4 - 2 x^5 + 8 x^6}{12 (1 - x)} L_2(x)
   + \frac{-6 + 28 x - 9 x^2 - 9 x^3 - 2 x^4 + x^5}{12 (1 - x)} L_3(x)
\nn\\* & \qquad
   + \frac{8 - 13 x + 9 x^2 - 3 x^3}{4 (1 - x)} \ln^3 x
   + \frac{195 - 405 x + 266 x^2 - 66 x^3 + 2 x^4 - x^5}{24 (1 - x)} \ln^2 x
\nn\\* & \qquad
   + \Bigl[\frac{725 - 572 x + 29 x^2 + 28 x^3 + 32 x^4}{48} + (7 - 7 x + x^2) \frac{\pi^2}{12}\Bigr] \ln x
\nn\\* & \qquad
   + \frac{451 - 1531 x + 1553 x^2 - 525 x^3 - 16 x^4 + 32 x^5}{96 (1 - x)}
   + \frac{59 - 149 x + 150 x^2 - 53 x^3 - 2 x^4 + x^5}{12 (1 - x)} \frac{\pi^2}{6} \biggr\}
\nn\\ & \quad
+ C_A \biggl\{
     \frac{-4 + 12 x + 8 x^2 - 11 x^3 + 3 x^4 + x^5}{12} L_2(x)
   + \frac{-3 - 10 x + 15 x^2 + 3 x^3 + 2 x^4 - x^5}{24 (1 - x)} L_3(x)
\nn\\* & \qquad
   - \frac{x^2}{8} \ln^3 x
   + \frac{24 + 9 x + 4 x^2 + x^3 - x^4}{48} \ln^2 x
\nn\\* & \qquad
   + \Bigl[ \frac{-110 + 188 x - 5 x^2 - 45 x^3 + 10 x^4 + 4 x^5}{48 (1 - x)} + (9 - 7 x + 3 x^2) \frac{\pi^2}{24} \Bigr] \ln x
\nn\\* & \qquad
   + \frac{-10 + 129 x - 65 x^2 + 4 x^3 + 2 x^4}{48}
   + \frac{22 - 67 x + 54 x^2 - 7 x^3 + 2 x^4 - x^5}{24 (1 - x)} \frac{\pi^2}{6} \biggr\}
\nn\\ & \quad
+ \beta_0 \biggl\{
\frac{2 + 2 x - x^2}{8x}\Bigl[\Li_2(1-x) - \frac{\pi^2}{6} \Bigr]
   + \frac{3(8 - 7 x + 2 x^2)}{16} \ln^2 x
   + \frac{1 - 45 x + 45 x^2 - 19 x^3}{48 (1 - x)} \ln x
\nn\\* & \qquad
   + \frac{-63 + 92 x - 41 x^2}{32} + (8 - 7 x + 2 x^2) \frac{\pi^2}{48} \biggr\}
\,.\end{align}
To write $w_{77}^{\nons\two}(x)$ we defined the following functions of $x$,
which diverge at most logarithmically for $x\to 0$
\begin{align}
L_1(x)
&= \frac1x \biggl\{
   \Li_3(1-x)
   + 2 \Li_3\Bigl(\frac{1}{1 + x}\Bigr)
   - 2 \Li_3\Bigl(\frac{1 - x}{1 + x}\Bigr)
   + \frac14 \Li_3\Bigl[\Bigl(\frac{1 - x}{1 + x}\Bigr)^2\Bigr]
   - \frac{1}{6} \bigl[2 \ln(1+x)^2 - \pi^2\bigr]\ln(1+x)
   - \frac{5\zeta_3}{4}
\biggr\}
\,,\nn\\
L_2(x) &= \frac{1}{2x^3} \Bigl\{\Li_2(x^2) + 2\bigl[x^2 + \ln(1 - x^2)\bigr] \ln x  - x^2 \Bigr\}
\,,\nn\\
L_3(x) &= \frac12 \Li_2(x^2) - \Li_2(x) + \ln(1+x)\ln x
\,.\end{align}

Above we mentioned that a factor of $E_\gamma^3 \propto (1-x)^3$ was universal
for the $77$ contributions. It follows that $(1-x)^3 W_{77}^\nons$ should also
vanish as $(1-x)^3$ as $x\to 1$, and hence that the $w_{77}^\nons(x)$
coefficients should vanish like $(1-x)^3$ for $x\to 1$ to cancel the overall
factor $1/(1-x)^3$ in \eq{W77nons}. Expanding the above results in the limit
$x\to 1$, we find
\begin{align}
w_{77}^{\nons\one}(x)
&=  \frac{9}{4} (1 - x)^3 + \mathcal{O}[(1-x)^4]
\,,\nn\\
w_{77}^{\nons\two}(x)
&= \biggl[
     C_F \Bigl(-\frac{4933}{1728} - \frac{3\pi^2}{8} + \frac{\zeta_3}{2} \Bigr)
   + C_A \Bigl(\frac{599}{576} - \frac{3\pi^2}{16} - \frac{\zeta_3}{4}\Bigr)
   + \beta_0 \Bigl(\frac54 - \frac{\pi^2}{24}\Bigr)
\biggr] (1 - x)^3 + \mathcal{O}[(1-x)^4]
\,.\end{align}
If we would expand the $E_\gamma^3$ in the singular SCET contribution
$W_{77}^\sing$, then this would modify the nonsingular contribution, such that
it would not vanish like $E_\gamma^3$ either. In this situation, as was also
noted in \refcite{Misiak:2008ss}, the proper $E_\gamma^3$ behavior of the
spectrum would be obtained only by nontrivial cancellations between the singular and
nonsingular contributions. Although formally the difference between these approaches
corresponds to a different treatment of nonsingular corrections, this difference
can be numerically important even to rather low values of $x$ because of the
third power, and the fact that the resummation in the singular terms can
potentially spoil the cancellation at small $x$. For this reason, our approach
of keeping the $E_\gamma^3$ prefactor unexpanded is preferred.

For the remaining nonsingular coefficient functions, the fixed-order $\overline C_7 \overline C_8$ result from
\refscite{Ewerth:2008nv, Asatrian:2010rq} allows us to extract
$w_{78}^{\nons\one}$ and $w_{78}^{\nons\two}$.  Although both of these coefficients are used in our analysis, for brevity of the presentation we only quote here the first-order term
\begin{align}
w_{78}^{\nons\one}(x)
&= \frac{x}{3 (1 - x)} \ln x + \frac{5 - 2 x + x^2}{12}
\,.
\end{align}
Finally, for the remaining nonsingular structures, the full theory results at one loop are
well known~\cite{Ali:1990tj, Ali:1990vp, Ali:1995bi, Pott:1995if}. They enable us to determine
the following $\ord{\alpha_s}$ nonsingular coefficient functions
\begin{align}
w_{88}^{\nons\one}(x)
&= \frac{1}{36(1-x)} \Bigl[2 (1 + x^2) \Bigl(\ln x - 2 \ln\frac{\mu}{m_b}\Bigr) - 3 - 7 x^2 + 2 x^3\Bigr]
\,,\nn\\
w_{72}^{\nons\one}(x)
&= - \frac{8}{3} \rho^2\, \cG_1\Bigl(\frac{1-x}{4\rho}\Bigr)
\,,\qquad
w_{71}^{\nons\one}(x)
= -\frac{1}{6} w_{72}^{\nons\one}(x)
\,,\qquad
w_{18}^{\nons\one}(x)
= -\frac{1}{6} w_{82}^{\nons\one}(x) = \frac{1}{18} w_{72}^{\nons\one}(x)
\,,\nn\\
w_{22}^{\nons\one}(x)
&= \frac{4}{9} \rho\, \cG_2\Bigl(\frac{1-x}{4\rho}\Bigr)
\,,\qquad
w_{11}^{\nons\one}(x)
= -\frac{1}{6} w_{12}^{\nons\one}(x)
= -\frac{1}{6} w_{21}^{\nons\one}(x)
= \frac{1}{36} w_{22}^{\nons\one}(x)
 \,,
\end{align}
where $\rho = \hm_c^2/\hm_b^2$ and the charm-loop functions $\cG_{1,2}(u)$ are given by
\begin{align}
\cG_1(u)
&= \int_0^u \!\df u'\, \mathrm{Re}\bigl[\cG(u') + u'\bigr]
\,, 
& \cG_2(u)
&= \int_0^u \!\df u'\, (1 - 4\rho\, u')\biggl\lvert\frac{\cG(u')}{u'} + 1\biggr\rvert^2
\,,
\nn\\
\cG(u) &= \begin{cases}
- \bigl[ \arctan\sqrt{u/(1-u)} \bigr]^2 \,,  \quad&  u\leq 1
\,, \\[0.5ex]
\bigl[\ln\bigl(\sqrt u + \sqrt{u-1}\bigr) -i\pi/2 \bigr]^2 \,, \quad&  u>1
\,.\end{cases}
\end{align}
The corresponding $\ord{\alpha_s^2}$ nonsingular coefficient functions
$w_{ij}^{\nons\two}$ are not yet fully known. However, we stress that analogous
to the $\bC_7 \bC_8$ contribution in \eq{78nons_example}, all singular
contributions as well
as a subset of the nonsingular contributions appearing at two loops that behave
$O_7$-like are already accounted for via the $\abs{C_7^\incl}^2 (W_{77}^\sing +
W_{77}^\nons)$ term. For the remaining two-loop contributions $w_{ij}^{\nons\two}$
we use the known results for the $\alpha_s^2 \beta_0$ terms obtained from the full theory results
of \refscite{Ligeti:1999ea, Ferroglia:2010xe, Misiak:2010tk}, and thus leave out
contributions with the color structure $C_A$. Again for brevity, we do not list
here the results for these coefficient functions. The nonsingular corrections
for $i,j = 3,4,5,6$ are known to be very small~\cite{Pott:1995if} and are neglected.

\subsubsection{Scale choices and estimation of perturbative uncertainties}
\label{sec:pertunc}

We now discuss our treatment of the central scales $\mu_i$ and their variations
used to estimate perturbative uncertainties. The soft and jet scales take different
forms in the three parametrically distinct regions of the spectrum:
\begin{alignat}{9}
1)\,\ & \text{SCET shape function region: }   & \lqcd & \sim (m_B-2 E_\gamma) \ll \hm_b \,,
  \nn \\
2)\,\ & \text{Shape function OPE: } & \lqcd &\ll (m_B-2 E_\gamma) \ll \hm_b \,,
  \nn \\
3)\,\ & \text{Local OPE: }   & \lqcd &\ll (m_B-2 E_\gamma) \sim \hm_b \,.
\end{alignat}
This can be properly accounted for by using profile functions, $\mu_S=\mu_S(E_\gamma)$ and
$\mu_J=\mu_J(E_\gamma)$, as discussed in \refcite{Ligeti:2008ac} (see also~\refcite{Abbate:2010xh}).
The hard scale $\mu_b\sim \hm_b$ is independent of $E_\gamma$. For the remaining
scales we use
\begin{align} \label{eq:profiles}
\mu_S(E_\gamma) &=
  \begin{cases}
   \mu_0 &  E_1 \le E_\gamma \ \\
   \mu_0 + (\mu_b-\mu_0) \frac{2(E_\gamma-E_1)^2}{(E_2-E_1)^2}
     &  \frac{1}{2}(E_1+E_2) \le E_\gamma < E_1 \ \\[5pt]
   \mu_b - (\mu_b-\mu_0) \frac{2(E_\gamma-E_2)^2}{(E_2-E_1)^2}
     &  E_2 \le E_\gamma < \frac{1}{2}(E_1+E_2) \ \\
   \mu_b &  E_\gamma < E_2
  \,,\end{cases}
\nn\\
\mu_J(E_\gamma) &= \bigl[\mu_S(E_\gamma) \bigr]^{(1 - e_J)/2}\, \mu_b^{(1 + e_J)/2}
 \,, \nn\\[5pt]
 \mu_{ns}(E_\gamma) &= \bigl[\mu_J(E_\gamma) \bigr]^{(1 - e_{ns})/2}\, \mu_b^{(1 + e_{ns})/2}
  \,.
\end{align}
The constant parameters $\mu_0$, $\mu_b$, $E_{1}$, $E_2$, $e_J$,
and $e_{ns}$ can be varied to assess perturbative uncertainties. In
\eq{profiles} the soft scale $\mu_S(E_\gamma)$ takes the value $\mu_S =
\mu_0\sim 1\,{\rm GeV}\gtrsim \Lambda_{\rm QCD}$ in the SCET region given by
$E_1 \le E_\gamma$.  In the local OPE region, $E_\gamma < E_2$,
all the scales become equal, $\mu_S = \mu_J = \mu_{ns} = \mu_b$, which turns off
the resummation and is crucial for the singular and nonsingular contributions to
properly recombine to reproduce the local OPE prediction for the spectrum.  In
between these two we have a transition region where we join the soft scales in a
smooth manner, as given by the quadratic functions of $E_\gamma$ shown in
\eq{profiles}.  Since the transition scales $E_1$ and $E_2$ are
not very widely separated, there is no need to separately implement a shape
function OPE scaling region for the soft scale, noting that it is anyway well
captured by the form of $\mu_S(E_\gamma)$ used in the transition.
The parameters $e_J$ and $e_{ns}$ provide a means to
independently vary the jet and nonsingular scales when assessing perturbative
uncertainties. By default we
have $e_J=e_{ns}=0$. For $\mu_J$ this gives the geometric mean of the soft and
hard scales as required. For $\mu_{ns}$ we choose our default as the geometric
mean between the hard and jet scales, and we will vary this choice up to
the hard scale and down to the jet scale. This allows us to capture the fact
that the nonsingular perturbative series are sensitive to lower scales than the
hard scale (as would be made explicit in subleading power factorization
theorems for these terms).

Taken together we consider a total of $3^5 = 243$ different variations for the profile
parameters to assess the perturbative uncertainty, given by the choices
\begin{align}
  \mu_b &= \{ 4.7, 2.35, 9.4\} \, {\rm GeV} \,,
 & \mu_0 & = \{ 1.3, 1.1, 1.8\} \, {\rm GeV} \,,
 & E_1 & = \{ 2.2, 2.1, 2.3 \} \, {\rm GeV} \,,
 & E_2 & = 1.6 \, {\rm GeV} \,,
  \nn\\
 e_J &= \{ 0, -1/3, +1/3 \} \,,
 & e_{ns} &= \{ 0, -1/2, +1/2 \} 
\,.
\end{align}
For each parameter, the first case in the list is the default central value, and
the next two are the variations. We do not vary $E_2$ since our fit analysis is
not sensitive to the uncertainty in the spectrum in the region $E_\gamma
\lesssim 1.6\,{\rm GeV}$.  To assess the theoretical uncertainty we separately
carry out the fit for each of these 243 cases and then consider the spread of
the results as giving the range of possibilities for the central values. The
results for these 243 fits are shown by the dark yellow shape function curves in
\fig{BtoXsgammaSF} and as the yellow scatter points in
\figs{BtoXsgammaresults}{la1}.  The theoretical uncertainty for a given quantity
is then obtained by using the largest absolute deviation of these results from
the default central value.

\subsection{Shape Functions}

\subsubsection{Shape-function basis}
\label{sec:basis}

We briefly summarize the functional basis used for expanding the shape function
in \eq{expdef}. For more details we refer to \refcite{Ligeti:2008ac}. The
orthonormal basis functions $f_n(x)$ are given by
\begin{equation}
f_n(x) = \sqrt{y'(x)}\,\phi_n[y(x)]
\,, \qquad
\phi_n(y) = \sqrt{\frac{2n+1}{2}}\, \frac{1}{2^n n!}\,\frac{\df^n}{\df y^n} (y^2 - 1)^n
\,,\end{equation}
where $\phi_n(y)$ is an orthonormal basis on $y\in[-1,1]$, given by the
normalized Legendre polynomials. The function $y(x)$ can be any variable
transformation that maps $x\in [0,\infty)$ to $y\in[-1,1]$, i.e., it has to
satisfy $y(0) = -1$, $y(\infty) = +1$, and $y'(x) > 0$.
Given any positive and normalized function $Y(x)$ on $x\in[0,\infty)$, we can
construct $y(x)$ from its integral
\begin{equation}
y(x) = -1 + 2 \int_0^x\!\df x'\, Y(x')
\,,\qquad
y'(x) = 2 Y(x)
\,.\end{equation}
With this construction we have
\begin{equation}
f_0^2(x) = y'(x)\, \phi_0^2[y(x)] = Y(x)
\,,\qquad
F_{00}(k) = \frac{1}{\lambda}\, Y\Bigl(\frac{k}{\lambda}\Bigr)
\,.\end{equation}
Hence, $Y(x)$ or equivalently $F_{00}(k)$ acts as the generating function for
the basis, for which we can use any suitable model function.

We consider the following functional forms
\begin{align}
Y_{\rm exp}(x, p) &= \frac{(p + 1)^{p+1}}{\Gamma(p+1)}\,x^p\, e^{-(p+1)x}
\,,\nn \\
Y_{\rm gauss}(x, p) &= \frac{2\,a^{p+1}}{\Gamma[(1+p)/2]}\, x^p\, e^{-a^2 x^2}
\,,\qquad
a = \frac{\Gamma(1+p/2)}{\Gamma[(1+p)/2]}
\,.\end{align}
where the parameter $p$ determines the behavior of $F_{00}(k) \sim k^p$ for
$k\to 0$. As explained in \refcite{Ligeti:2008ac}, for integer $p$ we need at
least $p \geq 3$ to ensure that after short-distance subtractions, which involve
taking two derivatives of $\widehat F(k)$, the spectrum vanishes at the
kinematic endpoint. We have tested the three functional forms $Y_{\rm exp}(x,
3)$, $Y_{\rm exp}(x, 4)$, $Y_{\rm gauss}(x, 3)$ in the pre-fit. Of these,
$Y_{\rm exp}(x, 3)$ provides the best pre-fits and is thus used as the default
functional form.

\subsubsection{Treatment of leading and subleading contributions to $\cF(k)$}
\label{sec:Fmoments}

Our definition of the shape function ${\cal F}(k)$, appearing in the leading power contributions to the cross section, absorbs the non-resolved subleading power shape functions appearing in $B\to X_s\gamma$. This induces corrections in the formulas for the moments of ${\cal F}(k)$ which are used in our analysis.

Taking a set of values $\{c_n\}$ as input, from a fit or otherwise, the $i$th moment of $\cF(k)$ is given by
\begin{equation}\label{eq:momdef}
M^i[\cF]
= \inte{k} k^i \cF(k) = \sum_{m,n} c_m c_n \inte{k} k^i F_{mn}(k)
\equiv \sum_{m,n} c_m c_n M_{mn}^i
\,.\end{equation}
Here in the second step we inserted the basis expansion for $\cF(k)$,
and in the last relation we defined the moment matrices $M^i_{mn} \equiv M^i[F_{mn}]$
as the moments of the $F_{mn}(k)$ basis functions defined in \eq{Fmn_def}.

Theoretically moments of $\cF(k)$ up to $\ord{\lqcd^3}$ are given in terms of HQET hadronic parameters by~\cite{Tackmann:2005ub, Ligeti:2008ac}
\begin{align} \label{eq:moments}
M^0[\cF] &= \sum_{n} |c_n|^2 = 1 + \ord{\alpha_s \lqcd^2/\hm_b^2}
\,,\\
M^1[\cF]  &= \sum_{m,n} c_m c_n M^1_{mn}
= m_B - \hm_b
  + \frac{-\hla_1 + 3\hla_2}{2\hm_b}
  + \frac{5\widehat\rho_1 - 3\widehat\rho_2}{6\hm_b^2}
  + \ord{\alpha_s \lqcd^2/\hm_b}
\,,\nn\\
M^2[\cF] &= \sum_{m,n} c_m c_n M^2_{mn}
 = - \frac{\hla_1}{3}
   + \frac{\widehat\rho_1 + 3\widehat\rho_2}{3\hm_b}
   - (m_B - \hm_b)^2
   + 2(m_B - \hm_b) M^1[\cF]
   + \ord{\alpha_s \lqcd^3/\hm_b}
\,,\nn\\
M^3[\cF] &= \sum_{m,n} c_m c_n M^3_{mn}
 = \frac{\widehat\rho_1}{3}
   + (m_B - \hm_b)^3
   - 3(m_B - \hm_b)^2\, M^1[\cF]
   + 3(m_B - \hm_b)\,M^2[\cF]
   + \ord{\alpha_s \lqcd^4/\hm_b}
  \nn
\,,\end{align}
where $\hm_b$ is defined in the $1S$ scheme, and
\begin{equation} \label{eq:hla_def}
\hla_1 = \lambda_1^i(R) + \frac{\Tau_1 + 3\Tau_2}{\hm_b}
\,, \qquad
\hla_2 = \lambda_2(\mu) + \frac{\Tau_3 + 3\Tau_4}{3\hm_b}
\,.\end{equation}
Here, $\lambda_1^i(R)$ is defined in the invisible scheme~\cite{Ligeti:2008ac} with $R=1\,\GeV$,
and $\lambda_2(\mu)$ is the usual chromomagnetic matrix element (defined in the \MSbar
scheme).  The $\widehat\rho_i$ are matrix elements of local
dimension-6 operators in HQET in a suitable short-distance scheme,
and the $\Tau_i$ are matrix elements of time-ordered products~\cite{Gremm:1996df}.

The $1/\hm_b$ corrections in \eq{moments} arise from absorbing the subleading
shape functions into $\cF(k)$. By doing so, the moment expansion of $\cF(k)$
in \eq{moments} reproduces the complete $\ord{\lqcd^3/\hm_b^3}$ local OPE corrections
for $B\to X_s\gamma$~\cite{Bauer:1997fe, Tackmann:2005ub}. Note that the normalization
of $\cF(k)$ does not receive $\ord{\lqcd^2/\hm_b^2}$ and $\ord{\lqcd^3/\hm_b^3}$ corrections.
At $\ord{\alpha_s}$ and beyond, the subleading shape functions will in general involve
different perturbative prefactors than the leading shape function. Since
$\ord{\alpha_s \lqcd/\hm_b}$ corrections are beyond the order we are working, they
are also effectively absorbed into $\cF(k)$, which means the moments
receive relative corrections of $\ord{\alpha_s\lqcd/\hm_b}$ as indicated in \eq{moments}, which we neglect.
The exception is the normalization of $\cF(k)$, which only receives relative
$\ord{\alpha_s\lqcd^2/\hm_b^2}$ corrections.
(The $\ord{\lqcd^4}$ corrections to the moments are not included and not
explicitly indicated.)

It turns out that the numerical effect of the included subleading shape functions
on the first moment
is significant. For typical values of the $\hla_i$ and $\widehat\rho_i$ parameters,
the $1/\hm_b$ corrections to the first moment
contribute about $70-80\,\MeV$ causing a corresponding $70-80\,\MeV$ shift in
the extracted value of $\hm_b$. In other words, without including these effects
we would obtain a value of $\hm_b$ that is $70-80\,\MeV$ too small.

Values for $\hla_2$ and $\widehat\rho_2$ are obtained from meson mass relations as discussed below in \sec{la2_rho2},
whereas values of $\hm_b$, $\hla_1$, and $\widehat\rho_1$ are obtained whenever necessary from
$M^{1}[\cF]$, $M^{2}[\cF]$, and $M^{3}[\cF]$ by inverting the moment relations in \eq{moments}.
In particular, the moment relations are inverted when the current value of $\hm_b$ is needed inside the fit.

\subsubsection{Resolved-photon contributions}
\label{sec:resolved}

Considerable attention has been paid to the so-called resolved photon contributions,
as they were estimated to yield a 5\% theoretical uncertainty in the total rate,
not reducible below 4\%~\cite{Benzke:2010js}.
More recently \refcite{Gunawardana:2019gep} estimated their impact to be substantially
smaller. Using somewhat different considerations, we also find that these
contributions are not as large as estimated in \refcite{Benzke:2010js}.
From our analysis we find that the only marginally
relevant contributions are those related to the calculable $\ord{\lambda_2/\hm_c^2}$ corrections to the total rate~\cite{Voloshin:1996gw, Ligeti:1997tc, Grant:1997ec}, which enter via
the subleading shape function $g_{27}(k)$ as discussed below.

The resolved-photon contributions coming from
$O_8 O_7$~\cite{Kapustin:1995fk, Lee:2006wn} and $O_2 O_7$ are expected to be most significant~\cite{Benzke:2010js}, while contributions from $O_2 O_2$,
$O_2 O_8$, and $O_8 O_8$ can be neglected.

As pointed out in \refcite{Misiak:2009nr},
the potentially relevant $O_8 O_7$ contribution can be constrained using the
measured isospin asymmetry in $B\to X_s\gamma$, defined by
\begin{equation}
\Delta_{0-} = \frac{\Gamma(\bar B^0\to X_s\gamma) - \Gamma(B^-\to X_s\gamma)}
{\Gamma(\bar B^0\to X_s\gamma) + \Gamma(B^-\to X_s\gamma)}
\equiv \frac{\Gamma^0 - \Gamma^-}{\Gamma^0 + \Gamma^-}
\,.\end{equation}
To see this, we decompose these contributions to $\Gamma^0$ and $\Gamma^-$,
denoted as $\delta \Gamma^-$ and $\delta \Gamma^0$, according to the quark to
which the photon couples (besides the $O_7$ operator),
\begin{align}
\delta \Gamma^- &= Q_u \delta \Gamma^a + (Q_d+Q_s) \delta\Gamma^b 
  = Q_u (\delta \Gamma^a - \delta \Gamma^b)
\,, \nn\\
\delta \Gamma^0 &= Q_d \delta \Gamma^a + (Q_u+Q_s) \delta\Gamma^b 
  = Q_d  (\delta \Gamma^a - \delta \Gamma^b)
\,,\end{align}
where for $\delta\Gamma^a$ the photon couples to the valence quark flavor,
and for $\delta \Gamma^b$ to any non-valence flavors. For the non-valence contribution
we used that $SU(3)$ flavor symmetry implies that $\delta \Gamma^b$ is universal at leading order.
Since $Q_u+Q_d+Q_s=0$, both contributions are proportional to $\delta \Gamma^a - \delta \Gamma^b$.
The isospin asymmetry is given by~\cite{Misiak:2009nr}
\begin{equation}
\Delta_{0-}
= \frac{\delta\Gamma^0 - \delta\Gamma^-}{\Gamma^0 + \Gamma^-}
= - \frac{\delta\Gamma^a - \delta\Gamma^b}{\Gamma^0 + \Gamma^-}
\,.\end{equation}
Hence, the relative impact of these contributions to the isospin-averaged rate
is given by
\begin{equation}
\frac{\delta\Gamma^-+\delta\Gamma^0}{\Gamma^0 + \Gamma^-}
= \frac{1}{3}\frac{\delta\Gamma^a - \delta\Gamma^b}{\Gamma^0 + \Gamma^-}
= - \frac{\Delta_{0-}}{3}
= (0.16 \pm 0.71)\%
\,.\end{equation}
where we used the latest Belle measurement $\Delta_{0-}
= -(0.48 \pm 2.12)\%$~\cite{Watanuki:2018xxg}
(for $m_{X_s} < 2.8\,\GeV$ or equivalently $E_\gamma > 1.9\GeV$),
which is nearly a factor of three more precise than earlier results.
Hence, the $O_7 O_8$ contribution is experimentally constrained to be much
smaller than the current sensitivity, and can be neglected.

Concerning the $O_2 O_7$ contribution, unlike
\refcite{Benzke:2010js}, we treat the charm quark as heavy in our analysis,
which amounts to expanding the charm loop in $\lqcd/m_c$.  The
resulting contribution to the spectrum is then given in terms of an unknown
$\ord{\lqcd^2}$ subleading shape function $g_{27}(k)$ as
\begin{align} \label{eq:SSF27}
\frac{\df\Gamma_{g27}}{\df E_\gamma}
&= 2\Gamma_0 \frac{(2E_\gamma)^3}{\hm_b^3}\,\frac{1}{\hm_b^2}
\int\!\df k\, \widehat P_{27}(k)\,g_{27}(m_B - 2E_\gamma - k)
\,, \nn \\
\widehat P_{27}(k)
&= 2\mathrm{Re}(C_7^\incl)\Bigl(\cC_2 - \frac{\cC_1}{6}\Bigr)
\Bigl(-\frac{\hm_b^2}{18\,\hm_c^2}\Bigr)\, U_{\rm NLL}(k)
\,.\end{align}
In a complete factorization analysis, this contribution would involve some
evolution between hard, jet, and soft contributions, which is currently not known.
To provide some reasonable Sudakov suppression in the peak region,
which is important to avoid artificially enhancing this contribution relative
to the leading, resummed $W_{77}^\sing$ in \eq{Wsing77_resummed},  we include in it
the NLL evolution factor of the leading contribution given by the product
$U_{\rm NLL} = [U_H(\mu_b, \mu_J)\, U_S(k, \mu_J, \mu_S)]_{\rm NLL}$
with $\mu_{b, J, S}$ fixed to their central scales.

The subleading shape function $g_{27}(k)$ is not known, but its moments can
be calculated in terms of local matrix elements. To parametrize it, we
expand it as
\begin{equation}  \label{eq:g27expansion}
g_{27}(k) = \hla_2\sum_{n = 0}^2\, d_n\, F_n(k)
\,, \qquad
F_n(k) = \frac{1}{\lambda} f_n\Bigl(\frac{k}{\lambda}\Bigr)
\,,\end{equation}
where we use our default $\la = 0.55\,\GeV$ and the functional basis $f_n(x)$ is
generated from $Y_{\rm exp}(x, p)$. For our central results we use $p = 2$,
corresponding to linear scaling $F_n(k) \sim k$ for $k\to 0$,
and for the uncertainties we also use $p = 4$. The $\df\Gamma_{g27,n}/\df E_\gamma$
in \eq{expand_detail} are obtained by inserting the basis expansion in
\eq{g27expansion} into \eq{SSF27}.

\begin{figure}[t!]
\includegraphics[width=0.45\textwidth]{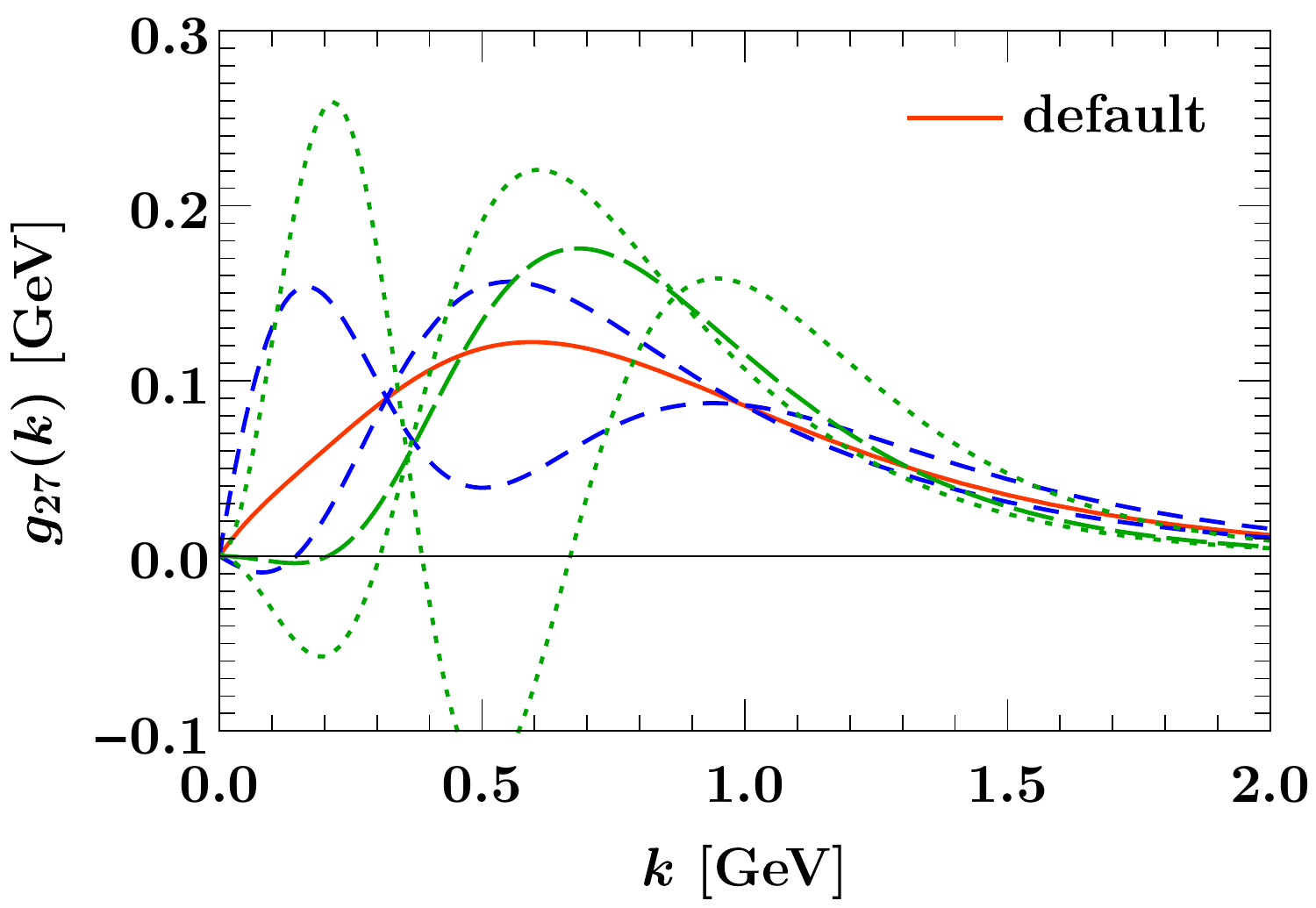}
\caption{Variations for $g_{27}(k)$ for fixed norm and first moment. The
solid orange line shows the default
choice ($p = 2$, $d_2 = 0$), the blue dashed lines the $d_2$ variations, and the
green lines the different basis ($p = 4$) for $d_2 = 0$ (long-dashed) and with
$d_2$ varied (dotted).}
\label{fig:g27}
\end{figure}

At present, we have no sensitivity to determine the basis coefficients $d_i$
from the data. Instead, we determine $d_0$ and $d_1$ for a given value of $d_2$
from the norm and first moment of $g_{27}(k)$, which are given by
\begin{equation}  \label{eq:g27moments}
M^0[g_{27}] = \hla_2
\,,\qquad
M^1[g_{27}] = \frac{\widehat\rho_2}{2}
\,.\end{equation}
For our central results we set $d_2 = 0$, and to estimate the uncertainties we
vary $d_2$ by an $\ord{1}$ amount to provide a reasonably large variation in the
shape of $g_{27}(k)$. The variations for $g_{27}(k)$ for fixed norm and first moment
are illustrated in \fig{g27}, with the solid orange line showing the default central choice.

The main impact of this contribution is due to the norm of $g_{27}(k) \sim
\hla_2$, which reproduces the well-known $\ord{\lambda_2/m_c^2}$ correction to
the total rate~\cite{Voloshin:1996gw, Ligeti:1997tc, Grant:1997ec}. The central
values and uncertainties used for $\hla_2$ and $\widehat\rho_2$ are discussed in
\sec{la2_rho2}. The uncertainties due to the unknown shape of $g_{27}(k)$ beyond
its norm and first moment are much smaller than the fit uncertainties. They
change the extracted $\abs{C_7^\incl}$ by at most $0.25\%$ and \mbS by
$4\,\MeV$, and are thus irrelevant at the present level of accuracy and can be
neglected. With more available data in the future, the $d_n$ coefficients could
also be included in the fit and constrained by the data.

\subsection{Numerical inputs}
\label{sec:inputs}

Here, we collect all numerical input values entering in our analysis.
The following values are taken from \refcite{pdg201819}:
\begin{alignat}{9}
\alpha_s^{(5)}(m_Z) &= 0.1181
\,, \qquad &
\alpha_{\rm em}(0) &= 1/137.036
\,, \qquad &
G_F &= 1.1663787\times 10^{-5}
\,, \nn \\
m_Z &= 91.1876\,\GeV
\,, \qquad &
m_W &= 83.379\,\GeV
\,, \qquad &
m_t &= 173\, \GeV
\,, \nn \\
\mcbar(\mcbar) &= (1.27\pm 0.02)\, \GeV
\,, \qquad &
\mbbar(\mbbar) &= (4.18^{+0.03}_{-0.02})\, \GeV
\,, \nn \\
\abs{V_{td}} &= 0.00896^{+0.00024}_{-0.00023}
\,, \qquad &
\abs{V_{ts}} &= 0.04133 \pm 0.00074
\,, \qquad &
\abs{V_{tb}} &= 0.999105 \pm 0.000032
\,, \nn \\
m_B &= 5.279\, \GeV
\,, \qquad &
\tau_B &= 1.581\, \mathrm{ps}
\,, \nn \\
\Delta m_B &= 45.22\, \MeV
\,, \qquad &
\Delta m_D &= 141.315\, \MeV
\,,\end{alignat}
where $m_B$, $\Delta m_B = m_{B^*} - m_B$, $\Delta m_D = m_{D^*} - m_D$ are averaged over charged and neutral
mesons.

\subsubsection{Wilson coefficients}

At and above the split-matching scale $\mu_0 = 4.7\,\GeV$, we always use the exact $4$-loop
running of $\alpha_s(\mu)$ with $n_f = 5$ flavors. To obtain the SM values of
the Wilson coefficients at $\mu_0$, we start from the full NNLO $\ord{\alpha_s^2}$ boundary
conditions~\cite{Bobeth:1999mk, Misiak:2004ew} at the weak scale
$\mu_\weak = 160\,\GeV$ and evolve them down to $\mu_0$ with the anomalous
dimensions up to $\ord{\alpha_s^3}$~\cite{Buras:1992tc, Ciuchini:1993vr, Chetyrkin:1996vx, Gambino:2003zm, Gorbahn:2004my, Gorbahn:2005sa, Czakon:2006ss}.
To perform the evolution, we use the exact numerical solution of the coupled RGE system.
For the boundary conditions we convert the above top-quark pole mass to
$\mtbar(160\,\GeV) = 163.3\,\GeV$.
For $\bC_{7,8}(\mu_0)$ we evolve $\mbbar(\mbbar)$ to $\mu_0$ with $4$-loop
running and for $\hm_b$ we use $\hm_b^{\rm fix} = 4.7\,\GeV$ (see \sec{as_masses} below).

The resulting NNLO Wilson coefficients $\cC_i$ are given by
\begin{alignat}{9}
\cC_1 &= -0.27189
\,, \qquad &
\cC_2 &= 1.00967
\,, \qquad &
\cC_3 &= -5.195\times 10^{-3}
\,, \qquad &
\cC_4 &= -8.060\times 10^{-2}
\,, \nn \\
\cC_5 &= 3.611\times 10^{-4}
\,, \qquad &
\cC_6 &= 9.375\times 10^{-4}
\,, \qquad &
\cC_7 &= -0.25594
\,, \qquad &
\cC_8 &= -0.13801
\,.\end{alignat}
For theory predictions at $\mu_0$, $\mu_b$, and below, we treat the coefficients as
fixed input values, i.e., we do not expand them in $\alpha_s$ against the
perturbative corrections they multiply. Varying $\mu_{\rm weak} = 160\,\GeV$
by a factor of two has very little impact on the $\bC_i(\mu_0)$.
In particular, the combinations of $\cC_{i\neq 7}$ that enter the theory
predictions for the photon energy spectrum via \eq{Phat} only vary below the percent level
when varying $\mu_{\rm weak}$, so we can safely neglect their uncertainties and
keep their values fixed in the fit. Similarly, their uncertainties are
irrelevant for the SM prediction of $C_7^\incl$.

\subsubsection{$\alpha_s$, $\widehat m_b$, $\widehat m_c$}
\label{sec:as_masses}

As discussed in \secs{C7incl}{pertspectrum}, we integrate out
both bottom and charm quarks at the scale $\mu_0 = \mu_b = 4.7\,\GeV$.
At $\mu_b$ and below we then always use the $3$-loop running for $\alpha_s(\mu)$,
consistent with the NNLL resummation, with $n_f = 3$ flavors. As the starting value
we use $\alpha_s^{(3)}(\mu_b) = 0.207$, which is obtained as follows.
We first use $\alpha_s^{(5)}$ to evolve $\mbbar(\mbbar)$ to $\mbbar(\mu_b)$
and use it to decouple the $b$ quark at $\mu_b$ to obtain $\alpha_s^{(4)}(\mu_b)$.
Then, we use $\alpha_s^{(4)}$ to evolve $\mcbar(\mcbar)$ to $\mcbar(\mu_b)$ and
use it to decouple the $c$ quark at $\mu_b$ to obtain $\alpha_s^{(3)}(\mu_b)$.
The decoupling and running of the \MSbar masses is performed at $4$ loops using
the \texttt{RunDec} package~\cite{Chetyrkin:2000yt}. The uncertainties on
$\mbbar(\mbbar)$ and $\mcbar(\mcbar)$ are negligible for this purpose, only
affecting the result in the 4th digit.

Several perturbative ingredients, such as the hard-matching coefficient $h_s$
and the nonsingular corrections $W_{ij}^\nons$, depend on the value of $\widehat
m_b$. As a result, the perturbative fit inputs $\df\Gamma_{ij,mn}/\df E_\gamma$
in \eq{expand_detail} have a mild dependence on $\hm_b$, which is subleading
compared to the dominant dependence entering through the shape function. To be
able to precompute the perturbative inputs, for simplicity we use a fixed value
$\hm_b = 4.7\,\GeV$ obtained from $\mbbar(\mbbar) = 4.18\,\GeV$ for their computation.
We have checked that changing this value by $\pm 50\,\MeV$, which also covers
our final fit result for \mbS, has a negligible impact on the fit. (For the
$\df\Gamma_{77}$ terms it changes the fit results for $\abs{C_7^\incl}$ by
$0.15\%$ and for \mbS by less than $1\,\MeV$.)

The corrections $\sim \cC_{1,2}$ also require a value for the charm-quark mass
$\hm_c$. In fact, the main dependence in the perturbative inputs on both
$\hm_b$ and $\hm_c$ comes from the dependence on $\rho =
\hm_c^2/\hm_b^2$ in $w^\nons_{i2}$ and $w^\nons_{22}$. (The sensitivity of $h_s$
on the precise value of $\rho$ is negligible.) While $\mcbar(\mcbar)$ is known
precisely, the perturbative scheme to use for $m_c$ is also relevant, and this
scheme dependence is only canceled by the still unknown non-$\beta_0$
$\ord{\alpha_s^2}$ corrections.
Since the difference between the bottom and charm pole masses,
$\delta_{bc}=m_b^{\rm pole}-m_c^{\rm pole}$, is free of renormalons, we use
$\widehat m_c \equiv \widehat m_b - \delta_{bc}$ as a suitable charm-mass
definition consistent with our treatment of the charm quark. We obtain a value
for $\delta_{bc}$ by converting $\mbbar(\mu_b)$ and $\mcbar(\mu_b)$
obtained above to the pole scheme and taking their difference. As expected,
while the individual values for $m_{b,c}^{\rm pole}$ strongly depend on the
order at which the conversion is performed, the resulting $\delta_{bc} = 3.4\,\GeV$ only
changes by about $15\,\MeV$ when the conversion is performed at two vs.\ three
loops and when using $\alpha_s^{(4)}(\mu_b)$ vs.\ $\alpha_s^{(5)}(\mu_b)$.
Accounting also for the uncertainties in $\mcbar(\mcbar)$ and $\mbbar(\mbbar)$,
we assign a conservative uncertainty of $50\,\MeV$ for $\delta_{bc}$.

To summarize, to compute all perturbative inputs for the fit, we use
\begin{align}
\alpha_s^{(3)}(\mu = 4.7\,\GeV) &= 0.207
\,, \qquad
\hm_b = (4.70 \pm 0.05)\, \GeV
\,,\qquad
\delta_{bc} = (3.40 \pm 0.05)\, \GeV
\,,\quad
\hm_c = \hm_b - \delta_{bc}
\,.\end{align}

\subsubsection{$\widehat\lambda_2$ and $\widehat\rho_2$}
\label{sec:la2_rho2}

The HQET parameters $\lambda_2$ and $\rho_2$ are needed in the moment relations
for $\cF(k)$ in \eq{moments}, and also in the moment constraints for $g_{27}(k)$
in \eq{g27moments}. Here, we discuss how we extract $\lambda_2$ and $\rho_2$ from
the measured heavy meson masses using relations that are
free of leading renormalon ambiguities.

The $B$ and $D$ meson masses can be expanded in $1/m_Q$,
following the notation of \refcite{Gremm:1996df}, as
\begin{align}\label{mesonmass}
m_M
&= m_Q + \bar\Lambda - \frac{\lambda_1 + d_M\, C_G(m_Q,\mu)\lambda_2(\mu)}{2m_Q}
 + \frac{\rho_1+d_M\rho_2}{4m_Q^2}
 - \frac{\Tau_1 + \Tau_3 + d_M(\Tau_2 + \Tau_4)}{4m_Q^2}
 + \ORD{\frac{\lqcd^4}{m_Q^3}, \frac{\alpha_s\lqcd^3}{m_Q^2}}
\,,\end{align}
where $m_M$ with $M= B, B^*, D, D^*$ are the masses of the lightest pseudoscalar and vector
mesons containing the heavy quark $Q$ with $d_{B,D}=3$ and $d_{B^*, D^*}=-1$ for the
pseudoscalar and vector mesons.
Here, $m_Q$ is the pole mass of the heavy quark $Q$. We have included
the \MSbar Wilson coefficient $C_G(m_Q,\mu)$ for the scale-dependent chromomagnetic
\MSbar matrix element $\lambda_2(\mu)$, but neglect Wilson coefficients for the
terms of higher order in $1/m_Q$.

Only three linear combinations of $\Tau_i$ appear in expressions for inclusive $B$
decays~\cite{Bauer:2002sh}, which are
\begin{align} \label{eq:taui_values}
\tau_1 &= \Tau_1 - 3\Tau_4 = (0.161 \pm 0.122)\, \GeV^3
\,, \nn \\
\tau_2 &= \Tau_2 + \Tau_4 = (-0.017 \pm 0.062)\, \GeV^3
\,, \nn \\
\tau_3 &= \Tau_3 + 3\Tau_4 = (0.213 \pm 0.102)\, \GeV^3
\,.\end{align}
The numerical values are taken from a global fit in the $1S$ scheme to semileptonic
$B\to X_c\ell\nu$ and radiative $B\to X_s\gamma$ moments~\cite{Amhis:2019ckw}.
The parameters are only weakly correlated and are insensitive to whether or not the
radiative moments are included in the fit~\cite{Phill}.

Denoting $\Delta m_B = m_{B^*} - m_B$ and $\Delta m_D=m_{D^*}-m_D$, we have
\begin{align} \label{eq:Deltam}
\Delta m_B
&=  2 C_G(m_b,\mu)\, \frac{\lambda_2(\mu)}{m_b} - \frac{\rho_2 - \tau_2}{m_b^2}
= 2 C_G^R(\hm_b,\mu,R)\, \frac{\lambda_2(\mu)}{\hm_b}
- \frac{\widehat\rho_2(R) - \tau_2}{\hm_b^2}
\,, \nn \\
\Delta m_D
&=  2 C_G(m_c,\mu)\, \frac{\lambda_2(\mu)}{m_c} - \frac{\rho_2 - \tau_2}{m_c^2}
= 2 C_G^R(\hm_c,\mu,R)\, \frac{\lambda_2(\mu)}{\hm_c}
- \frac{\widehat\rho_2(R) - \tau_2}{\hm_c^2}
\,.\end{align}
The first equality in each of these expressions involves
the \MSbar Wilson coefficient $C_G(m_Q, \mu)$, 
which has a $\ord{\lqcd/m_Q}$ renormalon ambiguity that is canceled by a corresponding ambiguity in $\rho_2$.
In the second equalities, we switched to renormalon-free quantities, where the Wilson coefficient $C_G^R(\hm_Q,\mu,R)$
and matrix element $\widehat \rho_2(R)$ are defined in the renormalon-free MSR scheme~\cite{Hoang:2009yr}.  To evaluate the Wilson coefficient $C_G$ or $C_G^R$ we use the fixed-order results evaluated at $\mu = R = \sqrt{\hm_b \hm_c}$ (which are known to 3-loops~\cite{Grozin:2007fh}) and evolve down to $\mu = R = 1\,\GeV$, using the \MSbar RGE or RRGE~\cite{Hoang:2008yj,Hoang:2009yr} respectively. To highlight the improvement obtained in the renormalon-free scheme, we note that in \MSbar we have $C_G(\hm_c,\sqrt{\hm_b \hm_c}) = \{1.26,\, 1.46,\, 1.69\}$ at 1, 2, and 3-loops respectively, whereas in MSR the results exhibit convergence with $C_G^R(\hm_c,\sqrt{\hm_b \hm_c},\sqrt{\hm_b \hm_c}) = \{0.991,\, 1.034,\, 1.045\}$ at 1, 2, and 3-loops.
Inverting the MSR results in \eq{Deltam}, we obtain
\begin{align}
\lambda_2(\mu)
&= \frac{1}{2}\, \frac{\hm_b^2\, \Delta m_B - \hm_c^2\, \Delta m_D}
{\hm_b C_G^R(\hm_b,\mu,R) - \hm_c C_G^R(\hm_c,\mu,R)}
\,, \nn \\
\widehat\rho_2(R) - \tau_2
&
= \frac{\hm_c C_G^R(\hm_c, \mu, R)\, \hm_b^2 \Delta m_B - \hm_b C_G^R(\hm_b, \mu, R)\, \hm_c^2 \Delta m_D} {\hm_b C_G^R(\hm_b,\mu,R) - \hm_c C_G^R(\hm_c,\mu,R)}
\,.\end{align}
With the input values for the meson and quark masses from above we then find
\begin{equation} \label{eq:la2_rho2_values}
\la_2(\mu = 1\GeV) = (0.128 \pm 0.005)\,\GeV^2
\,,\qquad
\widehat\rho_2(R = 1\GeV) - \tau_2 = (0.110 \pm 0.052)\,\GeV^3
\,.\end{equation}
Here the uncertainty in $\widehat\rho_2$ comes from varying the low scale down to $0.8\,\GeV$ and up to $1.3\,\GeV$, which in MSR provides an estimate of the size of neglected $\ord{\lqcd/\hm_{b,c}^3}$ corrections in \eq{Deltam}. We then combine this in quadrature with an estimate of $\ord{\alpha_s\lqcd^3/\hm_c^2}$ corrections to \eq{Deltam}. (The additional variation of the starting scale by a factor of two has a very small effect.)  These uncertainty estimates for higher-order terms are then propagated to obtain the uncertainty quoted for $\la_2(\mu=1\,\GeV)$ in \eq{la2_rho2_values}.

Finally, for the parameters appearing in the moment relations we take $\mu=\sqrt{\hm_b\hm_c}\simeq 2.5\,{\rm GeV}$ for $\lambda_2(\mu)$, still taking $R=1$ for $\widehat\rho_2(R)$, and hence will use
\begin{equation}
\hla_2 = \lambda_2(2.5\,\GeV) + \frac{\tau_3}{3\hm_b} = (0.135 \pm 0.009)\,\GeV^2
\,,\qquad
\widehat\rho_2(1\,\GeV) = (0.093 \pm 0.081)\,\GeV^3
\,,\end{equation}
where we added the uncertainties from \eqs{taui_values}{la2_rho2_values} in quadrature.

\end{document}